\documentclass[10pt,a4paper]{article}
\usepackage{subcaption}
\usepackage{xfrac}
\usepackage{wrapfig}
\usepackage{svg}
\usepackage[english]{babel}
\usepackage[latin1]{inputenc}
\usepackage{tipa}
\usepackage{amsfonts,amsbsy,bm,euscript,mathrsfs}
\usepackage{amssymb,stmaryrd,faktor}
\usepackage[tbtags]{amsmath}
\numberwithin{equation}{section}
\usepackage[title,titletoc]{appendix}
\usepackage[bookmarks=true,colorlinks=true,linkcolor=blue,citecolor=blue,urlcolor=blue,bookmarksnumbered]{hyperref}
\usepackage{tcolorbox}
\usepackage{braket}
\usepackage[nosort]{cite}
\usepackage{ulem}
\usepackage{yfonts}
\usepackage{multirow}
\usepackage{hhline}
\usepackage{scalerel}
\usepackage{mathtools}
\usepackage{xcolor}
\usepackage{cool}

\usepackage{comment}
\usepackage{enumitem}
\usepackage{mwe}

\usepackage{tikz}
\usetikzlibrary{decorations.markings}
\usepackage{caption}

\tikzstyle{layerpink} = [rectangle, rounded corners, minimum width=3cm, minimum height=1cm,text centered, draw=black, fill=red!30]

\tikzstyle{layerblue} = [rectangle, rounded corners, minimum width=3cm, minimum height=1cm,text centered, draw=black, fill=blue!30]

\tikzstyle{layergreen} = [rectangle, rounded corners, minimum width=3cm, minimum height=1cm,text centered, draw=black, fill=green!30]

\tikzstyle{arrow} = [thick,->,>=stealth]
\usepackage{listofitems} 
\usetikzlibrary{arrows.meta} 
\usepackage[outline]{contour} 
\contourlength{1.4pt}

\tikzset{>=latex} 
\usepackage{xcolor}
\colorlet{myred}{red!80!black}
\colorlet{myblue}{blue!80!black}
\colorlet{mygreen}{green!60!black}
\colorlet{myorange}{orange!70!red!60!black}
\colorlet{mydarkred}{red!30!black}
\colorlet{mydarkblue}{blue!40!black}
\colorlet{mydarkgreen}{green!30!black}
\tikzstyle{node}=[thick,circle,draw=myblue,minimum size=22,inner sep=0.5,outer sep=0.6]
\tikzstyle{node in}=[node,green!20!black,draw=mygreen!30!black,fill=mygreen!25]
\tikzstyle{node hidden}=[node,blue!20!black,draw=myblue!30!black,fill=myblue!20]
\tikzstyle{node convol}=[node,orange!20!black,draw=myorange!30!black,fill=myorange!20]
\tikzstyle{node out}=[node,red!20!black,draw=myred!30!black,fill=myred!20]
\tikzstyle{connect}=[thick,mydarkblue] 
\tikzstyle{connect arrow}=[-{Latex[length=4,width=3.5]},thick,mydarkblue,shorten <=0.5,shorten >=1]
\tikzset{ 
  node 1/.style={node in},
  node 2/.style={node hidden},
  node 3/.style={node out},
}
\def\nstyle{int(\lay<\Nnodlen?min(2,\lay):3)} 

\DeclareMathOperator{\swish}{swish}

\begin{document}

\begin{center}

{\Large\bfseries \vspace{0.2cm}
{\color{black} The R-mAtrIx Net }}
\vspace{1.5cm}

\textrm{\large 
Shailesh Lal${}^{b}$, \ 
Suvajit Majumder${}^{a}$\footnote{Corresponding Author} , \ 
Evgeny Sobko${}^{c}$
}

\vspace{2em}

\vspace{1em}
\begingroup\itshape
${}^{a}$ Centre for Mathematical Science, City, University of London
\\
Northampton Square, EC1V 0HB London, UK
\vspace{0.2cm}
\\
${}^{b}$ Yanqi Lake Beijing Institute of Mathematical Sciences and Applications (BIMSA)\\
Yanqi Island, Huairou District, Beijing 101408, China
\vspace{0.2cm}
\\
${}^{c}$ Laboratoire de Physique 
            de l'\'Ecole Normale Sup\'erieure, PSL University, CNRS\\
            24 rue Lhomond, 75005 Paris, France
\par\endgroup

\vspace{2em}

\begingroup\ttfamily
Email: suvajit DOT majumder AT city DOT ac DOT uk,
shaileshlal AT bimsa DOT cn,\\ evgenysobko AT gmail DOT com

\par\endgroup

\end{center}

\vspace{2em}

\begin{abstract}
 We provide a novel Neural Network architecture that can: i) output R-matrix 
 for a given quantum integrable
 spin chain, ii) search for an integrable Hamiltonian and the corresponding 
 R-matrix under assumptions of certain symmetries or other restrictions, iii) 
 explore the space of Hamiltonians around already learned models and 
 reconstruct the family of integrable spin chains which they belong to. The neural network training is done by minimizing loss functions encoding Yang-Baxter equation, regularity and other model-specific restrictions such as hermiticity. Holomorphy is implemented via the choice of activation functions. We demonstrate the work of our Neural Network on the two-dimensional spin chains of difference form. In particular, we reconstruct the R-matrices for all 14 classes. We also demonstrate its utility as an \textit{Explorer}, scanning a certain subspace of Hamiltonians and identifying integrable classes after clusterisation. The last strategy can be used in future to carve out the map of integrable spin chains in higher dimensions and in more general settings where no analytical methods are available.    
\end{abstract}

\newpage
\tableofcontents

\section{Introduction}
Neural Networks and Deep Learning 
have recently emerged as a competitive 
computational tool in many areas of theoretical physics and mathematics, in addition to their several impressive achievements in computer vision and natural language processing \cite{lecun2015deep}. In String Theory and Algebraic Geometry for instance, the application of these methods
was initiated in 
\cite{He:2017aed,carifio2017machine,krefl2017machine,ruehle2017evolving}.
Since then, deep learning has seen several interesting and remarkable applications
in the field, both on the computational front 
\cite{Brodie:2019dfx,Deen:2020dlf,He:2020lbz,Erbin:2020tks,Erbin:2021hmx,
Gao:2021xbs,
Ashmore:2021ohf,anderson2021moduli,douglas2022numerical,Larfors:2022nep} 
as well as 
towards the explication
of foundational questions \cite{He:2021eiu}. 
They have appeared in the context of Conformal Field Theory, critical phenomena, spin systems and Matrix Models 
\cite{morningstar2017deep,zhang2017machine,Chen:2020dxg,
Kuo:2021lvu,Basu:2022qaf,shiina2020machine,Han:2019wue}.
More generally, deep learning has found 
interesting applications in mathematics, ranging from  the solution of partial 
nonlinear differential equations \cite{e2017deep,raissi2017physics}, to
symbolic calculations \cite{lample2019deep} and to hypothesis generation 
\cite{davies2021advancing,He:2021oav}. Interestingly, deep learning is starting to play an increasingly important role in symbolic regression, i.e. the extraction of exact analytical expressions from numerical 
data \cite{udrescu2020ai}. 
While it is difficult to pin-point any one solitary reason for this confluence
of several fields into deep learning, there are some important themes that
do seem to play a recurring 
role. Firstly, deep neural networks are a highly flexible
parametrized class of functions and provide us an efficient way to
approximate various functional spaces and scan over them
\cite{cybenko1989approximation,hornik1989multilayer,lu2017expressive,
telgarsky2015representation}. The same neural network, as we shall shortly
see, can learn a Jacobi elliptic
function as easily as it does a trigonometric or an exponential function. Such approximations train well for a variety of loss landscapes, including non-convex ones. Secondly, over the previous many years,
robust frameworks for the design and optimization of neural networks have been
developed, both as an explication of best practices 
\cite{glorot2010understanding,hinton2012improving,srivastava2014dropout,ioffe2015batch,smith2018disciplined,zhang2019fixup}
and the development of 
standardized software for implementation \cite{chollet2018keras,
tensorflow2015-whitepaper, pytorch}. This has made it possible
to reliably train increasingly deeper networks which are optimized to carry out
increasingly sophisticated tasks such as the direct computation of Ricci flat
metrics on Calabi Yau manifolds 
\cite{Ashmore:2021ohf,anderson2021moduli,douglas2022numerical,Larfors:2022nep}
and the solution of differential equations without
necessarily providing the neural network data obtained from explicitly sampling
the solution. Further, in recent interesting developments, deep learning has been
applied to analyze various aspects of symmetry in physical systems
ranging from their classification to their automated detection
\cite{Chen:2020dxg, 
liu2022machine,liu2021machine,bondesan2019learning,melkosiamese,
forestano2023deep}.

The profound role played by symmetry in theoretical physics and mathematics
is hard to overstate.
Probably its most compelling expression in theoretical physics
is found in the bootstrap program
which rests on the idea that a theory 
may be significantly or even fully constrained just by the use of general 
principles and symmetries without analysis of the microscopic dynamics. 
For example, the S-matrix Bootstrap bounds the space of allowed S-matrices relying only on unitarity, causality, crossing, analyticity and global symmetries \cite{chew1962s,Eden:1966dnq,Kruczenski:2022lot}. 
This provides rigorous numerical bounds 
on the coupling constants 
and significantly restricts the space of self-consistent theories
\cite{Paulos:2016but,Paulos:2017fhb,He:2018uxa}. 
This line of 
consideration finds its ultimate realization in two dimensions 
once applied to integrable 
theories. Integrable Bootstrap complements the aforementioned constraints with 
one extra functional Yang-Baxter equation, manifesting the scattering 
factorization, which allows us to fix the S-matrix completely 
\cite{Zamolodchikov:1977nu}. The same Yang-Baxter(YB) equation appears in the 
closely related context of integrable spin chains. Now instead of S-matrix, it 
restricts the R-matrix operator whose existence allows one to construct a 
commuting tower of higher charges and prove integrability. Practically, one has 
to solve the functional YB equation in a certain functional space. There is no 
known general method to do so, and all existing approaches are limited in the 
scope of application and fall into three groups. The first class of methods is 
algebraic in nature, exploiting the symmetry of R-matrix  
\cite{Kulish:1981gi,Jimbo:1985ua}. The second approach aims to directly solve the 
functional equation or the related differential equation
\cite{vieira2018solving}. The 
third alternative utilizes the boost operator to generate higher charges and 
impose their commutativity \cite{de2019classifying,deLeeuw:2020ahe,deLeeuw:2020xrw}. 

In this paper we shall demonstrate how neural networks and deep learning provide
an efficient way to numerically solve the Yang-Baxter equation for integrable quantum 
spin chains. On an immediate front, we are motivated by recent interesting work on classical integrable systems using machine learning
\cite{liu2022machine,liu2021machine,bondesan2019learning,Krippendorf:2021lee}.
The approach taken in the work \cite{Krippendorf:2021lee} of learning 
classical Lax pairs for integrable systems by minimization of the loss 
functions encoding a flatness condition has a particularly close parallel 
to our approach. However, to the
best of our knowledge, the present work is the first attempt to apply machine
learning to quantum integrability, the analysis of  R-matrices and the 
Yang-Baxter equation.

Our analysis utilizes neural networks  
to construct an approximator for the R-matrix and thereby solve 
functional Yang-Baxter equation while also allowing for the imposition of 
additional constraints. 
We look into the  sub-class of all possible 
R-matrices, namely those that are regular and holomorphic, and incorporate the
Yang-Baxter equation 
into the loss function. Upon training for the given integrable 
Hamiltonian, we successfully learn the corresponding R-matrix to a prescribed 
precision. Using spin chains with two-dimensional space as a main playground we 
reproduce all R-matrices of difference form which was recently classified in 
\cite{de2019classifying}. Moreover, this \textit{Solver} can be turned into an \textit{Explorer} which scans the space (or a certain subspace) of all Hamiltonians looking 
for integrable models, which in principle allows us to discover new integrable 
models inaccessible to other methods.
Below we provide the summary of the Neural Network and its training, as well as an overview of the paper.
\paragraph{Summary of Neural Network and Training:}
The functional Yang-Baxter equation, see Equation \eqref{eq:ybequation} below,
is holomorphic in the spectral parameter $u\in\mathbb{C}$ and as such, holds 
over the entire complex plane. In this paper, we shall restrict our training to
the interval $\Omega = \left(-1,1\right)$ on the real line, but design our neural
network so that it analytically continues to a holomorphic function over the
complex plane. Each entry into the R-matrix is separately modeled by
multi-layer perceptrons (MLP) with two hidden layers of 50 neurons each, 
taking as input parameters the variable $u\in\Omega$. More details are available
in Section \ref{sec:mlqint} and Appendix \ref{app:nn_design}.
All the neurons are \texttt{swish}
activated \cite{ramachandran2017searching}, 
except for the output neurons which are \texttt{linear} 
activated. Training proceeds by optimizing the loss functions that
encode the Yang-Baxter equations \eqref{eq:lossybe}, regularity 
\eqref{eq:regularityloss}, and constraints on
the form of the spin chain Hamiltonian, for instance \textit{via}
\eqref{eq:hamiltonianloss}. Hermiticity of the 
Hamiltonian, if applicable, is imposed by the loss \eqref{eq:losshermiticity}. Optimization is done using \texttt{Adam}
\cite{kingma2014adam} with a starting learning rate of 
$\eta= 10^{-3}$ which is annealed $\eta= 10^{-8}$
in steps of $10^{-1}$ by monitoring the Yang-Baxter loss 
\eqref{eq:lossybe} on validation data for saturation. \texttt{Adam}'s
hyperparameters $\beta_1$ and $\beta_2$ are fixed to
$0.9$ and $0.999$ respectively.
In the 
following, we will refer to this learning rate policy as the
\textit{standard schedule}. We apply this framework to explore the 
space of R-matrices using the following strategies:

\begin{enumerate}
    \item \textbf{Exploration by Attraction}: The Hamiltonian loss 
    \eqref{eq:hamiltonianloss} is imposed by specifying target numerical values 
    for the two-particle Hamiltonian, or some ansatz/symmetries instead (like 
    6-vertex, 8-vertex, etc.). We also formally include here the ultimate case of 
    \textit{general search} when no restrictions are imposed on the Hamiltonian at 
    all. This strategy is predominantly used in our 
    Section~\ref{subsec:specific_search}.
    \item \textbf{Exploration by Repulsion}: We can 
    generate new solutions by \textit{repelling away} from an ansatz or a given 
    spin chain. This requires us to activate the loss function 
    \eqref{eq:replusionloss} for a few epochs in order to move from the specific Hamiltonian. This strategy is employed in Section \ref{subsec:Explorer}.
\end{enumerate}
Further, we also have two schemes for initializing training.
\begin{enumerate}
    \item \textbf{Random initialization}: We randomly initialize the weights of the neural network using He initialization \cite{he2016deep}. This samples
    the weights from either a uniform or a normal distribution centered around \(0\)
    but with a variance that scales as the inverse power of the layer width.
    \item \textbf{Warm-start}: we use the weights and biases for an already learnt solution\,.
\end{enumerate}
A brief overview of this paper is as follows. In section \ref{sec:integrability} we quickly introduce the R-matrix and other key
concepts from the quantum integrability of spin chains relevant to 
this paper. Particularly in subsection~\ref{subsec:YBE_sol_classes}, we 
review the classification program of 2-D  spin chains of difference form through 
the boost automorphism method \cite{de2019classifying}. 
Section \ref{sec:ML} contains a 
review of neural networks with a view towards machine learning the R-matrix 
given an ansatz for the 
two-particle Hamiltonian. Our methodology for this computation is provided in 
Section \ref{sec:mlqint}. 
We then present our results in section~\ref{sec:results}. Section~\ref{subsec:specific_search} focuses on hermitian XYZ and XXZ models (section~\ref{subsubsec:XYZ-XXZ-XXX}), and prototype examples from the 14 gauge-inequivalent classes of models in
\cite{de2019classifying}(section~\ref{subsubsec:fullclassification}). The latter sub-section also contrasts training behaviour for integrable and non-integrable models. Section~\ref{subsec:Explorer} presents a preliminary search strategy for new models which we illustrate within a toy-model setting: rediscovering the two integrable subclasses of 6-vertex Hamiltonians. Section \ref{sec:futuredirections} discusses ongoing 
and future research directions.

\section{An Overview of Spin Chains and Quantum 
Integrability}\label{sec:integrability}

Quantum integrability, like its classical counterpart, hinges on the presence of 
tower of conserved charges in involution, i.e. operators that mutually commute. 
In this paper we will consider quantum integrable spin chains and the goal of this 
section is to introduce such systems, and provide a brief overview of the 
R-matrix construction in their context.

The Hilbert space of the spin chain is 
a $L$-fold tensor product \(\mathbb{V}=V_1\otimes ... \otimes V_L\) of \(d\)-dimensional vector spaces \(V_i\sim V=\mathbb{C}^d\).
The Hamiltonian \(H\) of a spin chain with nearest-neighbour interaction 
is a sum of two-site Hamiltonians \(H_{i,i+1}\):
\begin{equation} \label{SpinChainHamiltonian}
H=\sum_{i=1}^{L}H_{i,i+1}\,,
\end{equation}
where we assume periodic boundary conditions : \(H_{L,L+1}\equiv H_{L,1}\).  The chrestomathic example of the integrable spin-chain is spin-1/2 XYZ model : 
\begin{equation}
    H = \sum_{i=1}^{L} \sum_{\alpha} J^\alpha S_i^{\alpha}S_{i+1}^{\alpha}\,,
\end{equation}
where $\alpha = \left\lbrace x,y,z\right\rbrace$ and \(S^\alpha_i\) are Pauli matrices acting in the two-dimensional space \(V_i=\mathbb{C}^2\) of \(i\)-th site. In particular case when \(J^x=J^y\) it reproduces XXZ model, while in the case of three equal coupling constants \(J^x=J^y=J^z=J\) the Hamiltonian reduces to the XXX spin chain. These famous magnet models are just a few examples of integrable spin chains and now we turn to the general construction.

The central element for the whole construction and proof of quantum 
integrability is the R-matrix operator \(R_{ij}(u)\) which acts in the tensor 
product \(V_i\otimes V_j\) of two spin sites \footnote{In general, 
the R-matrix is an analytic 
function of two complex arguments $u,v$, which can be viewed as momenta of two 
particles at the two sites.
Here and throughout the paper we shall exclusively focus our analysis to a 
restricted class of R-matrices of \textit{difference form} $R(u,v)=R(u-v)$  depending only on a single complex argument \(w=u-v\).}
and satisfies the Yang-Baxter equation:
\begin{equation}\label{eq:ybequation}
    R_{ij}(u-v)R_{ik}(u)R_{jk}(v)=R_{jk}(v)R_{ik}(u)R_{ij}(u-v)
\end{equation}
where the operators on the left and right sides act in the tensor product 
\(V_i\otimes V_j \otimes V_k\). The R-matrix is assumed to be an 
analytic function of the
spectral parameter \(u\). Further, in order to guarantee locality of the interaction in \eqref{SpinChainHamiltonian}, it must reduce to the permutation operator $P_{ij}$ when evaluated at $u=0$, i.e.
\begin{gather}\label{eq:regularity}
R_{ij}(0)=P_{ij}\,.
\end{gather}
This condition will be referred to as \textit{regularity} in the following sections.
We next turn to defining the monodromy matrix 
$\mathcal{T}_a(u)$.
This matrix, denoted by $\mathcal{T}_a(u)\in \mathrm{End}(V_a\otimes \prod_{i=1}^L
\otimes_i V_i)\times\mathbb{C}$, acts on the spin chain plus an auxiliary spin 
site labeled by $a$ with Hilbert space as $V_a\sim\mathbb{C}^d$. It is defined as 
a product of R-matrices $R_{a,i}(u)$ acting on the auxiliary site and one of the 
spin chain sites and is given by
\begin{equation}
    \mathcal{T}_a(u)= R_{a,L}(u)R_{a,L-1}(u)\dots R_{a,1}(u)\,.
\end{equation}
The transfer matrix $T(u)\in \mathrm{End}(\prod_{i=1}^L\otimes_i V_i)\times\mathbb{C}$ is obtained by taking a trace over the auxiliary vector space $V_a$ :
\begin{equation}
    T(u)=\mathrm{tr}_a(\mathcal{T}_a(u))\,.
\end{equation}
From the Yang-Baxter equation one can derive the following $R\mathcal{T}\mathcal{T}$ relation constraining monodromy matrix entries
\begin{equation}\label{eq:monodromy}
    R_{12}(u-v)\mathcal{T}_1(u)\mathcal{T}_2(v)=\mathcal{T}_2(v)\mathcal{T}_1(u)R_{12}(u-v)\,.
\end{equation}
This condition can be used to prove that the transfer matrices commute at different values of the momenta
\begin{equation}
    [T(u),T(v)]=0\,.
\end{equation}
The above condition implies that the transfer matrix $T(u)$ encodes all the commuting charges $\mathbb{Q}_i$ as series-expansion in $u$ :
\begin{equation}
    \log{T(u)}=\sum_{n=0}^{\infty} \mathbb{Q}_{n+1} \frac{u^n}{n!}\,.
\end{equation}
Hence we have \footnote{In practice, numerically it's more stable to work with the second formula on the right hand side than the first. }
\begin{equation}\label{eq:highercharges}
    \mathbb{Q}_{n+1} = \frac{d^n}{du^n}\log{T(u)}\vert_{u=0}\, = \frac{d^{n-1}}{du^{n-1}}\left(T^{-1}(u)\frac{d}{du}T(u)\right)\Bigg\vert_{u=0}\,.
\end{equation}
The Hamiltonian density \(H_{i,i+1}\) introduced earlier in equation \eqref{SpinChainHamiltonian} can be generated from the R-matrix using
\begin{equation}\label{eq:hamiltoniandensity}
    H_{i,i+1}=R_{i,i+1}^{-1}(0)\frac{d}{du}R_{i,i+1}(u)|_{u=0}=P_{i,i+1}\frac{d}{du}R_{i,i+1}(u)|_{u=0}
\end{equation}
where $P_{i,i+1}$ is the permutation operator between sites $i,i+1$.
Also, we emphasize that while the charges are conventionally computed in
Equation \eqref{eq:highercharges} at $u=0$, this computation can equally well
be done at generic values of $u$ to extract mutually commuting charges. The only
difference is we no longer recover the Hamiltonian directly as
one of the commuting charges.

Yang-Baxter equation \eqref{eq:ybequation} should be supplemented with certain 
analytical properties of R-matrix. For example, as was already mentioned, we assume that the R-matrix is a holomorphic function of 
spectral parameter \(u\) and equal to the permutation matrix  at \(u=0\) 
\eqref{eq:regularity}.
Furthermore, one can impose extra physical constraints like braided unitarity
\begin{equation} \label{eq:braidedunitarity}
    R_{12}(u)R_{21}(-u)=g(u)\mathbf{I}\,,\quad 
    g(u)=g(-u)\,,
\end{equation}
crossing symmetry \footnote{The explicit form of the crossing symmetry varies for 
the different classes of models.}, and possibly additional global symmetries. 
We shall also impose restrictions on the form of the resulting Hamiltonian. These
restrictions may follow from requirements such as hermiticity and from symmetries
of the spin chain. In addition, given a solution for the 
Yang-Baxter equation, one can generate a whole family of solutions by acting with the following transformations :
\begin{enumerate}
  \item Similarity transformation : \((\Omega\otimes \Omega) R(u) (\Omega^{-1}\otimes \Omega^{-1})\) where \(\Omega\in Aut(V)\) is a basis transformation. It transforms the commuting charges as \(\mathbb{Q}_{n}\rightarrow (\otimes^L \Omega)\mathbb{Q}_{n}(\otimes^L \Omega^{-1})\)
  \item Rescaling\footnote{For the general r-matrix of non-difference form \(R(u,v)\) there is a reparametrization freedom \(u\rightarrow f(u)\), however for the difference form \(R(u)\) it reduces just to rescaling} of the spectral parameter : \(u\rightarrow cu, \ \forall\, c\in \mathbb{C}\). This leads to a scaling in the charges as $\mathbb{Q}_{n}\rightarrow c^{n-1} \mathbb{Q}_{n}$
  \item Multiplication by any scalar holomorphic function \(f(u)\) preserving regularity condition  : \(R(u)\rightarrow f(u)R(u)\), \(f(0)=1\). This degree of freedom can be used to set one of the entries of \(R\)-matrix to one or any other fixed function.
  \item Permutation, transposition and their composition: \(PR(u)P,\ R(u)^T,\ PR^T(u)P\). They transform the commuting charges as well. The Hamiltonian $\mathcal{H}$ is transformed to $P\mathcal{H}P,P\mathcal{H}^TP,\mathcal{H}^T$ respectively.
\end{enumerate}
In general, one should always be careful 
of these redundancies when comparing a trained solution against analytic results.
Following \cite{de2019classifying}, we shall fix the above symmetries when 
presenting our results in section~\ref{subsubsec:fullclassification} and 
appendix~\ref{app:XYZ_nonXYZ_plots}. We look at gauge-equivalent solutions as 
well, by introducing similarity transformations 
in~\ref{subsubsec:fullclassification}. 

\subsection{Reviewing two-dimensional R-matrices of the difference form}\label{subsec:YBE_sol_classes}
We will illustrate the work of our neural network using two-dimensional spin chains as a playground. The regular difference-form integrable models in this context have recently been classified using the Boost operator in \cite{de2019classifying}. Here, we present a brief overview of the methods and results of this paper.   
Boost automorphism method allows one to find integrable Hamiltonians by reducing the problem to a set of algebraic equations. Let us focus on a spin chains with two-dimensional space \(V=\mathbb{C}^2\) and nearest-neighbour Hamiltonian  \eqref{SpinChainHamiltonian}. One formally defines the boost operator $\mathcal{B}$~\cite{tetel1982lorentz} as
\begin{equation}
    \mathcal{B}=\sum_{a=-\infty}^{\infty} a H_{a,a+1}\,,
\end{equation}
which generates higher charges $\mathbb{Q}_n$, $n\geq 3$, from the Hamiltonian $\mathbb{Q}_2$ via action by commutation:
\begin{equation}
    \mathbb{Q}_{r+1}=[\mathcal{B},\mathbb{Q}_r]\,.
\end{equation}
This  was used in \cite{de2019classifying} to successfully classify all 2-dimensional integrable Hamiltonians by solving the system of algebraic equations arising from imposing vanishing conditions on commutators between $\mathbb{Q}_i$, upto some finite value of $i$. Surprisingly it turns out that for the considered models, the vanishing of the first non-trivial commutator $[\mathbb{Q}_2,\mathbb{Q}_3]=0$ is a sufficient condition to ensure the vanishing of all other commutators. Then making an ansatz for the R-matrices and solving Yang-Baxter equation in the small \(u\) limit, the authors constructed the corresponding R-matrices and confirmed the integrability of the discovered Hamiltonians. 
The solutions can be organized into two classes: XYZ-type, and non-XYZ type,
distinguished by the non-zero entries appearing in the Hamiltonian.
\begin{equation}
    \label{eq:genclassification}
    H_{\mathrm{XYZ\, type}}=\begin{pmatrix}
        a_1&0&0&d_1\\
        0&b_1&c_1&0\\
        0&c_2&b_2&0\\
        d_2&0&0&a_2
    \end{pmatrix}\,,\qquad
    H_{\mathrm{non-XYZ\, type}}=\begin{pmatrix}
        a_1&a_2&a_3&a_4\\
        0&b_1&b_3&b_3\\
        0&c_1&c_2&c_3\\
        0&0&0&d_1
    \end{pmatrix}\,.
\end{equation}
Generically, all non-zero entries would be complex valued. Hermiticity, for the actual XYZ model and its XXZ and XXX limits, places additional constraints. Integrability also imposes additional algebraic constraints between the non-zero entries, none of which involve complex conjugation in contrast to hermiticity.
Amongst the XYZ type models, there are 8 distinct solutions each corresponding to  some set of algebraic constraints among the matrix elements of $H$. In particular, there is one 4-vertex model ($H_{4v}$) which is purely diagonal (see \eqref{eq:Hamiltonian4v}). Next,
there are two 6-vertex models ($H_{6v,1}$, $H_{6v,2}$) where $d_1$ and $d_2$
are constrained to vanish, among other conditions (see \eqref{eq:Hamiltonian6v,1}, \eqref{eq:Hamiltonian6v,2}). One of these, namely $H_{6v,1}$, is
a non-hermitian generalisation of the XXZ model. There are also
two 7-vertex 
models ($H_{7v,1}$, $H_{7v,2}$) where only $d_2$ vanishes (see \eqref{eq:Hamiltonian7v,1}, \eqref{eq:Hamiltonian7v,2}),
and three 8-vertex models ($H_{8v,1}$, $H_{8v,2}$, $H_{8v,3}$)
where all entries are non-zero(see \eqref{eq:Hamiltonian8v,1}, \eqref{eq:Hamiltonian8v,2}, \eqref{eq:Hamiltonian8v,3}). Here, $H_{8v,1}$ is the non-hermitian generalization of the XYZ model. Among
these classes the Hamiltonians are distinguished by additional algebraic 
constraints on the non-zero elements which we have enumerated in Appendix
\ref{app:XYZ_nonXYZ_plots}.
The corresponding R-matrices for these
models were obtained in \cite{vieira2018solving}.
The non-XYZ models are similarly divided into 6-classes with Hamiltonians 
$H_{class-1}\,,\dots H_{class-6}$ which have been explicitly enumerated in 
Equation \eqref{nonXYZhamiltonians}. Among these, the class 1 and class 2
Hamiltonians have rank less than four. 
For convenience, we also explicitly write down all of these R-matrices, both for the XYZ type and non-XYZ type models, 
in Appendix~\ref{app:XYZ_nonXYZ_plots}.
    
\section{Neural Networks for the R-matrix}\label{sec:ML}
This section reviews several essential facts about neural networks before
presenting our own network-design for deep-learning and the associated custom loss functions.
Further details regarding the network architecture and training schedule can be found in Appendix \ref{app:nn_design}.
\subsection{An overview of Neural Networks}
The central computation in this paper is the utilization of neural networks to construct R-matrices that
correspond to given integrable spin chain Hamiltonians. 
We therefore furnish a lightning overview of neural networks
in this section, along with the details of our
implementation of the neural network solver for Yang-Baxter equations. We will 
focus on \textit{dense
neural networks}, also known as \textit{multi-layer perceptrons} (MLPs), 
schematically displayed in Figure \ref{fig:densenet}. These networks consist of
an \textit{input layer} \(a^{in}\in \mathbb{R}^{n_0}\), followed by a series of \textit{fully connected layers} 
and terminate in an 
\textit{output layer} \(a^{out}\in \mathbb{R}^{n_{L+1}}\). Data is read in to the network at the input layer and the 
output is collected at the
output layer. There are $L$ fully connected layers in this network, where the $\ell$-th layer contains
$n_\ell$ neurons. 
Each neuron \(a_m^{(l)}\) in a \(l\)-th fully connected layer receives inputs from \textit{all} the neurons
in the previous $(l-1)$-th layer and the output of the neuron is in turn fed as an input to neurons in the succeeding
layer: 
\begin{figure}
    \centering
    \begin{tikzpicture}[x=2.2cm,y=1.4cm]
  \message{^^JNeural network with arrows}
  \readlist\Nnod{4,5,5,5,3} 
  
  \message{^^J  Layer}
  \foreachitem \N \in \Nnod{ 
    \edef\lay{\Ncnt} 
    \message{\lay,}
    \pgfmathsetmacro\prev{int(\Ncnt-1)} 
    \foreach \i [evaluate={\y=\N/2-\i; \x=\lay; \n=\nstyle;}] in {1,...,\N}{ 
      
      \node[node \n] (N\lay-\i) at (\x,\y) {$a_\i^{(\prev)}$};
      
      \ifnum\lay>1 
        \foreach \j in {1,...,\Nnod[\prev]}{ 
          \draw[connect arrow] (N\prev-\j) -- (N\lay-\i); 
        }
      \fi 
      
    }
    
  }
  
  \node[above=5,align=center,mygreen!60!black] at (N1-1.90) {input\\[-0.2em]layer};
  \node[above=2,align=center,myblue!60!black] at (N3-1.90) {hidden layers};
  \node[above=8,align=center,myred!60!black] at (N\Nnodlen-1.90) {output\\[-0.2em]layer};
\end{tikzpicture}
    \caption{The schematic for a Dense Neural Network, also known as a Fully Connected neural network. The 
    four-dimensional input $\left(a_1^{(0)}\ldots a_4^{(0)}\right)$ is fed \textit{via} the Input layer 
    (green) to a series of three Fully Connected layers (purple) containing four neurons each and finally
    feeds into the Output layer of three neurons (orange). Every neuron in a given layer receives inputs from
    all neurons in the preceding layer, and in turn, its output is passed as input to all neurons in the 
    next layer.}
    \label{fig:densenet}
\end{figure}
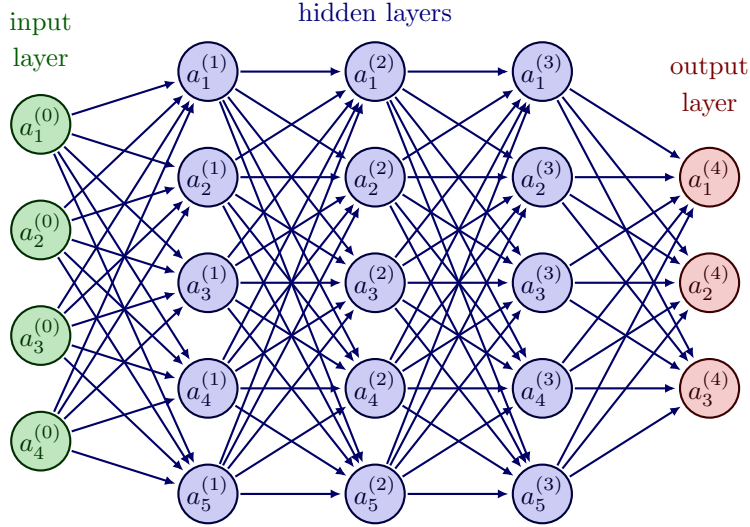

\begin{equation}\label{eq:fwdprop}
      \begin{pmatrix}
        a_{1}^{(\ell)} \\
        a_{2}^{(\ell)} \\
        \vdots \\
        a_{n_{\ell}}^{(\ell)}
      \end{pmatrix}
      =
       h \left[ 
      \begin{pmatrix}
        w^{(\ell)}_{1,0} & w^{(\ell)}_{1,1} & \ldots & w^{(\ell)}_{1,n_{\ell-1}} \\
        w^{(\ell)}_{2,0} & w^{(\ell)}_{2,1} & \ldots & w^{(\ell)}_{2,n_{\ell-1}} \\
        \vdots  & \vdots  & \ddots & \vdots  \\
        w^{(\ell)}_{n_{\ell},0} & w^{(\ell)}_{n_{\ell},1} & \ldots & w^{(\ell)}_{n_{\ell},n_{\ell-1}}
      \end{pmatrix}
      \begin{pmatrix}
        a_{1}^{(\ell-1)} \\[0.3em]
        a_{2}^{(\ell-1)} \\
        \vdots \\
        a_{n_{\ell-1}}^{(\ell-1)}
      \end{pmatrix}
      +
      \begin{pmatrix}
        b_{1}^{(\ell)} \\[0.3em]
        b_{2}^{(\ell)} \\
        \vdots \\
        b_{n_{\ell}}^{(\ell)}
      \end{pmatrix}
      \right],\, \ell =1,\ldots L+1\,,
\end{equation}
where $w^{(l)}\in \mathcal{M}(n_{l},n_{l-1},\mathbb{R})$ is a \textit{weight matrix}, $b^{(l)}\in \mathbb{R}^{n_l}$ - \textit{bias vector}, $h(z)$ is in general a non-linear, non-polynomial function 
known as the \textit{activation function} acting component-wise : 
\begin{equation}\label{eq:activation}
    h \begin{pmatrix}
        z_{1} \\[0.3em]
        z_{2} \\
        \vdots \\
        z_{n}
      \end{pmatrix}
    = \begin{pmatrix}
        h \left(z_{1}\right) \\[0.3em]
        h \left(z_{2}\right) \\
        \vdots \\
        h \left(z_{n}\right)
      \end{pmatrix}\,\,.
\end{equation}
In \eqref{eq:fwdprop} we also identify \(a^{(0)}=a^{in}\) and \(a^{(L+1)}=a^{out}\) with input and output layers respectively. Introducing shorthand notation for the affine transformations in equation~\eqref{eq:fwdprop} as \(A^{(\ell)}(a^{(\ell-1)})\equiv w^{(\ell)} a^{(\ell-1)}+b^{(\ell)}\), the neural network \(a^{out}(a^{in}):\ \mathbb{R}^{n_0}\rightarrow \mathbb{R}^{n_{L+1}}\) can be expressed as compositions of affine transformations, and activation functions:
\begin{gather}\label{NNcomposition}
a^{out}=h \circ A^{L+1} \circ h \circ A^{L} \circ ... \circ h \circ A^1 \circ a^{in}
\end{gather}
The output function of the neural network is tuned by tuning the
weights and biases. It is by now well established that 
such neural networks are a highly expressive framework capable of 
approximation of extremely complex
functions, and indeed there exists a series of 
mathematical proofs which attest to their
\textit{universal approximation} property, 
e.g. 
\cite{cybenko1989approximation,hornik1989multilayer,
lu2017expressive,hoffman2019robust,park2020minimum}. This property,
along with the feature learning capability of deep neural networks
is the key driver to the automated search for R-matrices which
we have implemented here.

Neural networks learn a target function $f(x)$ of the input data $x$
by optimizing a \textit{cost function} $\mathcal{L}$ which provides a measure of the discrepancy between the
actual and desired properties of the function $f$. The parameters $w$ and $b$ are then tuned to 
minimize this discrepancy. 
The best known and the most canonical 
examples of this are the \textit{supervised learning} problems where the
neural network is supplied with
data $\mathcal{D}=\left\lbrace\left(x,y\right)\right\rbrace$ consisting of
pairs of input vectors $x$ with
their expected output values $y$. The neural network then tunes its weights
and biases to minimize the cost function. Having done so, the output
function $f$ thus learned by the neural network obeys
\begin{equation}
    f\left(x\right) \approx 
    y \qquad \forall \qquad \left(x,y\right)\in \mathcal{D}\,,
\end{equation}
while allowing for the possibility of outliers.
A popular class of loss functions are 
\begin{equation}\label{eq:supervisedloss}
    \mathcal{L}\left(\left\lbrace w,b\right\rbrace \right) = \sum_{\left(x,y\right)\in\mathcal{D}}
    \left\vert y - y_{pred}\left(x\right)\right\vert^q\,,\qquad 
    y_{pred}\left(x\right) = f\left(x;w,b\right)\,,
\end{equation}
where $q=1$ corresponds to the \textit{mean average error} and $q=2$ to the 
\textit{mean square error}, respectively. 
We will shortly see that 
in contrast to the above classic supervised learning set-up, 
our loss functions impose constraints on the neural network output functions
rather than train on a dataset of input/output values for the functions
$\mathcal{R}\left(u\right)$ 
directly sample $\mathcal{R}\left(u\right)$ at various
values of $u$ for training.

\subsection{Machine Learning the R-Matrix}\label{sec:mlqint}
We are now ready to describe our proposed methodology for constructing 
R-matrices 
$\mathcal{R}\left(u\right)$ by optimizing a neural network using appropriate loss functions. An R-matrix has elements
$R_{ij}\left(u\right)$ at least
some of which are non-zero. In the following, we shall focus solely on the $R_{ij}\left(u\right)$ which are not identically zero as functions of
$u$. We also restrict the training to the real values of spectral parameter \(u\in
\Omega=(-1,1)\) and exclusively 
use holomorphic activations function in order to guarantee the holomorphy of the 
resulting R-matrix $\mathcal{R}\left(u\right)$. The matrix elements 
$\mathcal{R}_{ij}\left(u\right)$ of this R-matrix are modeled 
by neural networks as
\begin{equation}\label{eq:rvariabledesign}
\mathcal{R}_{ij}\left(u\right) = a_{ij}\left(u\right)+i\  
b_{ij}\left(u\right)\,:\,
\begin{tikzpicture}[baseline=(current  bounding  box.center)]
\coordinate (x1) at (-1.5,0.0);
    \coordinate (x2) at (0,1.5);
    \coordinate (xl2) at (-0.33,1.5-0.33);
    \coordinate (xn2) at (0,-1.5);
    \coordinate (xnu2) at (-0.33,-1.5+0.33);
    \coordinate (x4) at (1.5,1.5);
    \coordinate (xn4) at (1.5,-1.5);
    \coordinate (x5) at (3,0);
\draw [-] (x1) + (0.33,0.33) -- (xl2);
\draw [-] (x1) + (0.33,-0.33) -- (xnu2);
\draw [-] (x2) + (0.5,0) edge (1.0,1.5);
\draw [-] (x2) + (0.33,-0.33) edge (1.5-0.33,-1.5+0.33);
\draw [-] (xn2) + (0.33,0.33) edge (1.5-0.33,1.5-0.33);
\draw [-] (xn2) + (0.5,0.0) edge (1.0,-1.5);
\draw [-] (x4) + (0.33,-0.33) edge (3-0.33,0.33);
\draw [-] (xn4) + (0.33,0.33) edge (3-0.33,-0.33);
\draw (x1) circle (0.5);
\draw (x1) node {$u$};
\fill[ green,
nearly transparent]
{(x1) circle (0.5)};
\draw (x5) circle (0.5);
\draw (x5)+(0.05,0) node {$r_{ij}$};
\fill[red,
nearly transparent]
{(x5) circle (0.5)};
\draw (x2) circle (0.5);
\draw (x2)+(0.05,0) node {$a^{(1)}_{ij;1}$};
\fill[blue,
nearly transparent]
{(x2) circle (0.5)};
\draw (xn2) circle (0.5);
\draw [densely dotted] (0.0,0.75) -- (0.0,-0.75);
\draw (xn2)+(0.05,0) node {$a^{(1)}_{ij;50}$};
\fill[blue,
nearly transparent]
{(xn2) circle (0.5)};
\draw (x4) circle (0.5);
\draw (x4)+(0.05,0) node {$a^{(2)}_{ij;1}$};
\fill[blue,
nearly transparent]
{(x4) circle (0.5)};
\draw (xn4) circle (0.5);
\draw [densely dotted] (1.5,0.75) -- (1.5,-0.75);
\draw (xn4)+(0.05,0) node {$a^{(2)}_{ij;50}$};
\fill[blue,
nearly transparent]
{(xn4) circle (0.5)};
\end{tikzpicture}\quad; \quad 
r_{ij}=\left\lbrace a_{ij},b_{ij}\right\rbrace
\,.
\end{equation}
We have decomposed the matrix element \(\mathcal{R}_{ij}(u)\) into 
\(a_{ij}(u)+i\  b_{ij}(u)\) in order to learn complex-valued functions 
\(\mathcal{R}_{ij}\) while
training with real MLPs on the real interval $\Omega$. 
In this paper, purely for uniformity, we
have modeled each such \(a_{ij}(u)\) and \(b_{ij}(u)\) using an MLP 
containing two hidden layers of 50 neurons each and one \texttt{linear} activated
output neuron. We emphasize that the identification of \(a_{ij}(u)\) 
and \(b_{ij}(u)\) to 
real and imaginary parts of  \(\mathcal{R}_{ij}(u)\) is only valid over the
real line, and these functions separately continue into holomorphic functions
over the complex plane whose sum \(\mathcal{R}_{ij}(u)\) is holomorphic
by construction. Now, \(\mathcal{R}_{ij}(u)\) is required to solve the 
the Yang-Baxter equation \eqref{eq:ybequation} subject to \eqref{eq:regularity}.
We may also place constraints on the corresponding two-particle $H$ given by
\eqref{eq:hamiltoniandensity}. 
These criteria are encoded into loss functions which
the R-matrix \(\mathcal{R}_{ij}(u)\) aims to minimize by training. For example, in order to train \(\mathcal{R}_{ij}(u)\) to satisfy Yang-Baxter 
equation \eqref{eq:ybequation} for all values of spectral parameter \(u\) from 
the set \(\Omega \subset \mathbb{C}\) we introduce the following loss function :
\begin{equation}\label{eq:lossybe}
\mathcal{L}_{YBE} =
    \vert\vert 
    \mathcal{R}_{12}(u-v)\mathcal{R}_{13}(u)\mathcal{R}_{23}(v)-\mathcal{R}_{23}(v)\mathcal{R}_{13}(u)\mathcal{R}_{12}(u-v)\vert\vert
    \,,
\end{equation}
where \(||...||\) is a matrix norm defined as \(||A||=
\sum\limits_{\alpha,\beta=1}^n |A_{\alpha\beta}|\) for an 
complex-valued \(n\times n\)  matrix \(A\). During the forward propagation we sample a mini-batch of $u$ and $v$
values, from which the corresponding $u-v$ is constructed. Along this paper, the spectral parameters \(u\) and \(v\) run over the discrete set of 20000 points randomly chosen from the interval \(\Omega\). \footnote{The number of points used during the training bounds the precision which one can reach and in our case it will be of order \(10^{-4}\).}  The loss function \(\mathcal{L}_{YBE}\) is 
positive semi-definite, vanishing only when $\mathcal{R}\left(u\right)$ solves the Yang-Baxter equation. In principle, one may imagine a scan across the space
of all functions in which case, solutions of the Yang-Baxter equation would
minimize the loss \eqref{eq:lossybe} to zero.

In practice of course, one cannot scan across the space
of all functions and is restricted to a hypothesis class. Here the hypothesis
class is implicitly defined by the design of the neural network, the choice
of numbers of layers, number of neurons in each layer, as well as the
activation function. Varying the weights and biases of the neural network
allows us to scan across this hypothesis space. While in general
the exact R-matrix may not belong to this hypothesis class, 
and the loss function would
then be strictly positive, deep learning may allow us to approach the 
desired functions $\mathcal{R}\rightarrow R$ to a high degree of accuracy. In summary, if we
restrict to a hypothesis class which does not include an actual solution
of the Yang-Baxter equation then
\begin{gather}
\mathcal{L}_{YBE}\geq \epsilon=\min\limits_{\left\lbrace w',b'\right\rbrace} \mathcal{L}_{YBE}\left(\left\lbrace w',b'\right\rbrace \right) >0\,,
\end{gather}
where ideally \(\epsilon\) would be small, indicating that we have obtained
a good
approximation to the true solution.
We expect that scanning across wider and wider hypothesis classes 
would bring \(\epsilon\) closer and closer to zero. 
Further, while
the RTT equation \eqref{eq:monodromy}
follows from the Yang-Baxter equation \eqref{eq:ybequation},
it can also be imposed separately as a loss function on the network in order to improve the training :
\begin{equation}\label{eq:lossrtt}
\mathcal{L}_{RTT} =
    \vert\vert 
    \mathcal{R}_{12}(u-v)\mathcal{T}_1(u)\mathcal{T}_2(v)-\mathcal{T}_2(v)\mathcal{T}_1(u)\mathcal{R}_{12}(u-v)\vert\vert
    \,.
\end{equation}
Next, we have constraints that must be imposed on the R-matrix at $u=0$.
Following equation~\eqref{eq:regularity} and equation~\eqref{eq:hamiltoniandensity} in previous section, we require that \footnote{It is tempting to consider a variation of our
method
which involves \textit{residual learning} \textit{a la} the ResNet
family of networks \cite{he2016deep}. 
As opposed to learning deviations from identity, 
which is typically the approach adopted in the ResNet architecture,
we may define
\begin{equation}
    R\left(u\right) = P + \Tilde{R}\left(u\right)\,,
\end{equation}
where $\Tilde{R}\left(u\right)$ is the target function of the neural network,
which we design to identically output $\Tilde{R}\left(0\right) = 0$. While this is possible in principle, in practice it turns out
that since the neural network is learning a function in the vicinity of
$P$, which trivially minimizes the Yang-Baxter equation and all other 
constraints imposed, it almost invariably collapses 
to the trivial solution and learns $\Tilde{R}\left(u\right) = 0$
across all $u$. It would nonetheless be interesting to identify such 
architectures that successfully learn non-trivial R-matrices and this is
in progress.
}
\begin{equation}
    R\left(0\right) = P\,,\qquad  P\frac{d}{du}R(u)|_{u=0} = H\,,
\end{equation}
where $H$ is the two particle Hamiltonian. They both can be encoded in the loss function as
\begin{gather}\label{eq:regularityloss}
    \mathcal{L}_{reg} = || \mathcal{R}\left(0\right) -P
    ||, \\
    \mathcal{L}_{H} = || P\frac{d}{du}\mathcal{R}(u)|_{u=0} -H  || \,.\label{eq:hamiltonianloss}
\end{gather}
Here, we should mention that we have some flexibility in the manner in which we 
implement the Hamiltonian constraint  $\mathcal{L}_{H}$. 
Firstly, one can fix the exact numerical values for the entries of $H$ and learn 
corresponding R-matrix. We will also consider extensions of this loss function
where we supply only algebraic constraints restricting the search space for target
Hamiltonians to those with certain symmetries or belonging to 
certain gauge-equivalence classes.
In general, such Hamiltonian constraints give us
the requisite control to converge to the different classes of integrable
Hamiltonians, and we will name such regime as a \textit{exploration by attraction}. 

In the same spirit, when working with the XYZ spin chain or its XXZ and XXX 
limits, we 
also require
that the two-particle Hamiltonian computed from $R\left(u\right)$ is 
hermitian, i.e.,
\begin{equation}
    H=H^{\dagger}\,,
\end{equation}
We impose this condition by means of the loss function
\begin{equation}\label{eq:losshermiticity}
    \mathcal{L}_{\dagger} = \left\vert\left\vert H_{ij}-H^\dagger_{ij}\right\vert\right\vert\,,
\end{equation}
where $H_{ij}$ are the matrix elements of $H$.
We shall therefore train our neural 
network with the loss function 

\begin{equation}\label{eq:total_loss}
        \mathcal{L} = \mathcal{L}_{YBE} + 
        \mathcal{L}_{reg}+\lambda_{RTT}\mathcal{L}_{RTT}  +\lambda_H \mathcal{L}_{H}
        + \lambda_{\dagger} \mathcal{L}_{\dagger}\,,      
\end{equation}
where putting the coefficients $\lambda_{\alpha}$, for $\alpha=\{RTT, H,\dagger\}$, to zero removes the  corresponding loss term from being trained.

The loss function \eqref{eq:total_loss} produces a very complicated landscape and 
the NN should approach its minimum during the training. Usually, this search is 
performed with gradient based optimization methods. One might be skeptical about 
being stuck in some local minimum instead of finding the global minimum of such 
complicated loss function in a very high dimensional hypothesis space. However, 
recent analysis revealed that deep NNs end up having all their local minima 
almost at the same value as the global one \cite{choromanska2015loss}, 
\cite{dauphin2014identifying}. In other words, there are many different 
configurations of weights and biases resulting into a function of similar 
accuracy as the one corresponding to the global minimum. There are also many 
saddle points and some of them have big plateau and just a  small fraction of 
descendent directions, making them practically indistinguishable from the local 
minima. However, most of their losses are close to the global minimum as well. 
Those with significantly higher losses have a
bigger number of descendent directions and thus can be escaped during the 
learning \cite{choromanska2015loss}, \cite{dauphin2014identifying}.

We find that the training converges to yield simultaneously low values for
each of the above losses as applicable.
Further, while the hyper-parameters $\left\lbrace \lambda\right\rbrace$ are 
tunable experimentally, setting them all to 1 is a useful default. 
However, for fine-tuning the training it is 
also useful to tune these parameters to reflect the specific task at hand. 
We provide the requisite details in 
Section \ref{sec:results} where we discuss specific training methodologies
and the corresponding results. We will also discuss there a new loss
function 
\begin{equation}\label{eq:replusionloss}
\mathcal{L}_{repulsion}=\exp{(-||H-H_o||/\sigma)}\,,
\end{equation}
which is useful to fine-tune the training to access new integrable Hamiltonians
$H$
in the neighbourhood of previously known integrable Hamiltonians $H_o$, we will call such regime as a \textit{exploration by repulsion}.

As a final observation on the choice of activation functions, we note that
at the level of the discussion above, any holomorphic activation function
such as \texttt{sigmoid}, \texttt{tanh}, and the \texttt{sinh}
would suffice. In practice we find that the training converges faster and
more precisely using the \texttt{swish} activation
\cite{ramachandran2017searching}.
This is given by
\begin{equation}
    \swish\left(z\right) = z\,\sigma\left(z\right)\,,\qquad
    \sigma\left(z\right) = \frac{1}{1+e^{-z}}\,.
\end{equation}
We have provided some comparison tests across activation functions in
Appendix \ref{app:compareActivations}.

\section{Results}\label{sec:results}

We present our results for learning R-matrices within the restricted setting of two dimensional spin chains of difference form. Our analysis will be divided into three parts. First, we will learn hermitian XYZ model and its well-known XXZ and XXX limits, comparing our deep-learning results against the analytic plots. Then we remove hermiticity and reproduce all 14 classes of solutions from \cite{de2019classifying}. The last set of experiments demonstrates how our Neural Network in the Explorer mode can search for Integrable models exploring the space of Hamiltonians. 
\subsection{Specific integrable spin chains}\label{subsec:specific_search}
In this sub-section we look at specific physical models, by imposing tailored conditions on the Hamiltonian derived from the training R-matrix. This includes constraints on the Hamiltonian entries at $u=0$, and hermiticity of the Hamiltonian.

\subsubsection{Hermitian models: XYZ spin chain and its isotropic limits}\label{subsubsec:XYZ-XXZ-XXX}
\begin{figure}[ht]
  \centering
    \includegraphics[width=.8\textwidth]{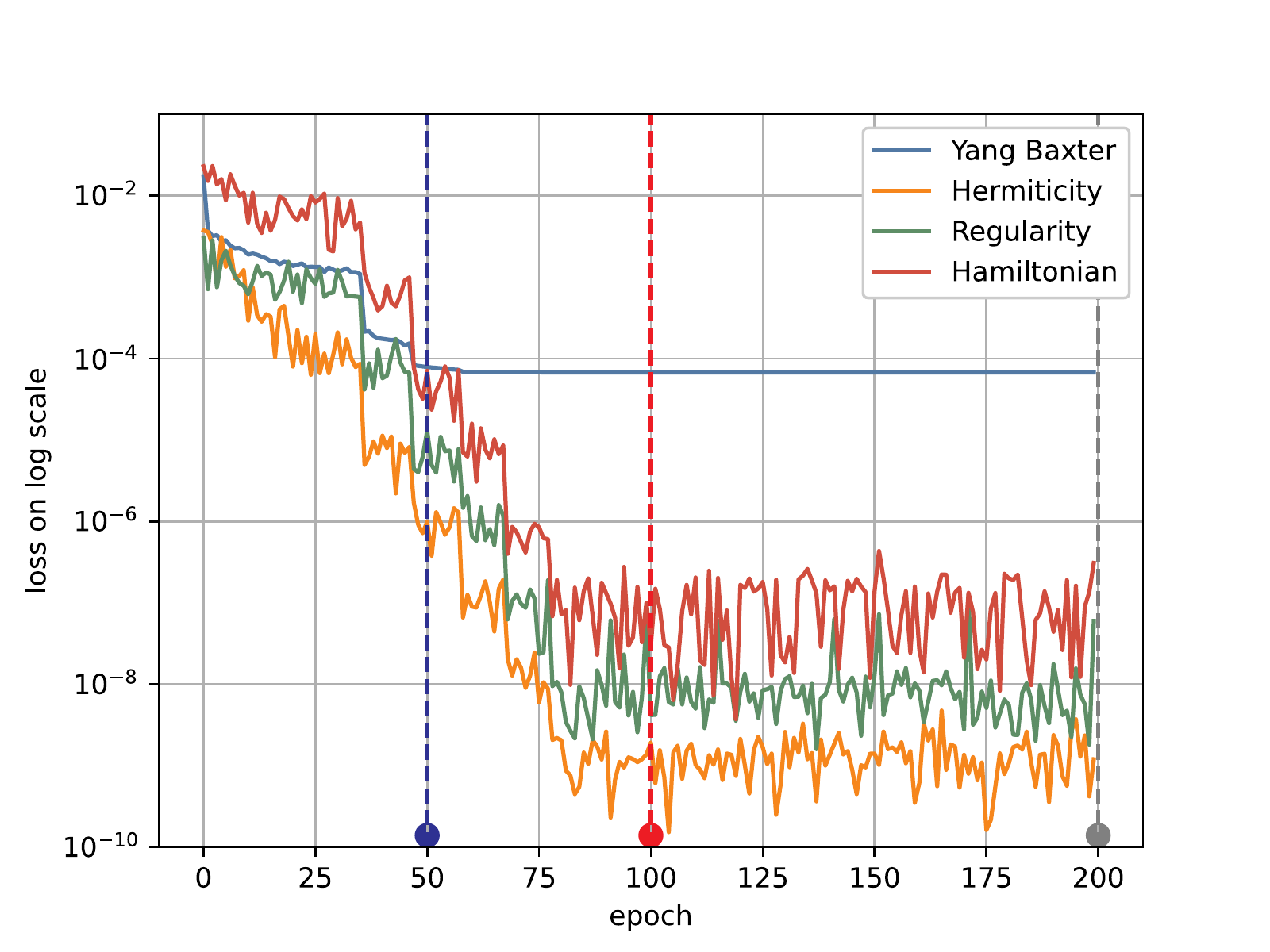}
    \caption{The evolution of training losses for the XYZ model, shown on 
    the log scale. The losses tend to fall in a step-wise manner, which
    corresponds approximately to the learning rate schedule the network is
    trained with.}
  \label{fig:xyztraining}
\end{figure}

\begin{figure}
  \centering
  \captionsetup[subfigure]{position=top}
  \subfloat[]{\includegraphics[width=0.5\linewidth,height=.5\textwidth]{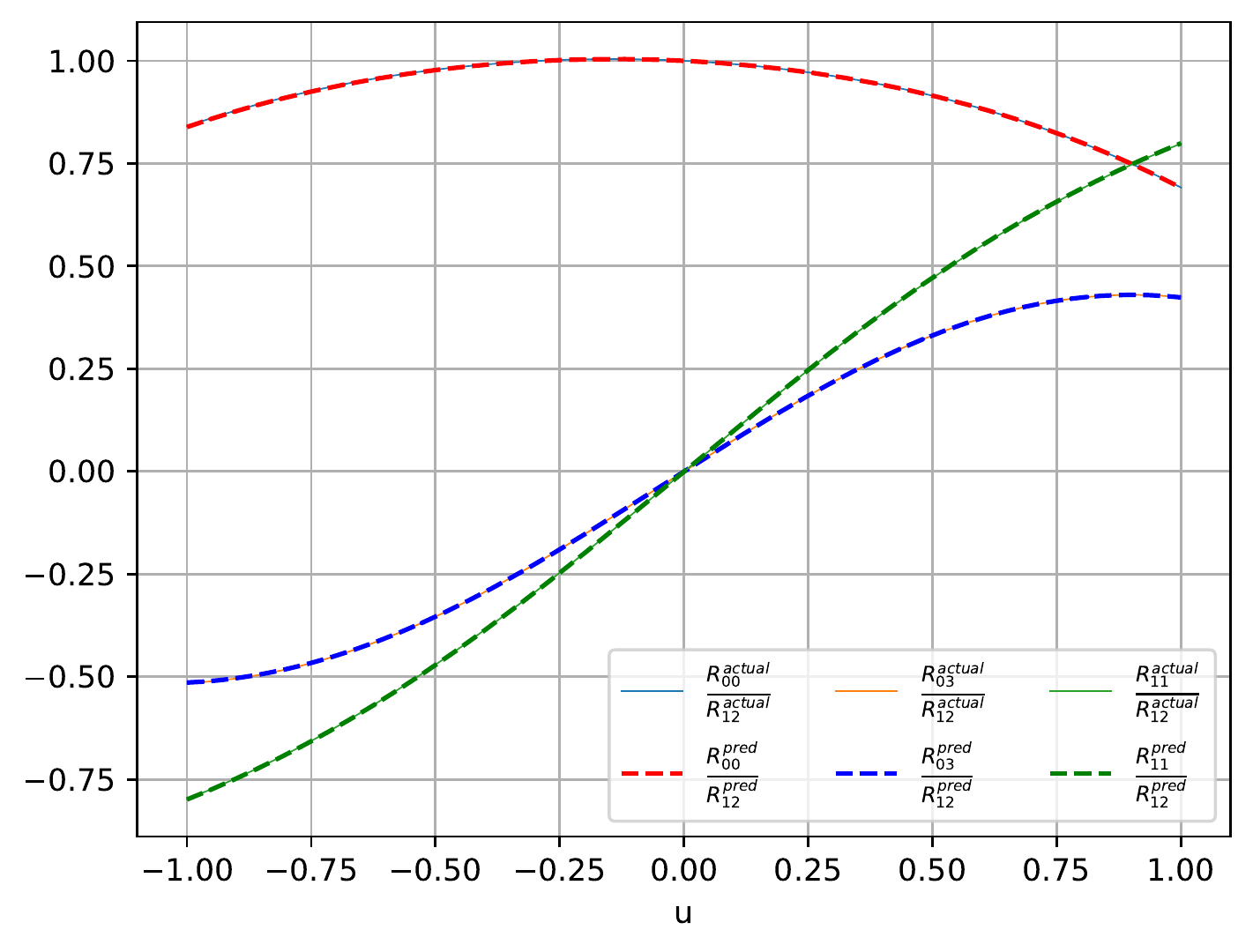}}
  \subfloat[]{
  \begin{minipage}[]{0.4\textwidth}
    \centering
    \includegraphics[width=0.8\linewidth,height=0.56\textwidth]{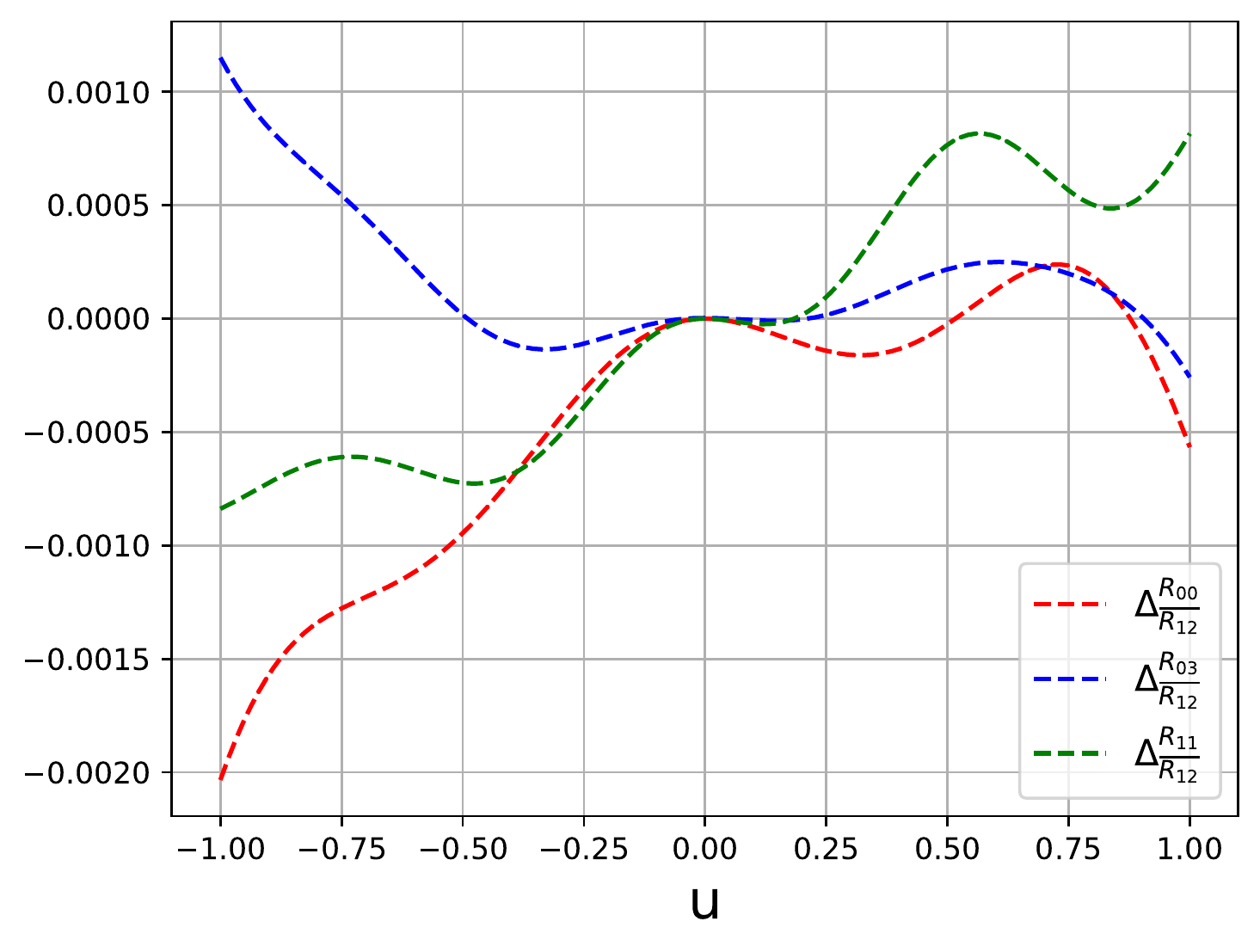}
    
    \vspace{10pt}
    \includegraphics[width=0.8\linewidth,height=0.56\textwidth]{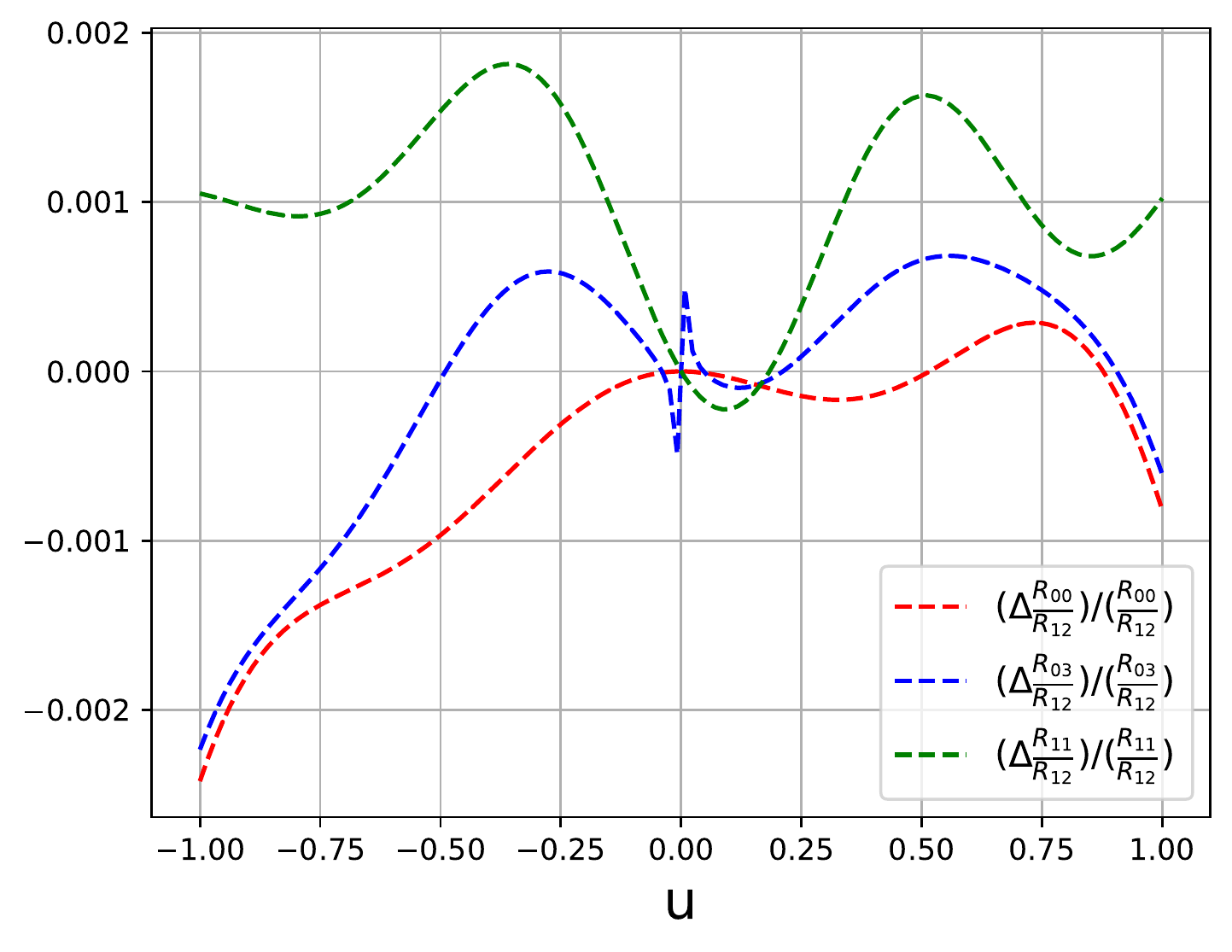}
    \end{minipage}
    }
  \caption{(a) XYZ R-matrix as ratios with respect to the (12) component for $\eta=\pi/3,m=0.6$,  (b) Relative and absolute errors for R-matrix. }
  \label{fig:XYZ_R_matrix}
\end{figure}

Imposing hermiticity on the 8-vertex Hamiltonian, we learn the classic XYZ integrable spin chain and its symmetric XXZ limit. We start with the following 8-vertex model ansatz for the R-matrix
\begin{equation}
    R(u)=\begin{pmatrix}
        a&0&0&d\\
        0&b&c&0\\
        0&c&b&0\\
        d&0&0&a
    \end{pmatrix}
\end{equation}
and impose the loss functions for YBE, hamiltonian constraint, regularity, and hermiticity (see equation~\ref{eq:total_loss}). The target Hamitonians comprise of a 2-parameter family $H_{XYZ}(J_x,J_y,J_z)$ given by
\begin{equation}
    H_{XYZ}(J_x,J_y,J_z)=J_x S^x_1S^x_2+J_y S^y_1S^y_2+J_z S^z_1S^z_2=\begin{pmatrix}
        J_z&0&0&J_x-J_y\\
        0&-J_z&2&0\\
        0&2&-J_z&0\\
        J_x-J_y&0&0&J_z
    \end{pmatrix}\,,
\end{equation}
where we have set $J_x+J_y$ to be equal to 2.
The symmetric limit of XXZ model is realised for $J_x=J_y=1$. A useful reparametrisation for these models is in terms of $(\eta,m)$ \cite{takhtadzhan1979quantum}
\begin{equation}
    J_x=1+\frac{\sqrt{m}\JacobiSN{2\eta}{m}}{2}\,,\quad
    J_y=1-\frac{\sqrt{m}\JacobiSN{2\eta}{m}}{2}\,,\quad
    J_z=\JacobiCN{2\eta}{m}\JacobiDN{2\eta}{m}
\end{equation}
The analytic solution for the XYZ R-matrix is given in terms of Jacobi elliptic functions as 
\begin{equation}\label{eq:xyz_rmat_analytic}
\begin{split}
    &a(u)=\frac{\JacobiSN{2\eta+\omega u}{m}}{\JacobiSN{2\eta}{m}}\exp\left(-\frac{\JacobiCN{2\eta}{m}\JacobiDN{2\eta}{m}}{2\JacobiSN{2\eta}{m}}\omega u\right)\,,\\
    &b(u)=\frac{\JacobiSN{\omega u}{m}}{\JacobiSN{2\eta}{m}}\exp\left(-\frac{\JacobiCN{2\eta}{m}\JacobiDN{2\eta}{m}}{2\JacobiSN{2\eta}{m}} \omega u\right)\,,\\
    &c(u)=\exp\left(-\frac{\JacobiCN{2\eta}{m}\JacobiDN{2\eta}{m}}{2\JacobiSN{2\eta}{m}}\omega u\right)\,,\\ 
    &d(u)=\sqrt{m}\JacobiSN{\omega u}{m}\JacobiSN{2\eta+\omega u}{m}\exp\left(-\frac{\JacobiCN{2\eta}{m}\JacobiDN{2\eta}{m}}{2\JacobiSN{2\eta}{m}}\omega u\right),  
\end{split}
\end{equation}
where $\omega=2\JacobiSN{2\eta}{m}$, and $m$ is the elliptic modular parameter\footnote{Usually, these expressions are written in terms of the elliptic modulus $k$ instead of the modular parameter $m=k^2$, e.g. as in 
\cite{vieira2018solving}. We have expressed them in terms of the modular
parameter following the implementation in both Python and Mathematica.}. Our model consistently learns the R-matrices for the XYZ model for generic values of the free parameters $\eta,m$. 
\noindent Figure \ref{fig:xyztraining} gives the time evolution of the different loss terms during training.
Figure~\ref{fig:XYZ_R_matrix} plots the R-matrix component ratios with respect to $R_{12}$ in terms of the spectral parameter, and compares them with the corresponding analytic functions for a generic choice of deformation parameters $\eta=\frac{\pi}{3}$ and $m=0.6$. Letting $m=0$, we recover the XXZ models for generic values of $\eta$.

\subsubsection{Two-dimensional  classification}\label{subsubsec:fullclassification}
Here, we lift the hermiticity constraint on the Hamiltonian, thus allowing for 
more generic integrable models. As we shall see below, the neural network 
successfully learns all the 14 classes\cite{de2019classifying} of difference-form 
integrable (not necessarily Hermitian) spin chain models with 2-dimensional space 
at each site. The R-matrices corresponding to each of these classes are written 
down explicitly in appendix~\ref{app:XYZ_nonXYZ_plots}. Towards the end of this 
sub-section, we also present results for learning solutions in generic gauge 
obtained by similarity transformation of integrable Hamiltonians from the 
aforementioned 14 classes. We shall discuss the results in two parts: XYZ type 
models, and non-XYZ type models.

The first set of Hamiltonians under consideration are generalisations of the XYZ model (discussed in the previous sub-section), with at most 8 non-zero elements in its Hamiltonian density
\begin{equation}\label{eq:8vertexhamiltonian}
    H_{8-\text{vertex}}=\begin{pmatrix}
        a_1&0&0&d_1\\
        0&b_1&c_1&0\\
        0&c_2&b_2&0\\
        d_2&0&0&a_2
    \end{pmatrix}
\end{equation}
where the coefficients can take generic complex values. The XYZ model corresponds to the subset with $a_1=a_2,b_1=b_2,c_1=c_2,d_1=d_2$. As discussed in section~\ref{subsec:YBE_sol_classes}, these models can be further sub-divided into four, six, seven and eight vertex models. On the other hand, there are 6 distinct classes of non-XYZ type solutions. Here we will discuss the training results for one example each from the XYZ and non-XYZ type models, since the training behaviour is similar within these two types. Rest of the models will be  presented in Appendix~\ref{app:XYZ_nonXYZ_plots}. Figure~\ref{fig:XXZtype_class1} plots the R-matrix components as ratios with respect to $R_{00}$ for a generic 6-vertex model with $d_1=d_2=0$, and $a_1=a_2$. The figure also includes the absolute and relative errors with respect to the corresponding analytic R-matrix (see equation~\eqref{eq:Hamiltonian6v,1}). 
\begin{figure}
  \centering
  \captionsetup[subfigure]{position=top}
  \subfloat[]{\includegraphics[width=0.55\linewidth,height=.5\textwidth]{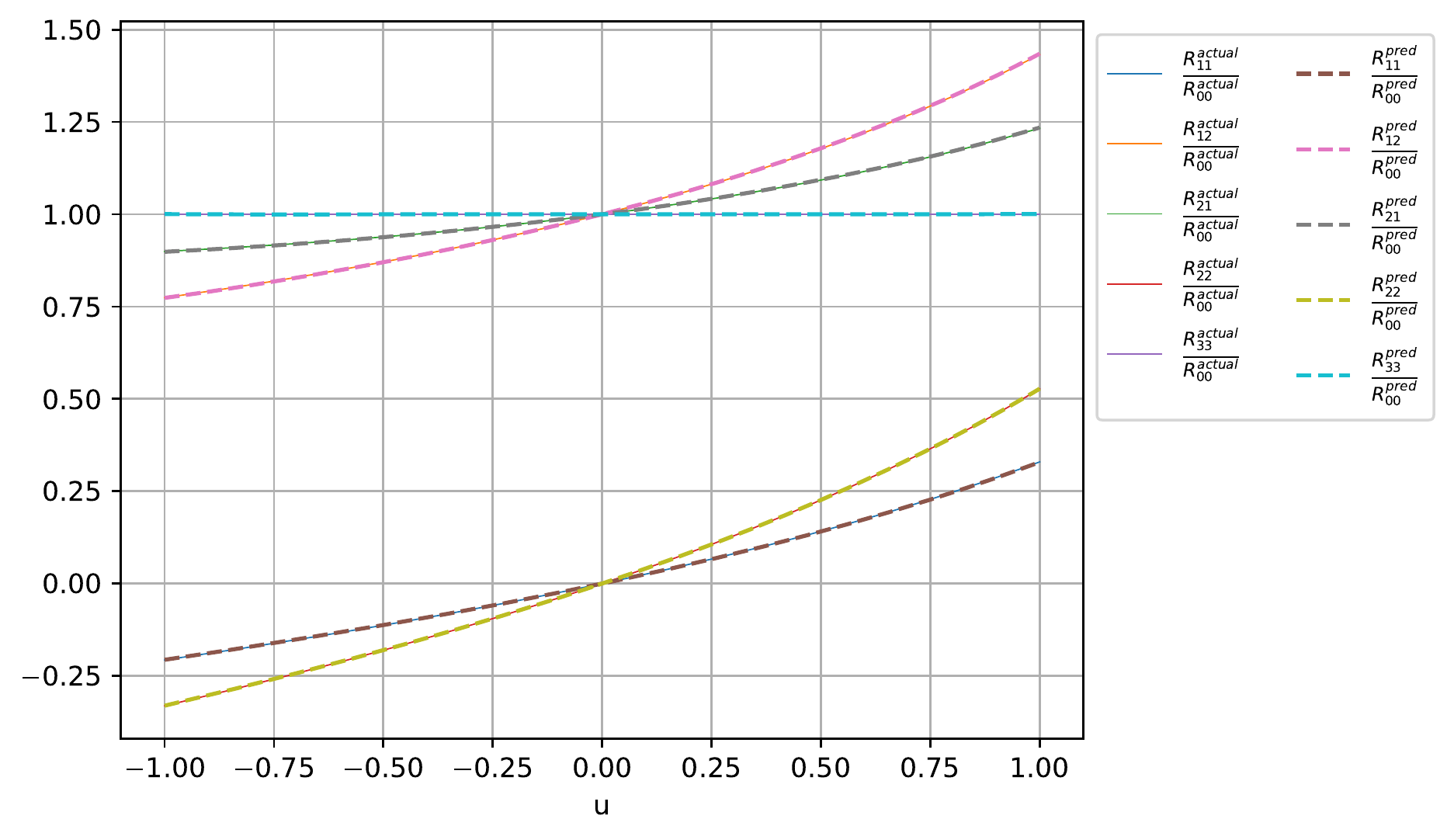}}
  \subfloat[]{\includegraphics[width=0.4\linewidth,height=.4\textwidth]{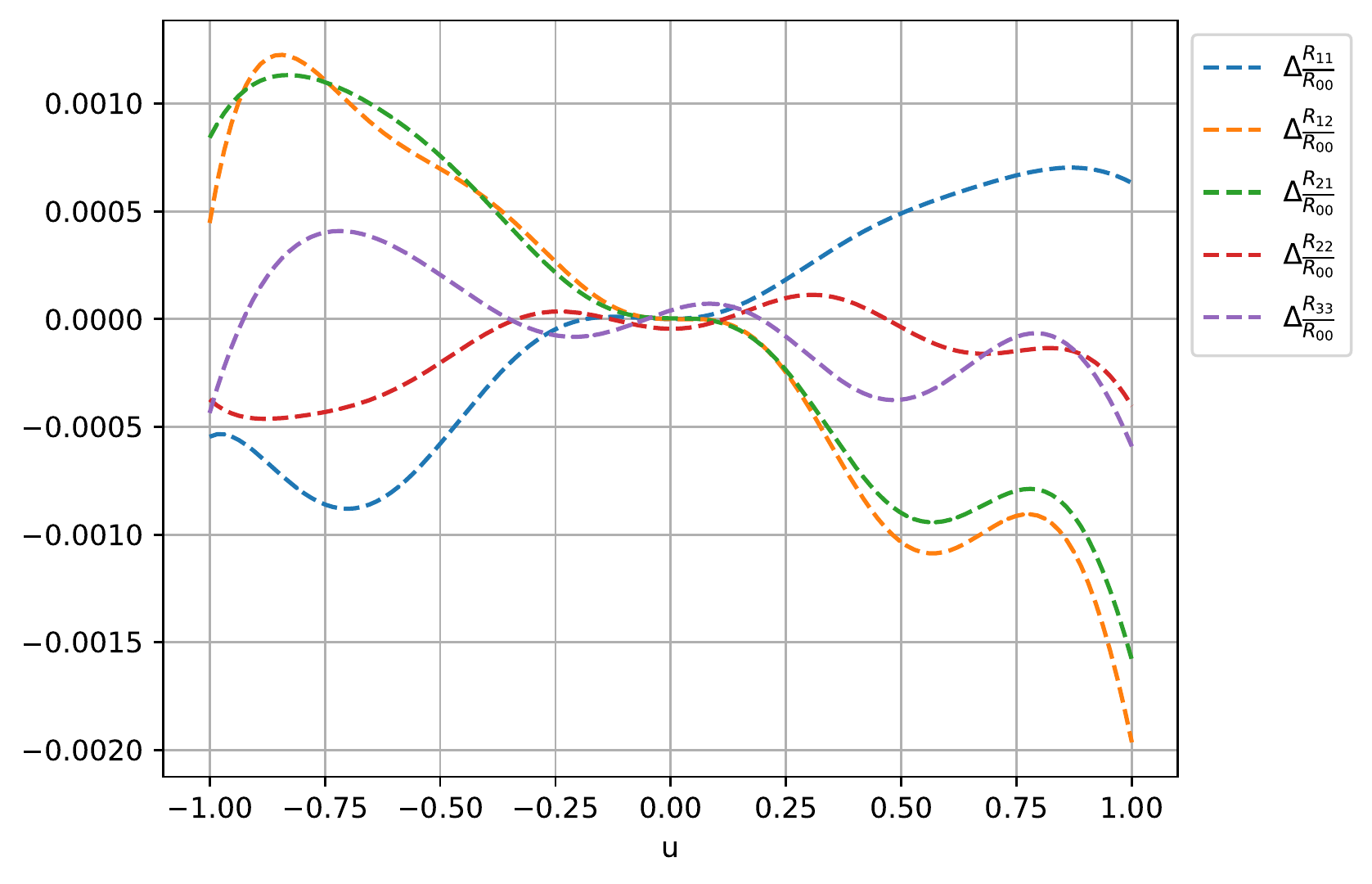}}
  \caption{(a) Predicted vs actual R-matrix component ratio w.r.t. $R_{00}$ for XXZ-type model with $a_1=a_2=0.3,b_1=0.45,b_2=0.6,c_1=0.4,c_2=0.25$, (b)  Absolute error between predicted and actual R-matrix component ratios}
  \label{fig:XXZtype_class1}
\end{figure}

\begin{figure}
  \centering
  \captionsetup[subfigure]{position=top}
  \subfloat[]{\includegraphics[width=0.5\linewidth,height=.4\textwidth]{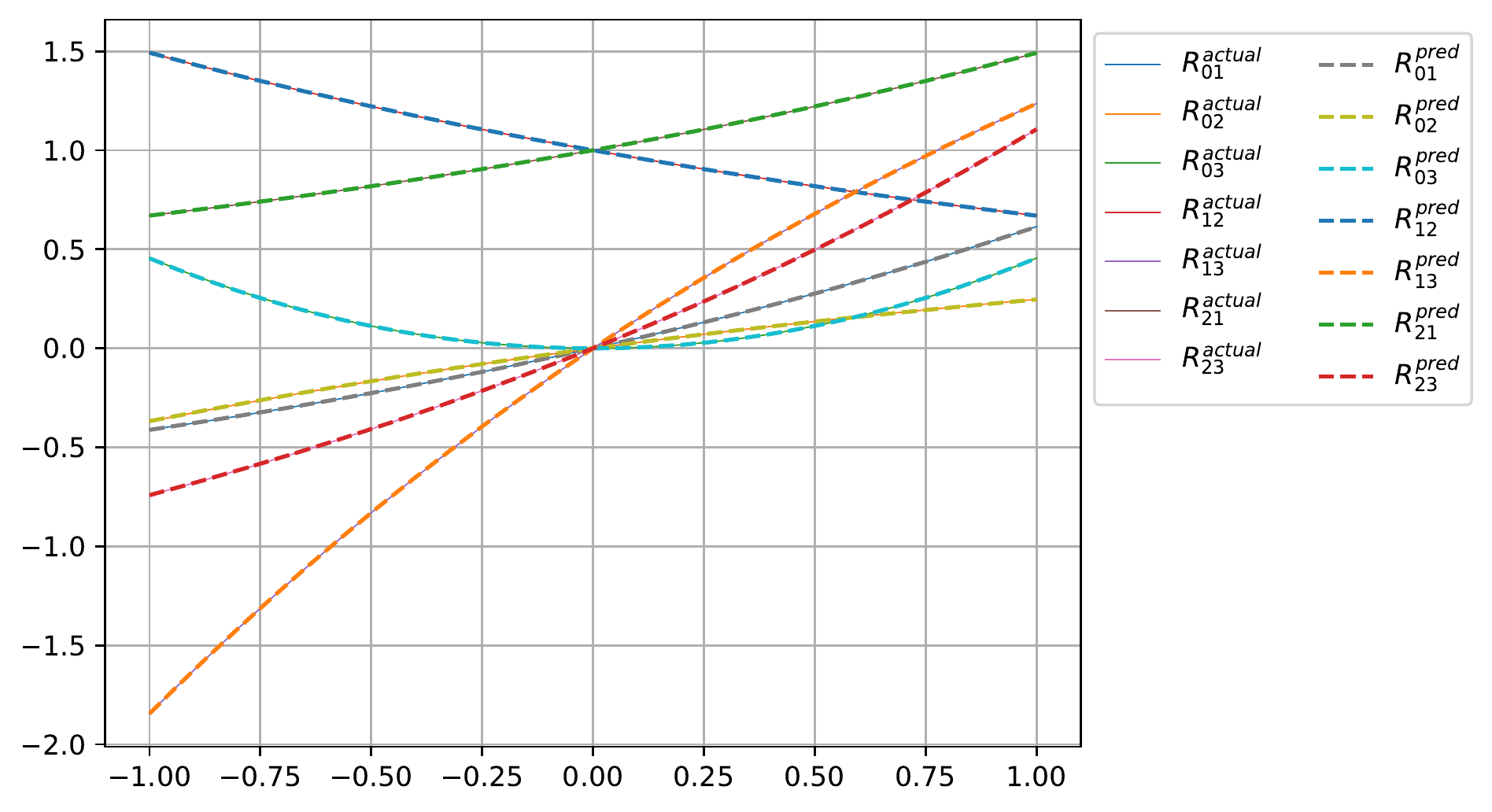}}
  \subfloat[]{\includegraphics[width=0.45\linewidth,height=.25\textwidth]{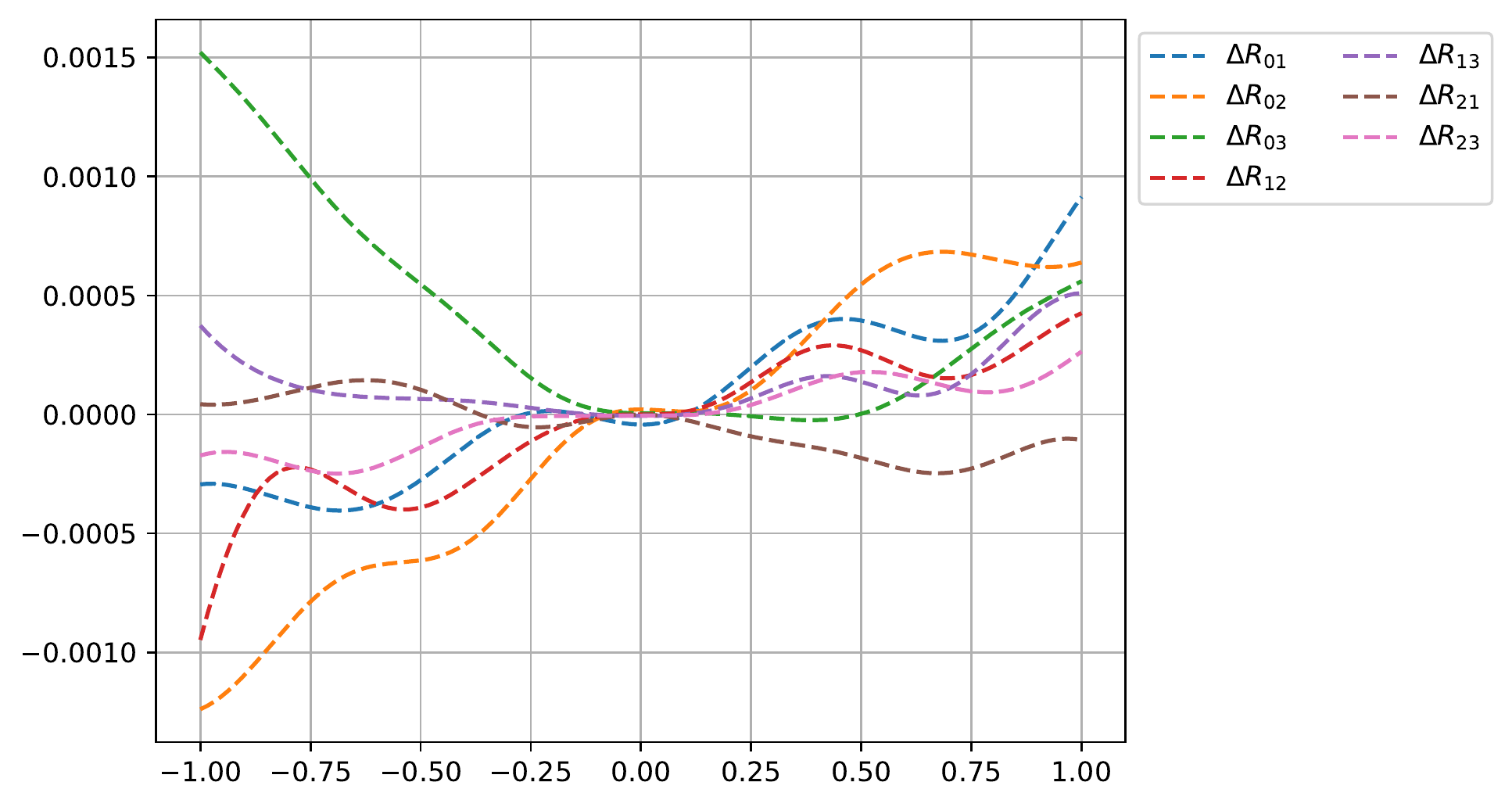}}
  \caption{(a) Predicted vs actual R-matrix components for the Hamiltonian of class-1, with coefficients $a_1=.5,\, a_2=.3, \, a_3=.9, \,a_4=1.5, \,a_5=.4$. Note here the R-matrices are automatically normalised since $R_{00}$ was fixed to the constant value of 1, (b)  Absolute error between predicted and actual R-matrix components }
  \label{fig:non_XYZ_class1}
\end{figure}

\begin{figure}[h!]
  \centering
  \captionsetup[subfigure]{position=top}
  \subfloat[]{\includegraphics[width=0.6\linewidth,height=.5\textwidth]{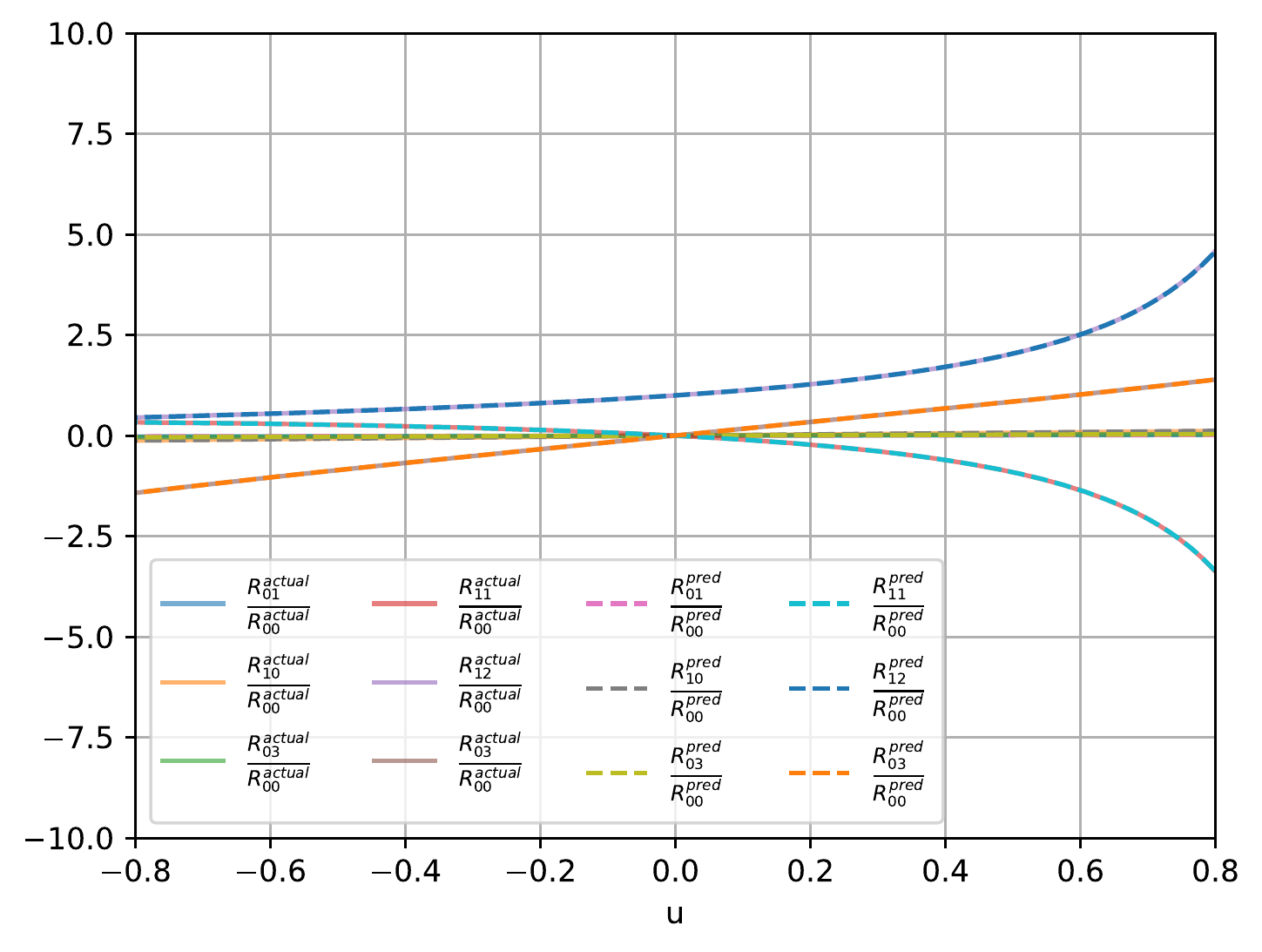}}
  \subfloat[]{\includegraphics[width=0.35\linewidth,height=.35\textwidth]{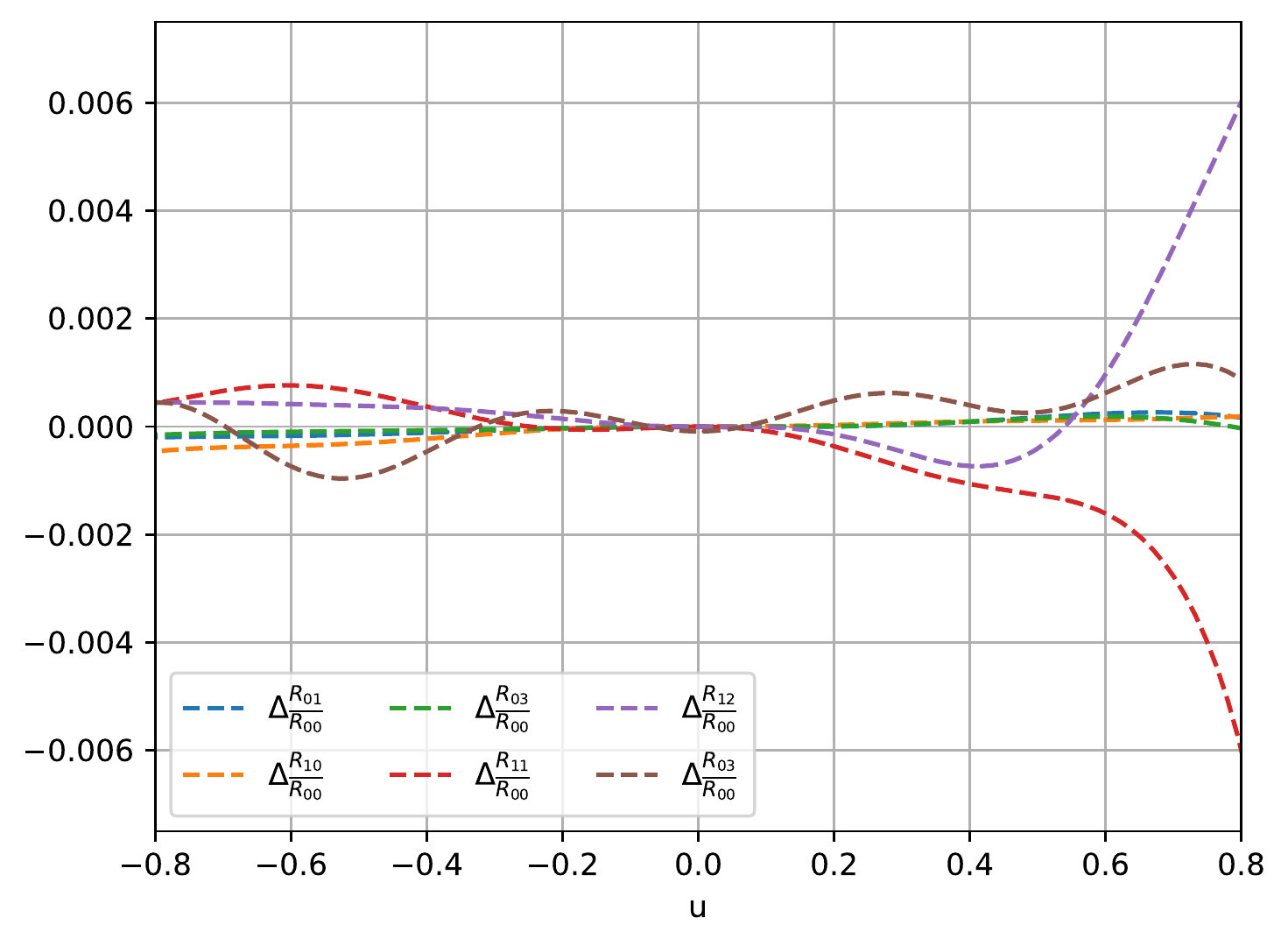}}
  \caption{(a) Predicted R-matrix component ratios w.r.t. $R_{00}$, for conjugated XXZ model with $\eta=\frac{\pi}{3}$ and similarity matrix $\Omega=\small \begin{pmatrix}
    0.4&0.5\\-1.2&1
    \end{pmatrix}$, (b) Absolute error between predicted and actual R-matrix ratios}
  \label{fig:twistedXXZ}
\end{figure}

From the non-XYZ classes, we will focus on the following 5-vertex Hamiltonians
\begin{equation}
    H_{class-1}=\begin{pmatrix}
        0&a_1&a_2&0\\
        0&a_5&0&a_3\\
        0&0&-a_5&a_4\\
        0&0&0&0
    \end{pmatrix}
\end{equation}
For integrability, we require the additional condition 
\begin{equation}
    a_1a_3=a_2a_4\,.
\end{equation}
Training the Hamiltonian constraint~\eqref{eq:hamiltonianloss} for generic values 
$a_1=0.5,a_2=0.3,a_3=0.9,a_4=1.5,a_5=0.4$ satisfying the above integrability 
condition, we get over \(0.1\%\) accuracy for training over $\sim$100 epochs. 
Figure~\ref{fig:non_XYZ_class1} plots the trained R-matrix components and 
absolute errors with respect to the analytic R-matrices in 
equation~\eqref{eq:nonXYZRmatrices}, for the above choice of target Hamiltonian.
We have also surveyed more general solutions beyond the representative solutions 
of the 14 classes \textit{a la} \cite{de2019classifying}, by changing the gauge 
of the R-matrix as well as the corresponding Hamiltonian. As noted earlier in 
section~\ref{sec:integrability}, we can act with a $2\times 2$ similarity  
matrix $\Omega$ on the R-matrix :
\begin{equation}
    R(u)\rightarrow R^{\Omega}(u)=(\Omega\otimes \Omega)R(u)(\Omega^{-1}\times \Omega^{-1})
\end{equation}
\begin{equation}
    H\rightarrow H^{\Omega}=(\Omega\otimes \Omega)H(\Omega^{-1}\times \Omega^{-1})
\end{equation}
If $R(u)$ satisfies Yang-Baxter equation, so does $R^{\Omega}(u)$. A generic similarity
matrix $\Omega$
\begin{equation}
    \Omega=\begin{pmatrix}
        v_{11}&v_{12}\\
        v_{21}&v_{22}
    \end{pmatrix}
\end{equation}
with non-zero off-diagonal entries $v_{12},v_{21}\neq 0$, results in conjugated R-matrices and Hamiltonians with all 16 non-zero entries. We trained 16-vertex Hamiltonians resulting from XYZ model in the general gauge and recovered the corresponding  R-matrix with a relative error of order $\mathcal{O}(0.1\%)$. Generic XYZ 
type models, as well as non-XYZ type models gave similar results for different gauges. Figure~\ref{fig:twistedXXZ} plots the learnt R-matrix components for XXZ 
model with $\eta=\frac{\pi}{3}$ conjugated by the  matrix \(\Omega\) =$\begin{pmatrix}
    0.4&0.5\\-1.2&1
\end{pmatrix}$. For comparison with analytic formulae, we normalised our results 
by taking ratios with respect to a fixed component $R_{00}$, i.e. we plot 
$\frac{R_{ij}}{R_{00}}$. As a result of starting from the XXZ model, the 
R-matrix $R_{XXZt}$ in the general gauge has following highly symmetric form
    \begin{equation}
        R_{XXZg}=\begin{pmatrix}
            R_{00}&R_{01}&R_{01}&R_{03}\\
            R_{10}&R_{11}&R_{12}&-R_{01}\\
            R_{10}&R_{12}&R_{11}&-R_{01}\\
            R_{30}&-R_{10}&-R_{10}&R_{00}
        \end{pmatrix}
    \end{equation}
Thus we only plot the entries $R_{00},R_{01},R_{03},R_{10},R_{11},R_{12},R_{30}$. 
Since there exists overall normalisation ambiguity, we should only compare ratio 
of R-matrix entries with the analytic solution written in the same gauge.

Next we discuss the difference in the training of integrable vs non-integrable models with our neural network. We will focus on two representative examples : 6-vertex model with Hamiltonian $H_{6v,1}$, and class 4 models with Hamiltonian $H_{class-4}$. Similar results hold across all the 
14 classes.

For 6-vertex models with Hamiltonians following equation~\eqref{eq:8vertexhamiltonian} with $d_1=d_2=0$, generic values of the coefficients $a_i,b_i,c_i,d_i$ for $i=1,2$ leads to non-integrable models, unless
\begin{equation}
    a_1=a_2\,,\quad a_1+a_2=b_1+b_2\,.
\end{equation}
These are the models with Hamiltonian $H_{6v,1}$, $H_{6v,2}$ in 
appendix~\ref{app:XYZ_nonXYZ_plots}. 
Figure \ref{fig:NintComparisonType6} 
compares the training for a generic Hamiltonian with coefficients satisfying
none of the above conditions against the training for $H_{6v,1}$-type model. 
We see that while the Hamiltonian constraint \eqref{eq:hamiltonianloss}
saturates to similarly low values in both cases, the Yang-Baxter loss saturates
at approximately one order of magnitude higher. 
\begin{figure}
     \centering
     \begin{subfigure}[b]{0.5\textwidth}
         \centering
         \includegraphics[width=1\textwidth]{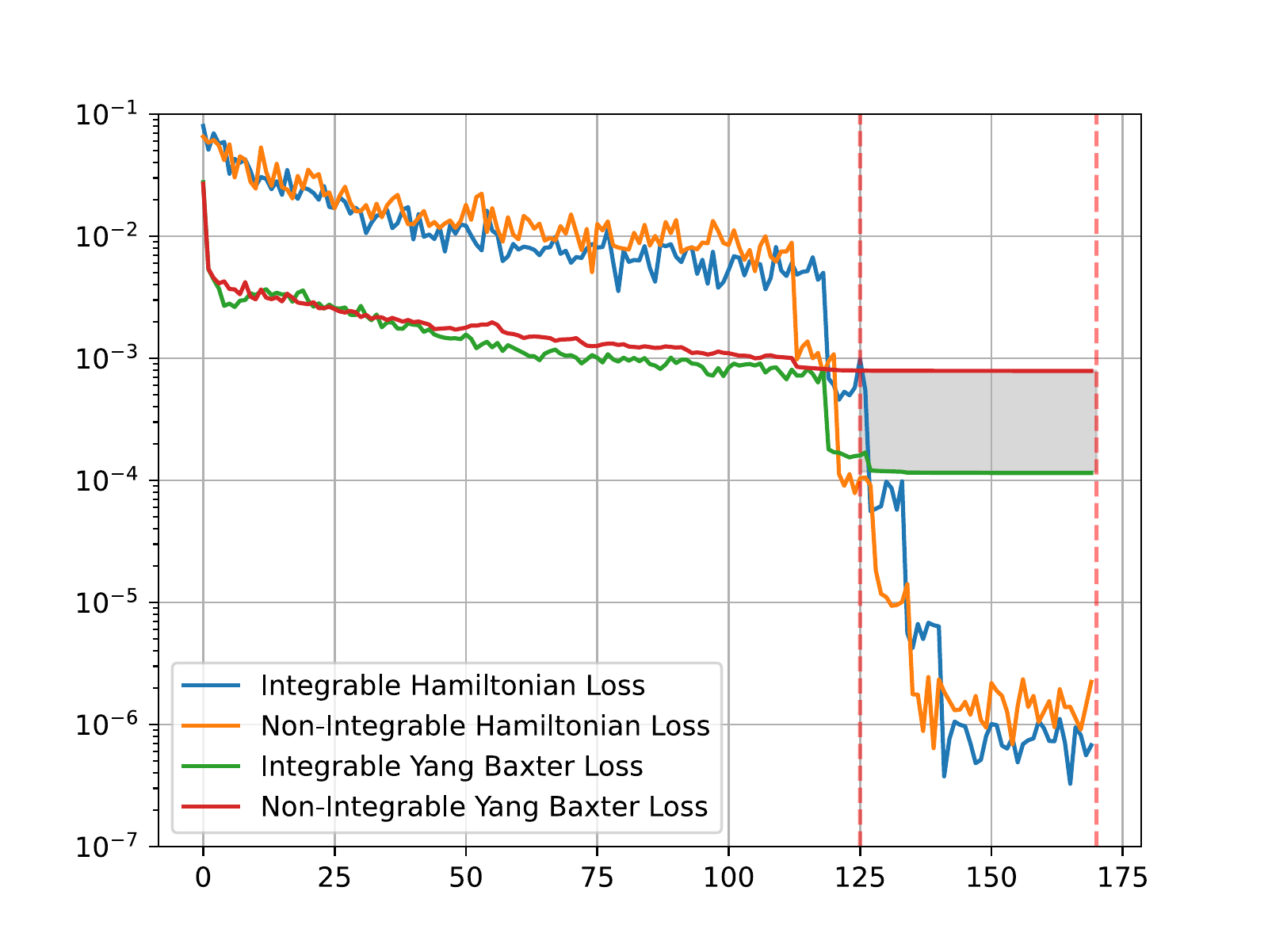}
         \caption{Hamiltonian $H_{6v,1}$  vs non-integrable.}
         \label{fig:NintComparisonType6}
     \end{subfigure}%
     \begin{subfigure}[b]{0.5\textwidth}
         \centering
         \includegraphics[width=1\textwidth]{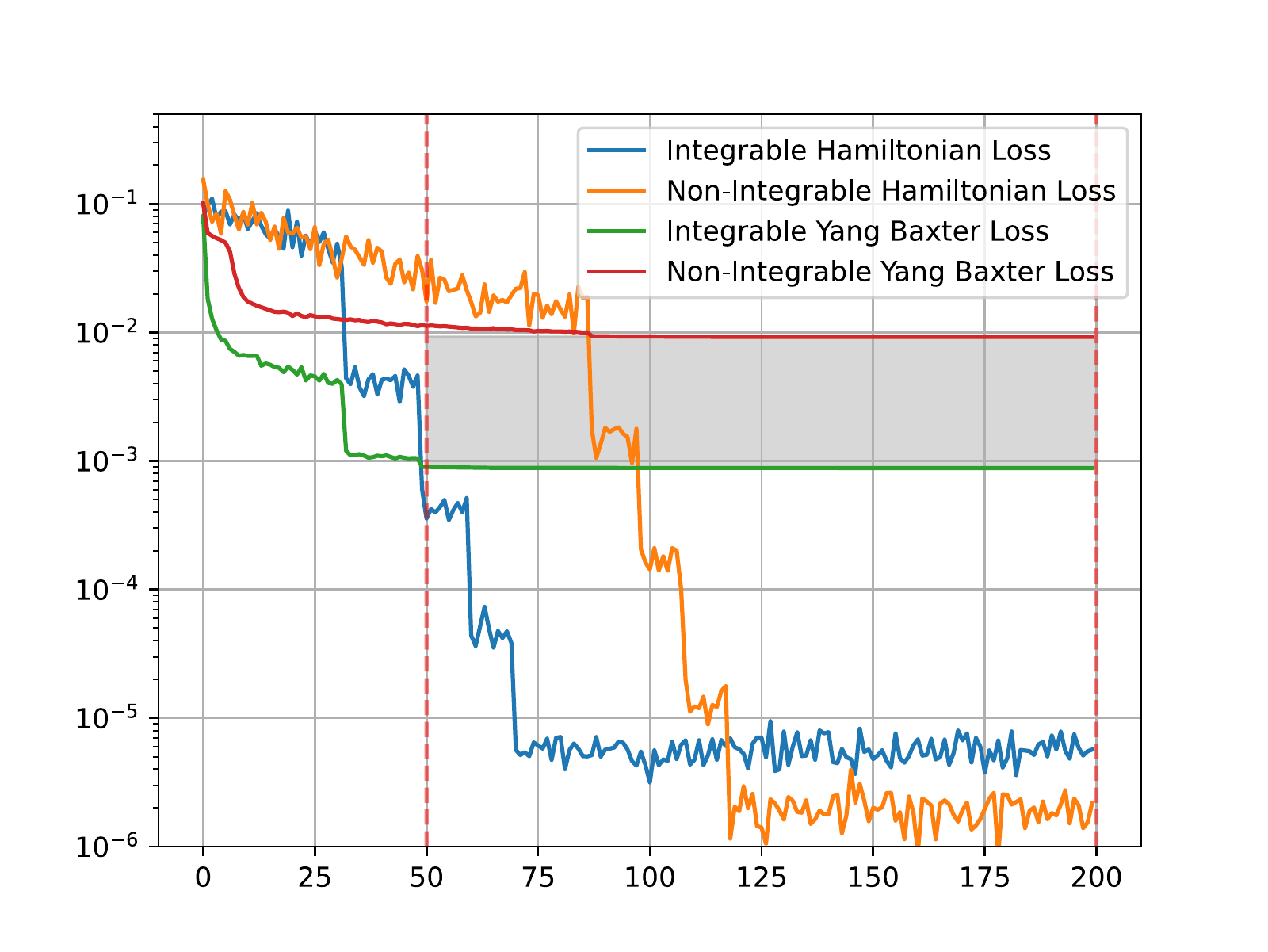}
         \caption{Hamiltonian $H_{\mathrm{class}-4}$ vs non-integrable.}
         \label{fig:NintComparisonClass4}
     \end{subfigure}%
        \caption{Comparing the training history of the Type XYZ and non-XYZ
        models against corresponding non-integrable Hamiltonians. There is 
        approximately an 
        order of magnitude difference between the Yang-Baxter losses for the 
        integrable
    case vs the non-integrable case after the training saturates, indicated
    by the gray region in the graph. The step-wise drops in the loss
    functions approximately correspond to the learning rate schedule. The presented Hamiltonians are the same as on Fig.\ref{fig:yberr6v1} and Fig.\ref{fig:yberrclass4}}
        \label{fig:intvsNintTraining}
\end{figure}
Similar behavior holds for the non-XYZ type models as well. The training for
a generic class-4 Hamiltonian with 
coefficients $a_1=0.5, a_2=0.3, a_3=0.4, a_4=0.9$ (see 
Equation~\ref{nonXYZhamiltonians}) and a non-integrable deformation
is shown in Figure \ref{fig:NintComparisonClass4}.

\begin{figure}
     \centering
     \begin{subfigure}[b]{0.5\textwidth}
         \centering
         \includegraphics[width=1\textwidth]{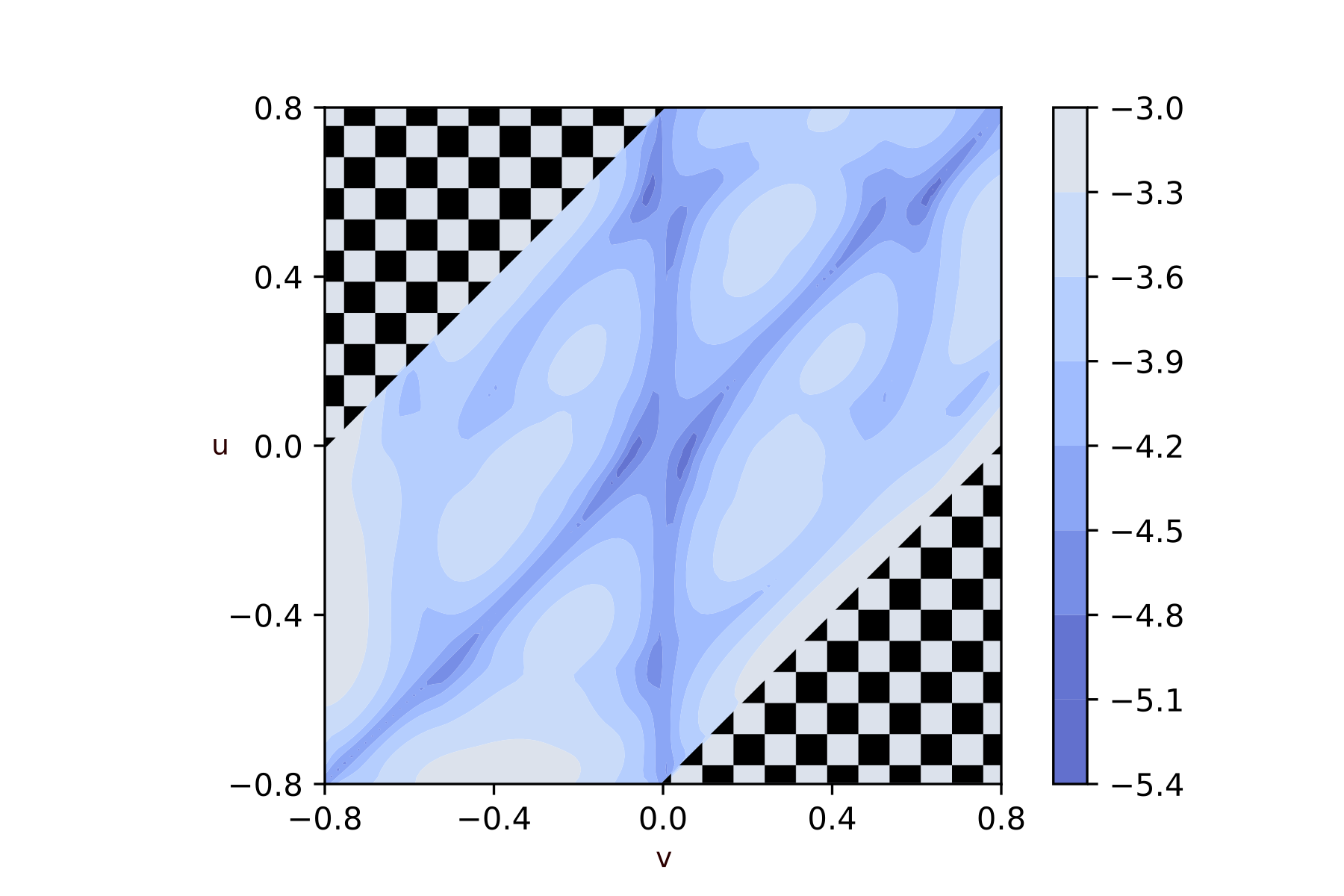}
         \caption{$H_{6v,1}$}
         \label{fig:h6v1normybe}
     \end{subfigure}%
     \begin{subfigure}[b]{0.5\textwidth}
         \centering
          \includegraphics[width=1\textwidth]{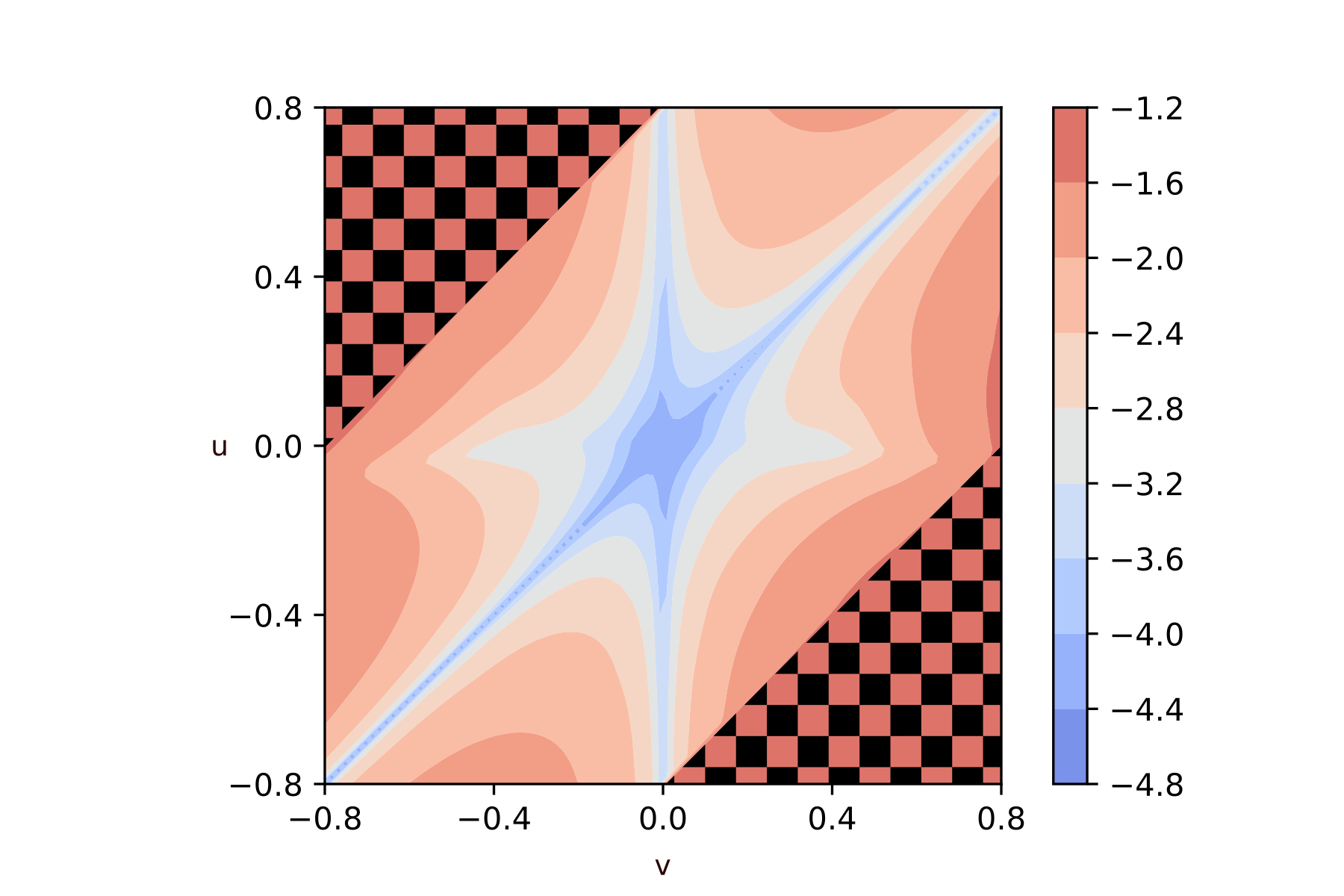}
         \caption{$H_{6v,1}$ deformation}
         \label{fig:h6v1nintnormybe}
     \end{subfigure}%
        \caption{(a) The normalized Yang-Baxter error \eqref{eq:ybmetric} plotted in the logarithmic scale 
        at the end of training for the Hamiltonian $H_{6v,1}$ with $a_1=0.3, a_2=0.3,b_1=0.45,b_2=0.6, c_1=0.4,c_2=0.25$, and (b) its non-integrable
        deformation with $a_1=-1.3,a_2=1.3$ and other parameters kept constant. In order to keep all three arguments appearing in YB equation inside the same inteval \(|u|,|v|,|u-v|<0.8\) we cut out the area \(|u-v|>0.8\) with chess-pattern triangles.}
        \label{fig:yberr6v1}
\end{figure}

\begin{figure}
     \centering
     \begin{subfigure}[b]{0.5\textwidth}
         \centering
         \includegraphics[width=\textwidth]{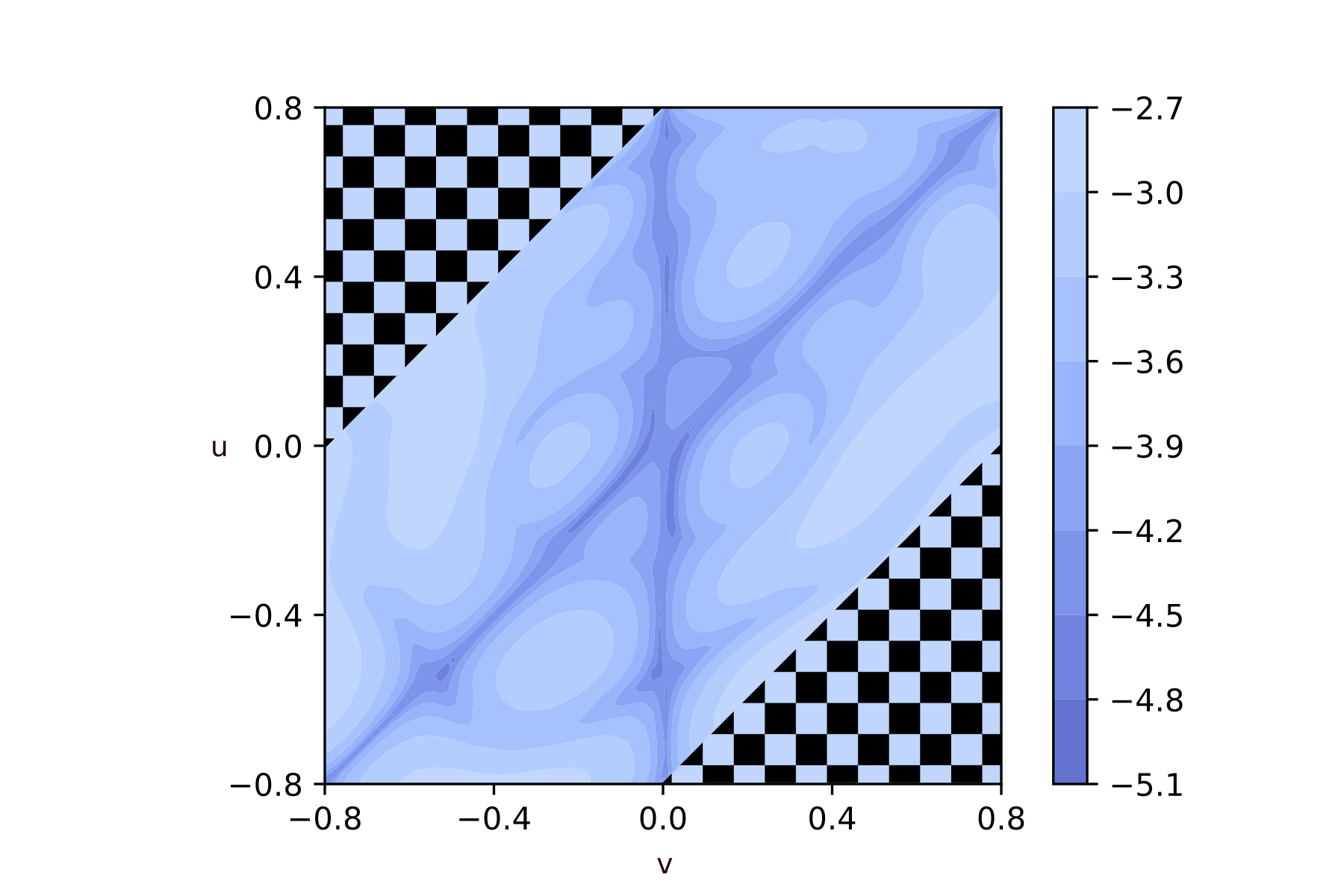}
         \caption{$H_{\mathrm{class}-4}$}
         \label{fig:hclass1normybe}
     \end{subfigure}%
     \begin{subfigure}[b]{0.5\textwidth}
         \centering
         \includegraphics[width=\textwidth]{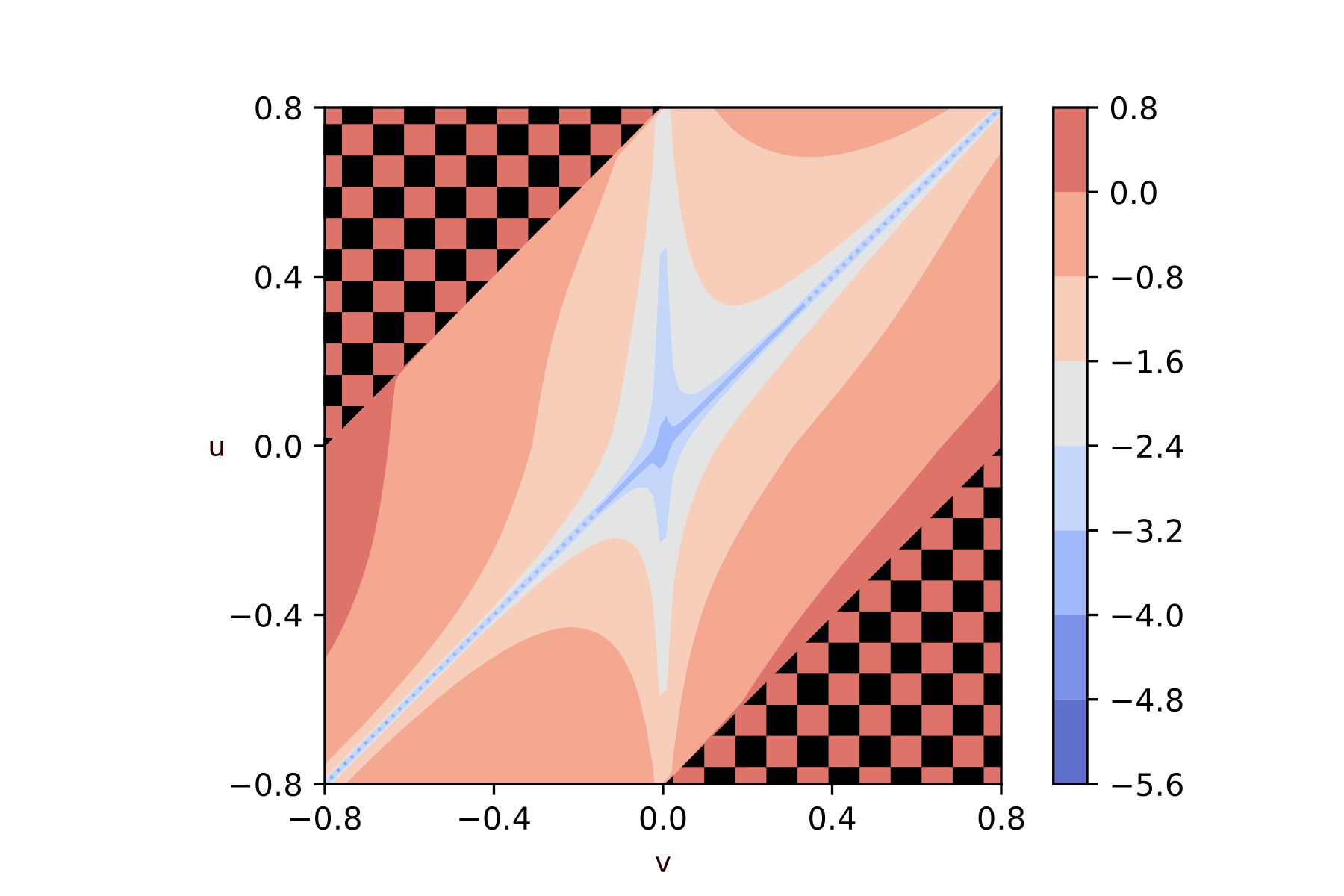}
         \caption{$H_{\mathrm{class}-4}$ deformation}
         \label{fig:hclass1nintnormybe}
     \end{subfigure}%
        \caption{(a) The normalized Yang-Baxter error \eqref{eq:ybmetric} plotted in the logarithmic scale 
        at the end of training for the Hamiltonian $H_{class-4}$ from \eqref{nonXYZhamiltonians}, with $a_1=0.5, a_2=0.3, a_3=0.4, a_4=0.9$, (b) Non-integrable deformation with same Hamiltonian parameters as in the integrable case, except for $H_{13}=-0.9$.}
        \label{fig:yberrclass4}
\end{figure}

One can further discriminate
between integrable and non-integrable models by checking the point-wise values of the Yang-Baxter losses in the two cases. Let us define the metric
\begin{equation}\label{eq:ybmetric}
    \Tilde{\mathcal{L}} = 
    \frac{\vert \vert
    \mathcal{R}_{12}(u-v)\mathcal{R}_{13}(u)\mathcal{R}_{23}(v)-\mathcal{R}_{23}
    (v)\mathcal{R}_{13}(u)\mathcal{R}_{12}(u-v)\vert\vert}
    {\vert \vert
    \mathcal{R}_{12}(u-v)\mathcal{R}_{13}(u)\mathcal{R}_{23}(v)\vert\vert}\,,
\end{equation}
which measures the relative error in the approximate solution of the Yang-Baxter equation. This metric is evaluated for the trained R-matrix for both integrable and non-integrable models in Figure \ref{fig:yberr6v1} (for $H_{6v,1}$ model), and Figure \ref{fig:yberrclass4} (for $H_{class-4}$ model). We see that the normalized error can be up to two orders of magnitude larger for the non-integrable case. Note that irrespective of the choice of Hamiltonian, there are two lines along $u=v$ and $v=0$ on which the Yang-Baxter equation is trivially satisfied, due to regularity. This metric also can detect anomalous situations when the learned solution once satisfied the Hamiltonian constraint at \(u=0\) quickly evolves to a true solution of Yang-Baxter equation producing relatively small YB loss \eqref{eq:lossybe}. In this case we will see the big spike in \eqref{eq:ybmetric} around zero which will indicate the fakeness of the found solution. 

The above consideration shows that one can define the metrics which together indicate the closeness of the given system to the integrable Hamiltonian. However, the final conclusion in the binary form of ``integrable/nonintegrable" regarding the given spin chain can be made only asymptotically, namely increasing the number of neurons, density of points and training time one can get the normalized YB loss \eqref{eq:ybmetric} uniformly decreasing to zero for integrable Hamiltonians while for nonintegrable case it will be bounded from below by some positive value. Also let's stress that such problem is specific for the solver mode once we stick to a given Hamiltonian, while in the case of relaxed Hamiltonian restrictions as we will see in the next section, the neural network moves to the true solution of the Yang-Baxter equation.

\subsection{Explorer: new from existing}\label{subsec:Explorer}

\begin{figure}[h]
    \centering
    \includegraphics[width=1\textwidth]{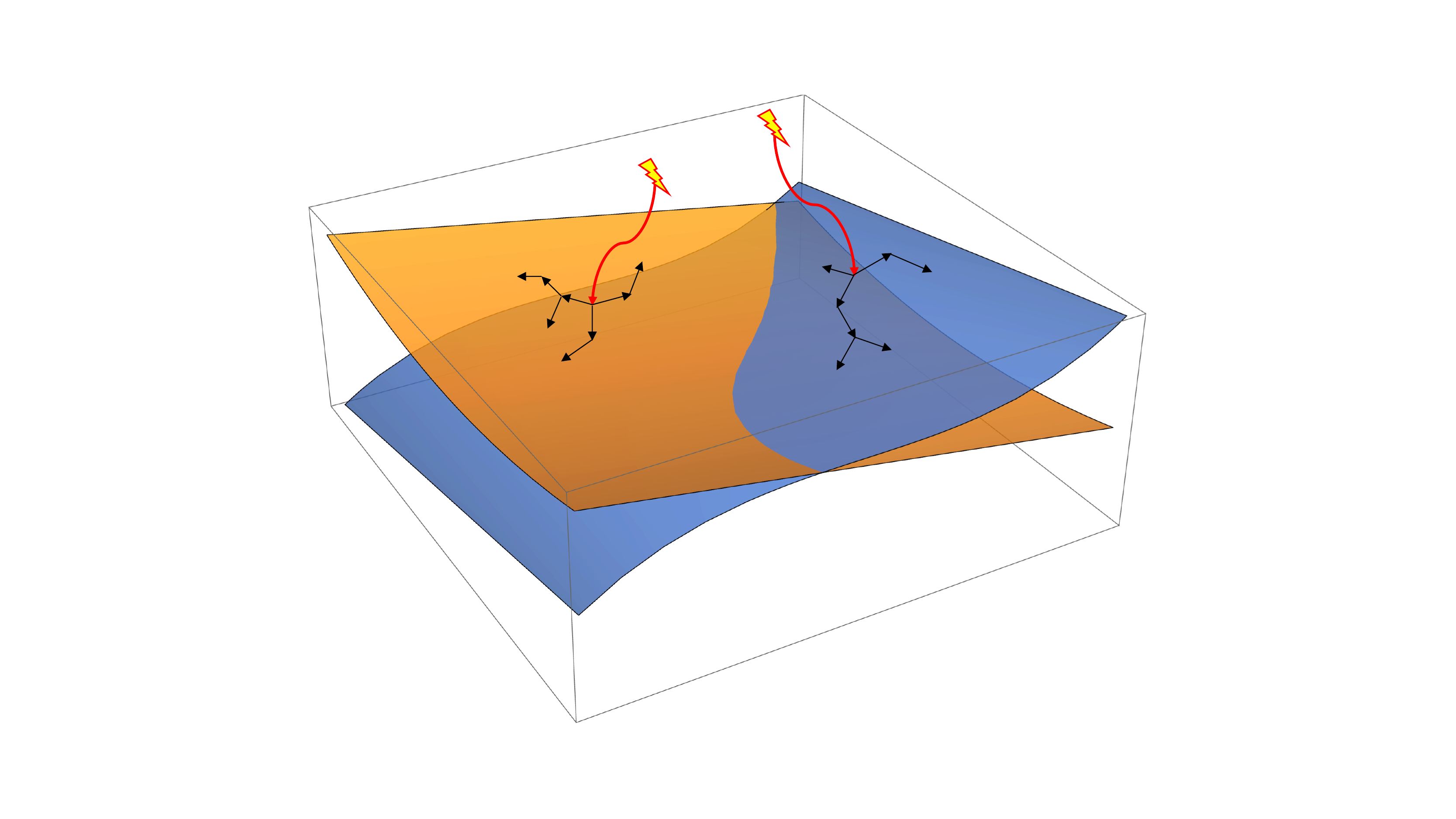}
    \caption{Visualizing the Explorer scheme. We start with random
    initializations, marked by lightning symbols, and perform 
    solver learning represented by red curve arrows.    
    Once we reach an submanifold of integrable Hamiltonians, we explore it using repulsion 
    to identify new integrable models.
    }
    \label{fig:ExplorerScheme}
\end{figure}

\begin{figure}
    \centering
    \begin{subfigure}[b]{0.5\textwidth}
        \centering
        \includegraphics[width=\textwidth]{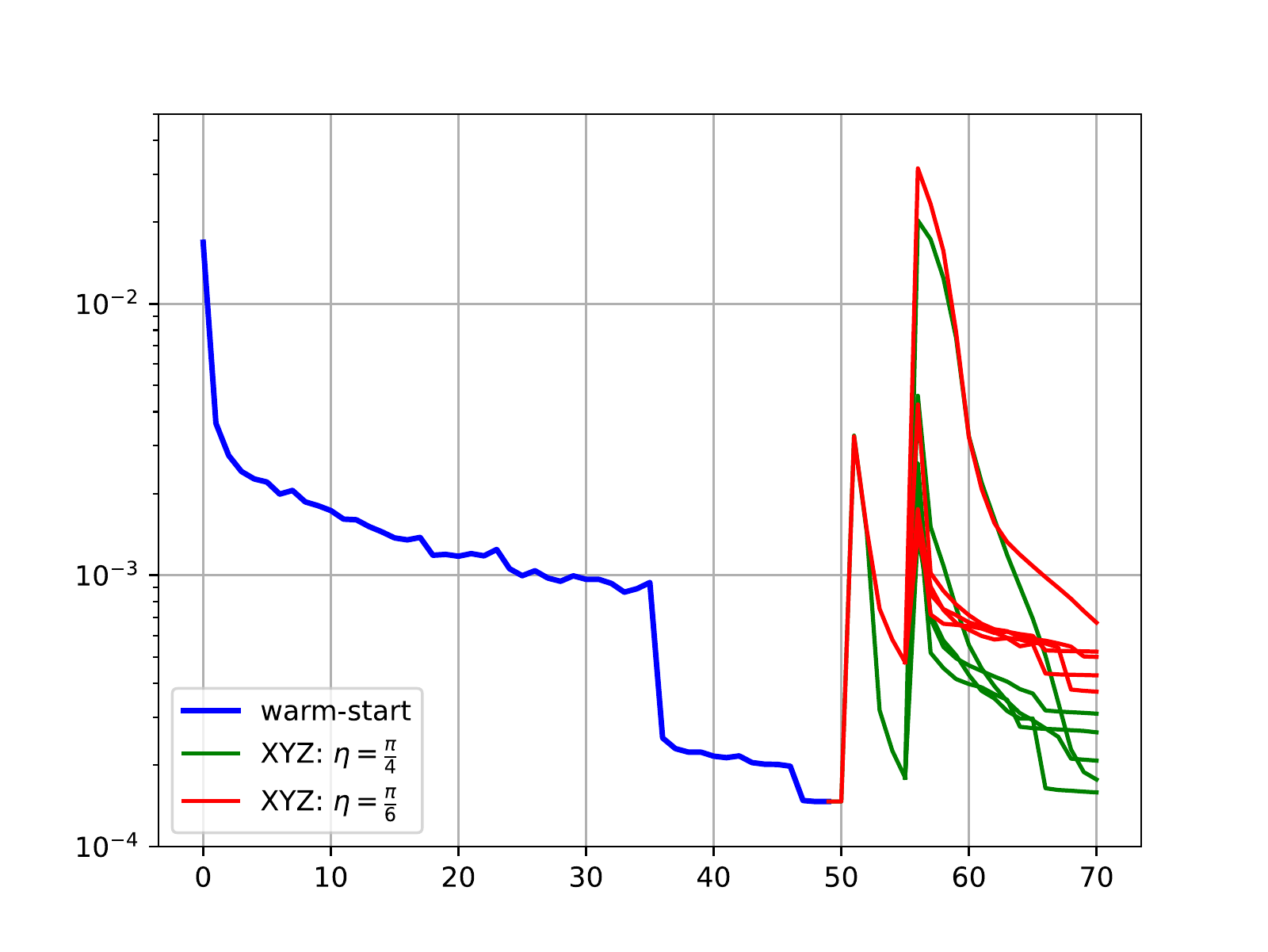}
         \caption{Evolution of Yang-Baxter Loss}
         \label{fig:xyz_exp_loss}
     \end{subfigure}%
     \begin{subfigure}[b]{0.5\textwidth}
         \centering
         \includegraphics[width=\textwidth]{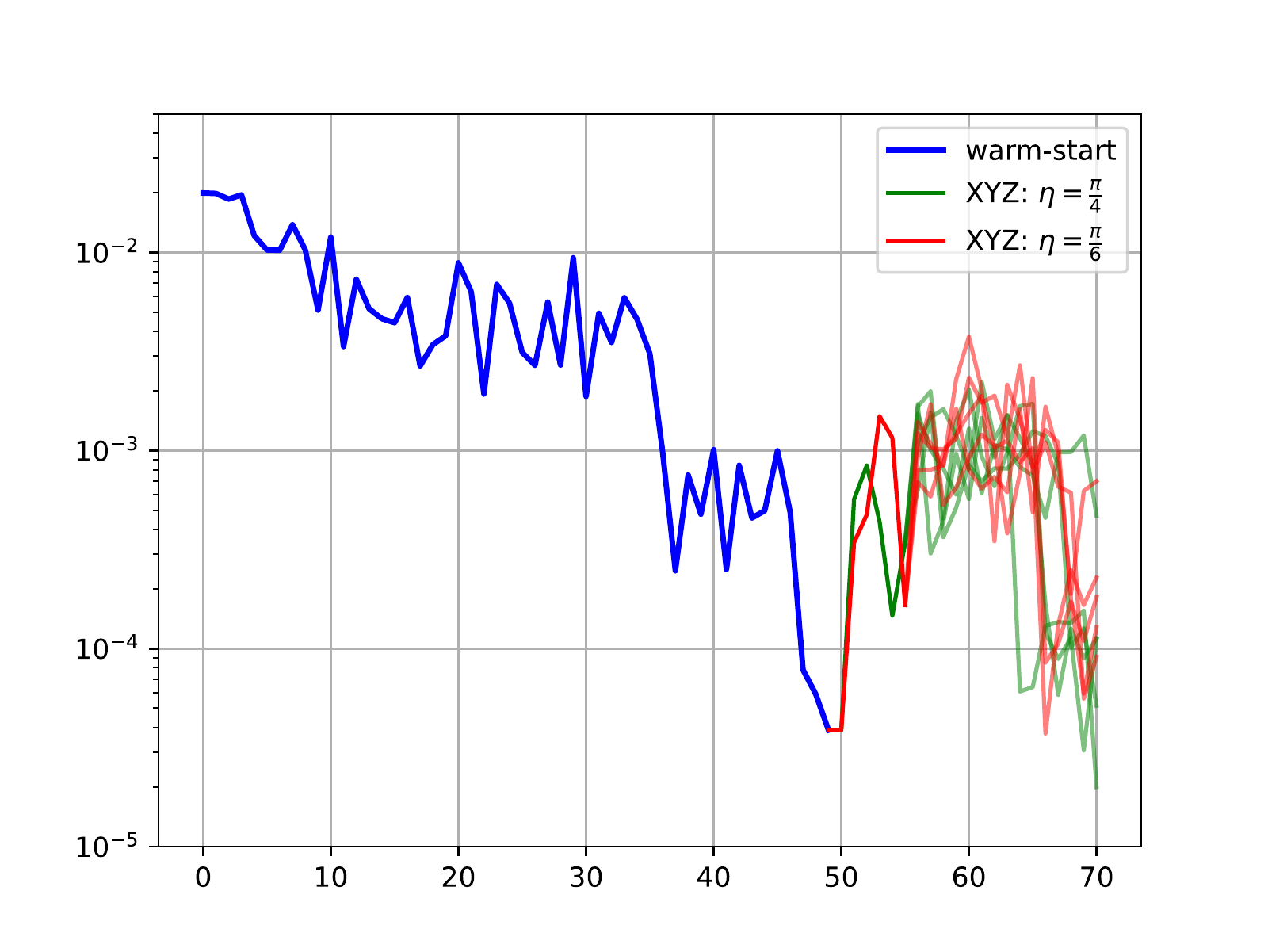}
         \caption{Evolution of Hamiltonian Loss}
         \label{fig:xyz_exp_hloss}
     \end{subfigure}%
    \caption{The convergence to XYZ models from XXZ models trained with different parameters. XXZ was trained for 50 epochs at $\eta=\frac{\pi}{3}$ and $m=0$. Then, it was trained for 5 more epochs at $\eta=\frac{\pi}{4}$ and $\eta=\frac{\pi}{6}$, still with $m=0$. After that, 5 non-zero values of $m$ were used for each XXZ model, and we trained for another 15 epochs. Loss spikes occurred when the target hamiltonian values were reset. The final training was run in parallel for convenience, but it can be run sequentially.}
    \label{fig:xyz_from_xxz}
\end{figure}

\begin{figure}
    \centering
    \includegraphics[width=0.8\textwidth,height=0.6\textwidth]{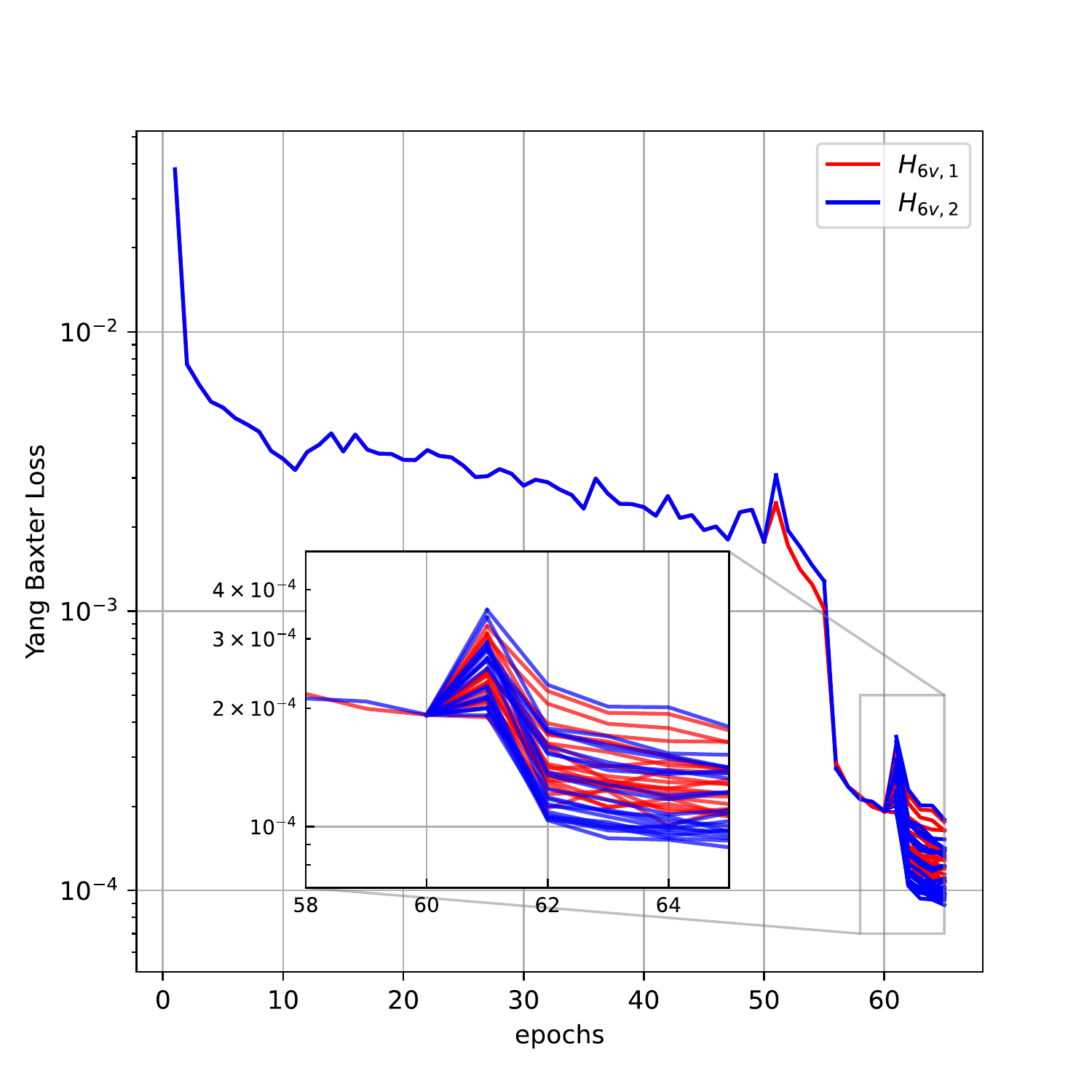}
    \caption{Time evolution of the Yang Baxter loss as the neural network explores
    the space of integrable Hamiltonians of 6-vertex models $H_{6v1,6v2}$ by repulsion. The loss evolves together until the $50^{th}$ epoch after which it fragments slightly as the training converges to the two warm-start points on the 60th epoch. For the remaining epochs the losses fragment completely as the neural network seeks out different new Hamiltonians
    and is terminated when the loss reaches the neighborhood of 
    $1\times 10^{-4}$.}
    \label{fig:6vexploretraining}
\end{figure}

\begin{figure}
    \centering
    \begin{subfigure}[b]{0.5\textwidth}
        \centering
        \includegraphics[width=\textwidth]{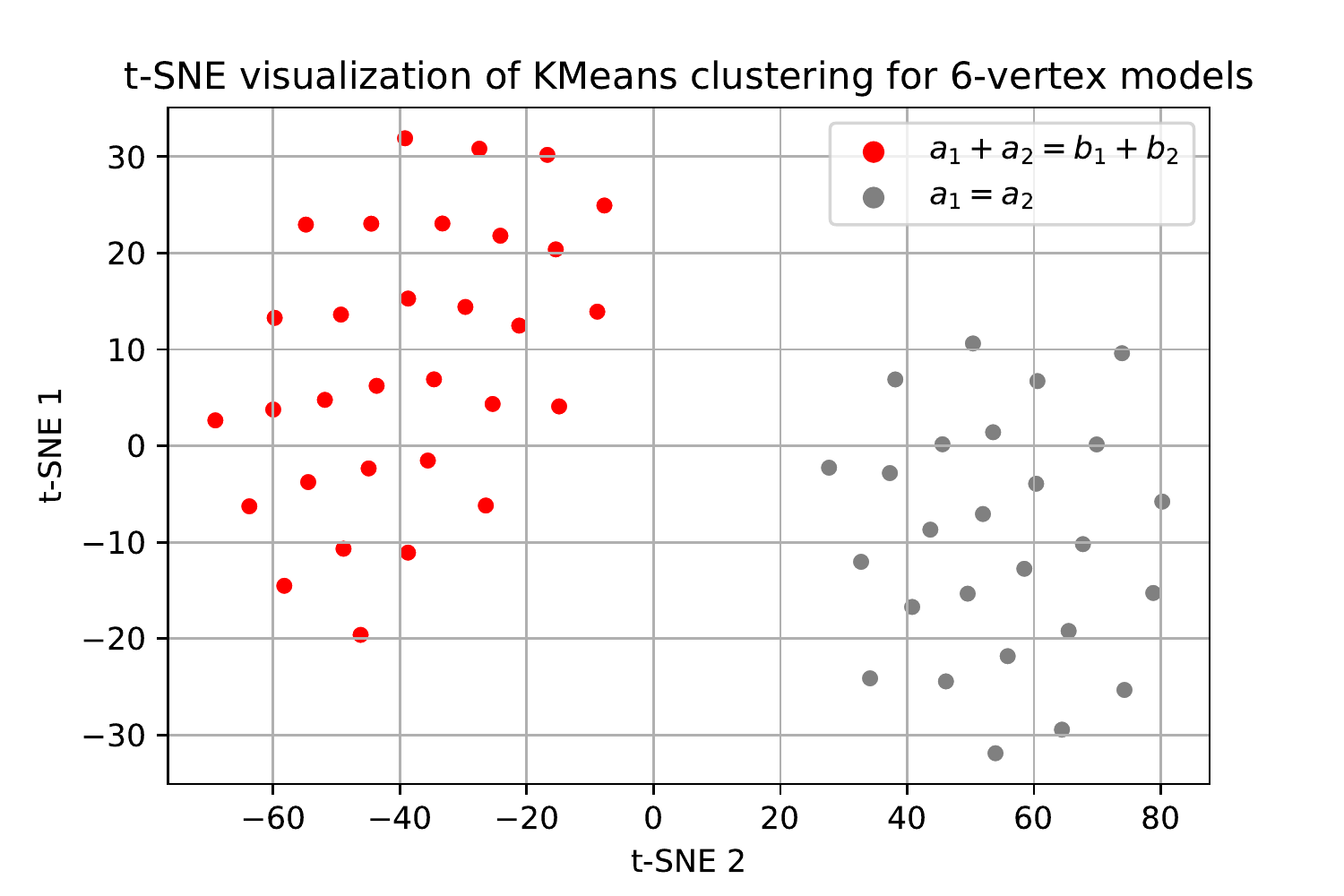}
         \label{fig:6v_tsne}
     \end{subfigure}%
     \begin{subfigure}[b]{0.5\textwidth}
         \centering
         \includegraphics[width=\textwidth]{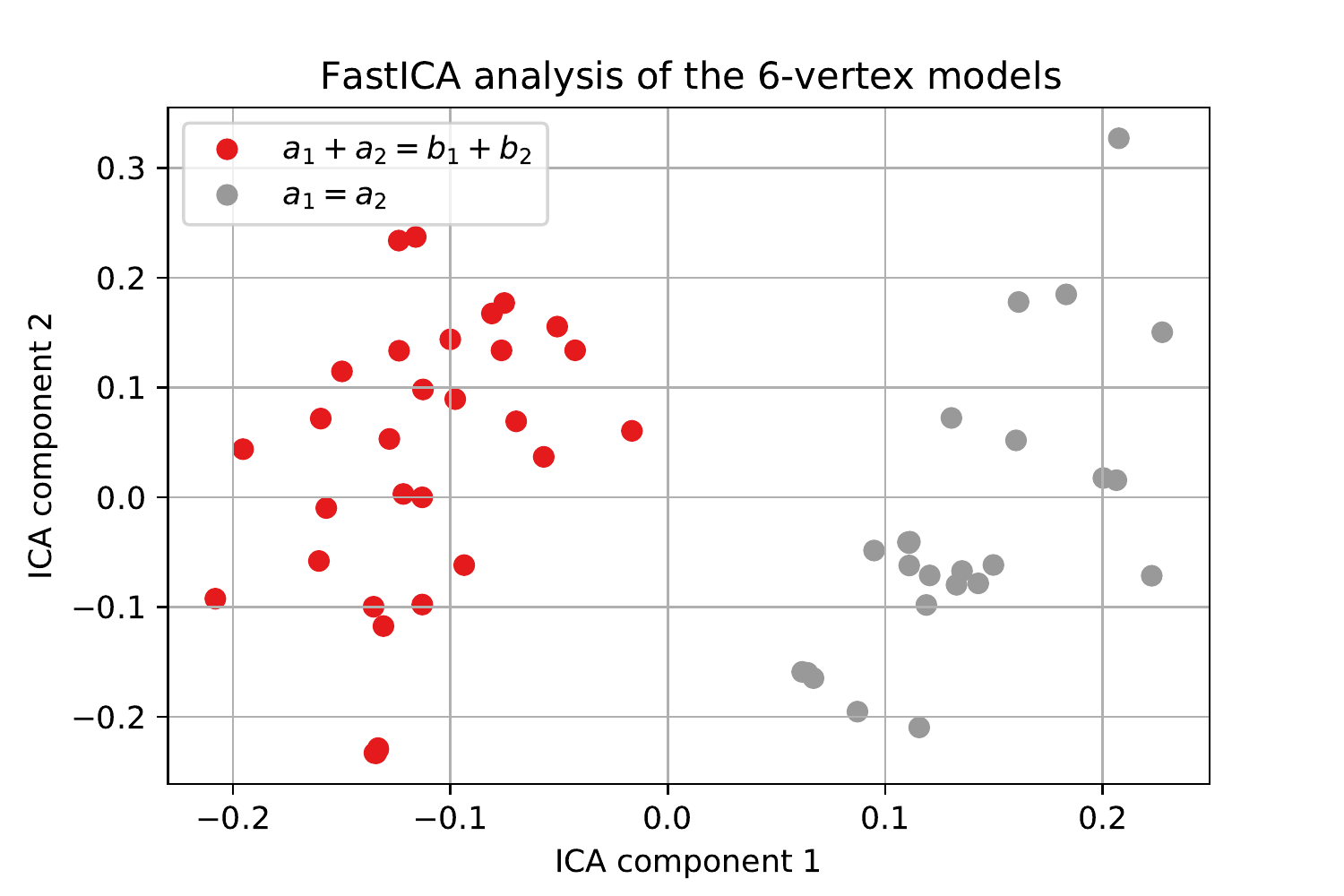}
         \label{fig:6v_fastICA}
     \end{subfigure}%
    \caption{Clustering of Hamiltonians from the 2 classes of gauge-inequivalent 6-vertex models obtained by Explorer using repulsion from solution at intersection of both classes.}
    \label{fig:K-means_clustering-6v}
\end{figure}

\begin{figure}
    \centering
    \includegraphics[width=.9\textwidth]{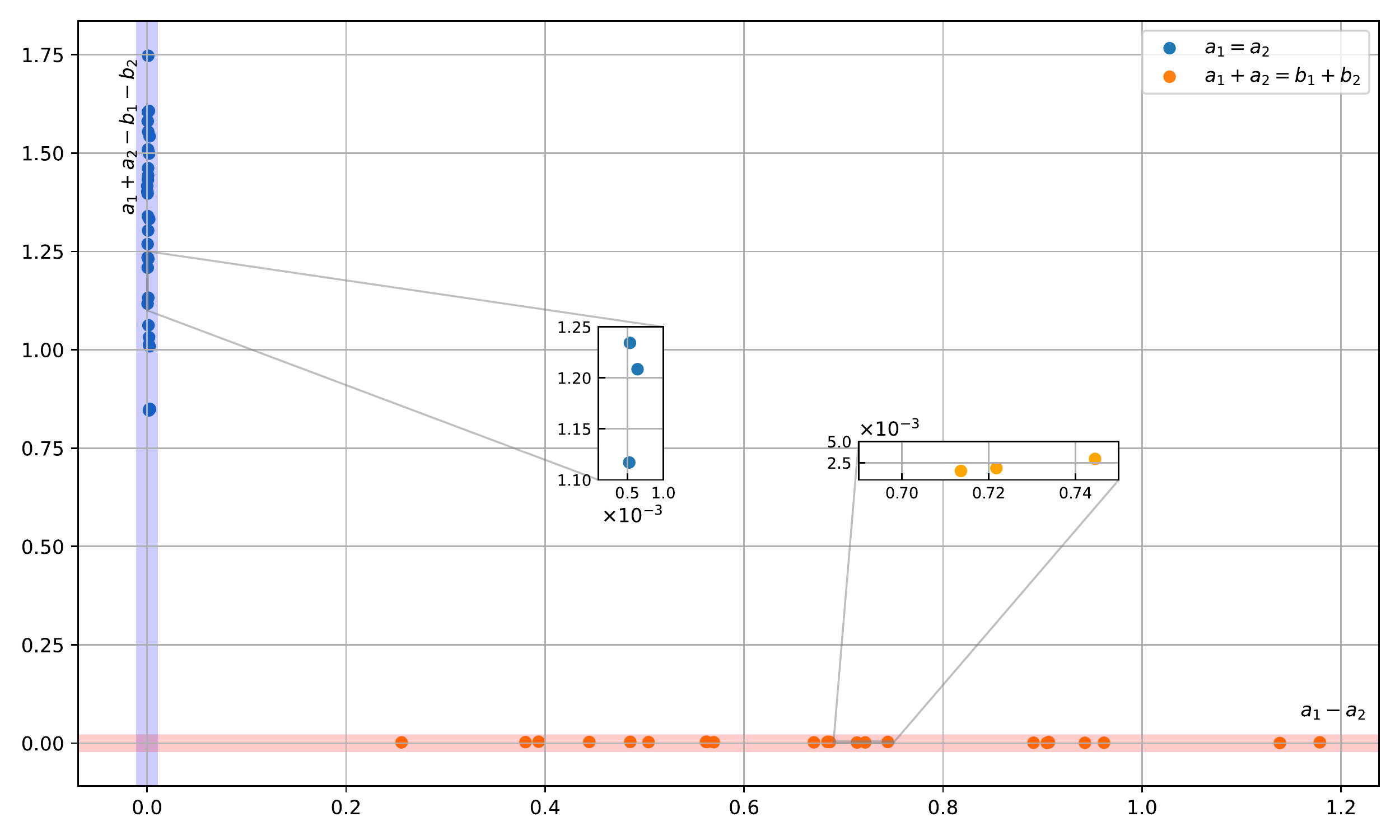}
    \caption{The 6-vertex models learnt by exploration. The graph visualizes 
    the obtained Hamiltonians by plotting their values along the 
    $a_1+a_2-b_1-b_2$ and the $a_1-a_2$ axes. The models $H_{6v,1}$
    lie along the $y$-axis and the models $H_{6v,2}$ along the $x$-axis with
    an error margin of order $10^{-3}$ as shown in the telescoped inset plots.}
    \label{fig:2-leaf}
\end{figure}
In this section we will present two kinds of experiments that illustrate how the
neural network presented above can be used to scan the landscape of two-dimensional spin-chains for integrable models. The training schedule adopted
in this section is visualized in Figure \ref{fig:ExplorerScheme} and
relies essentially on two new ingredients
which distinguish it from the previous \textit{solver} framework. These are
\textit{warm-start} and \textit{repulsion}. We will illustrate each by an
example. In the first case we shall simply use warm-start, and in the second, we 
shall combine warm-start with repulsion. Finally, we shall use unsupervised 
learning methods such as t-SNE and Independent Component Analysis to identify
distinct classes of Hamiltonians within the set of
integrable models thus discovered. Collectively, these strategies make up our
\textit{explorer} framework.

The first key new ingredient is a \textit{warm-start initialization}. As mentioned
previously, the standard solver framework of the previous section uses He 
initialization \cite{he2016deep}  to instantiate the weights and biases of the
neural network. In warm-start initialization, we use the knowledge of 
integrable systems previously discovered by the neural network to find 
new systems in its vicinity. The idea, at least intuitively, is that it should be
possible to find new integrable systems more efficiently than with the random initialization by exploring the vicinity in weight-space
of previously determined solutions using an iterative procedure such as gradient-based optimization. On doing so, we find a significant acceleration in 
training convergence, with new solutions being discovered typically in about 
5 epochs of training after warm-start initialization. For definiteness, we consider
the hermitian XYZ model discussed earlier in 
Section~\ref{subsubsec:XYZ-XXZ-XXX}. This has a two-parameter family of solutions,
corresponding to independent choices for the parameters $\eta$ and $m$ of the
Jacobi elliptic function, as seen from Equation \eqref{eq:xyz_rmat_analytic}.
The XXZ model is embedded into this space as the $m=0$ subspace of solutions.

We now describe how the above strategy can be used to quickly generate the cluster of 
XYZ R-matrices starting from a particular one which we choose from XXZ subclass. We begin with 
pre-training our neural network using the solver mode of the previous section,
but with the learning rate of the \texttt{Adam} optimizer set to $10^{-3}$. The
pre-training is stopped when all losses saturate below $\mathcal{O}\left(10^{-3}
\right)$, which typically requires about 50 epochs of training. We carried out
this pre-training setting arbitrary reference values of $\eta$, but with $m$
fixed to zero. The results shown here correspond to $\eta=\frac{\pi}{4}$.
The weights thus obtained correspond to our warm-start values. Then we shift the target Hamiltonian values to correspond to 
$\eta \rightarrow \eta +\delta\eta$, where $\delta\eta$ are randomly chosen
$\mathcal{O}\left(10^{-1}\right)$ numbers, and $m$ can take on non-zero values as
well. We then retrain the model with a smaller learning rate, $10^{-4}$
for a few epochs until all loss terms fall to 
$\mathcal{O}\left(10^{-4}\right)$, which typically takes about 5 epochs, upon
which we update the target Hamiltonian by updating $\eta$ and $m$ and continue
training. This strategy generates about 10 
XYZ models within the same time-scale (i.e. about 100 to 200 
epochs of training) as we 
earlier needed for a single model. For best results, while we randomly update
$\eta$, we systematically anneal the modular parameter $m$ to upwards of zero.
A sample of this training is visualized in Figure \ref{fig:xyz_exp_hloss}.

Our next key new ingredient for the Explorer mode  is \textit{repulsion}, which is
added to the previous strategy of warm-start initialization. In principle, it should allow us to rediscover all 14 classes of integrable spin chains. However, for sake of simplicity, we will illustrate it now with a toy-model example and return to the general analysis later \cite{3dAIintegrability}. Namely,
we consider the class of 6-vertex Hamiltonians with unrestricted \(a_1\) and \(a_2\). It includes both integrable 6-vertex classes $H_{6v,1},H_{6v,2}$ (\ref{eq:Hamiltonian6v,1}, \ref{eq:Hamiltonian6v,2}) as well as nonintegrable models. In order to mimic the general situation when all integrable classes intersect at zero, we begin by pre-training the neural network to a Hamiltonian belonging to the intersection
of the classes $H_{6v,1}$ and $H_{6v,2}$, i.e. whose matrix element satisfy the constraints $a_1=a_2$ and $a_1+a_2=b_1+b_2$ simultaneously. The results mentioned in this paper correspond to setting 
\begin{equation}
    a_1=a_2=\frac{b_1+b_2}{2}\,;\qquad
    b_1=0.6,\,b_2=0.8,\,c_1=0.5,\, c_2=0.9\,.    
\end{equation}
Having arrived at this model, we would like to navigate to neighboring 
models not by specifying target values of the Hamiltonian, but by scanning the
neighborhood of the current model. To do so, we employ a two step strategy.
First, we navigate to two\footnote{We stop the scanning once we found a representative from each of two classes because we know that there are only two integrable families here. In general case one of course should generate sufficiently many points in order to find all classes. We will return to this subtle point later in \cite{3dAIintegrability}} new 6-vertex integrable Hamiltonians by
random scanning the vicinity of the current model without giving specific target
values. We shall use these new models as our warm-start points. 
From each of them, 
we navigate away by using the 
$repulsion$ loss term \eqref{eq:replusionloss} for 1 epoch, followed by training for another 5 epochs. 
Note in this step, we still train within the restricted class of 
6-vertex models by fixing the corresponding entries of the R-matrix to zero. We repeat this 
schedule 25 times starting from either of the saved models. This way, 
we generate fifty 6-vertex integrable Hamiltonians
with over 1\% accuracy\footnote{If we further train the individual models for
more epochs, we can improve the accuracy of the obtained solution to 
similar levels as obtained in the examples presented in 
Section~\ref{subsec:specific_search}.}. 
The training curve displaying how the Yang-Baxter loss evolves is shown in
Figure \ref{fig:6vexploretraining}.

The learnt models are classified into two classes using 
clusterisation methods as shown in 
Figure~\ref{fig:K-means_clustering-6v}. Figure~\ref{fig:2-leaf} 
plots the trained models in terms of coordinates defined by the integrability conditions of the Hamiltonians $H_{6v,1},H_{6v,2}$. Models lying near the two 
axes were classified correctly into the two classes in 
Figure~\ref{fig:K-means_clustering-6v} with $100\%$ accuracy.

\section{Conclusions and Future directions}\label{sec:futuredirections}

In this paper we constructed a neural network for solving the Yang-Baxter equation
\eqref{eq:ybequation} in various contexts. Firstly, it can learn the
R-matrix corresponding to a given integrable Hamiltonian or search for an integrable spin chain and the corresponding R-matrix
from a certain class specified by imposed symmetries or other restrictions. We refer to this as the solver mode.
Next, in the explorer mode, it can search for new integrable models by
scanning the space of Hamiltonians.

We demonstrated the use of our neural network on two-dimensional spin chains of difference form. In the solver mode, the network successfully learns all fourteen distinct classes of R-matrices identified in \cite{de2019classifying}
to accuracies of the order of $0.01-0.1\%$. 
We demonstrated the work of the Explorer mode, restricting the search to the space of spin chains containing both classes of 6-vertex models as well as  nonintegrable Hamiltonians. Starting from the hamiltonian at the intersection of two classes , Explorer found 50 integrable Hamiltonians which after clusterisation clearly fall into two families corresponding to two integrable classes of 6-vertex model.
Working in the explorer mode, we find that warm-starting our
training from the vicinity of a previously learnt integrable model greatly speeds
up convergence, allowing us to identify typically about 50 new integrable models
in the same time that random initialization takes to converge to a single model. 

The main focus of this paper was creating the neural network architecture and
demonstrating its robustness in various solution generating frameworks 
using known integrable models as a testing ground. 
However, we expect that
this program can be extended to various scenarios such as the exploration and
classification of the space of integrable Hamiltonians in dimensions greater than
two. This would be of great interest since the general classification of models
is currently limited to two dimensions. Our experiments with exploration and
clustering are a promising starting point in this regard. In our setup the strategy is quite straightforward \cite{3dAIintegrability}. Because all integrable families of Hamiltonians can be multiplied by arbitrary scalar, we should only scan the Hamiltonians on the unit sphere which is compact. Scanning over sufficiently dense set of points on the sphere will allow us to identify integrable Hamiltonians from various classes. Then we can use the Explorer to reconstruct the whole corresponding families and perform clusterisation in order to identify them. On another footing, it would also be interesting to extend our study to R-matrices of non-difference
form as these are particularly relevant to the AdS/CFT correspondence \cite{beisert2003n,borsato2013all,majumder2021protected,frolov2022mirror}.

While our network learns a numerical approximation to the R-matrix, it can also 
be useful for the reconstruction of analytical solutions using 
symbolic regression \cite{udrescu2020ai,schmidt2009distilling}.
Alternately, one may try to use the learnt numerical solution 
for the reconstruction of the symmetry algebra such as the Yangian 
and then arrive at the analytical solution. Remarkably, machine learning 
is already proving helpful in the analysis of symmetry in physical systems. 
In particular, one may verify the presence of a conjectured symmetry or even
automate its search using machine learning 
\cite{Chen:2020dxg,liu2022machine, liu2021machine, bondesan2019learning,
melkosiamese,forestano2023deep,quessard2020learning}. It would be very
interesting to explicate the interplay of our program in this broader 
line of investigation.

In addition, the flexibility of our approach would also allow us to implement
various additional symmetries or other restrictions, both at the level of the
R-matrix and the Hamiltonian. It would therefore be very interesting to 
develop an `R-matrix bootstrap' in the spirit of the two-dimensional S-matrix
bootstrap and analyze the interplay between various symmetries. For example,
all 14 families of R-matrices considered in this paper 
satisfy the condition of braided unitarity \eqref{eq:braidedunitarity} and it would be interesting to rediscover them from the use of braided unitarity and other symmetries without imposing the Yang-Baxter condition, similar to how integrable two-dimensional S-matrices have been identified in the S-matrix Bootstrap approach \cite{Paulos:2016but,He:2018uxa,Paulos:2018fym}.

With mild modifications, we can adapt our architecture to the analysis of Yang-Baxter equation for the integrable S-matrices in two dimensions. The only new feature to implement is the analytic structure in the  s-plane. It can be naturally realized with the use of holomorphic networks. 

Learning solutions for different classes with the same architecture, we noticed that the number of epochs needed to reach the same precision varies for different classes while being roughly the same for the Hamiltonians from the same classes. Thus, it would be very tempting to use the training of losses to define the complexity of spin chains. Ideally, we should be able to go beyond the class of integrable models and see that they sit at the minima of complexity, matching common beliefs that the integrable models are the ``simplest" ones.

\section*{Acknowledgements}
We thank Jiakang Bao, Jim Halverson, Ed Hirst, Vladimir Kazakov, Sven 
Krippendorf, Praneeth Netrapalli, Hongfei Shu, and Hao Zhou for interesting 
discussions. We especially thank Yang-Hui He for initial collaboration and 
several helpful discussions. SM thanks participants at String Data 2022, where 
initial version of the work was presented. SM is grateful to NORDITA for 
hospitality, during which part of the work was completed.

\appendix

\section{Classes of 2D integrable spin chains of difference form}\label{app:XYZ_nonXYZ_plots}
\begin{figure}[htbp]
  \centering
  \captionsetup[subfigure]{}
  \subfloat[]{\includegraphics[width=0.5\linewidth,height=.4\textwidth]{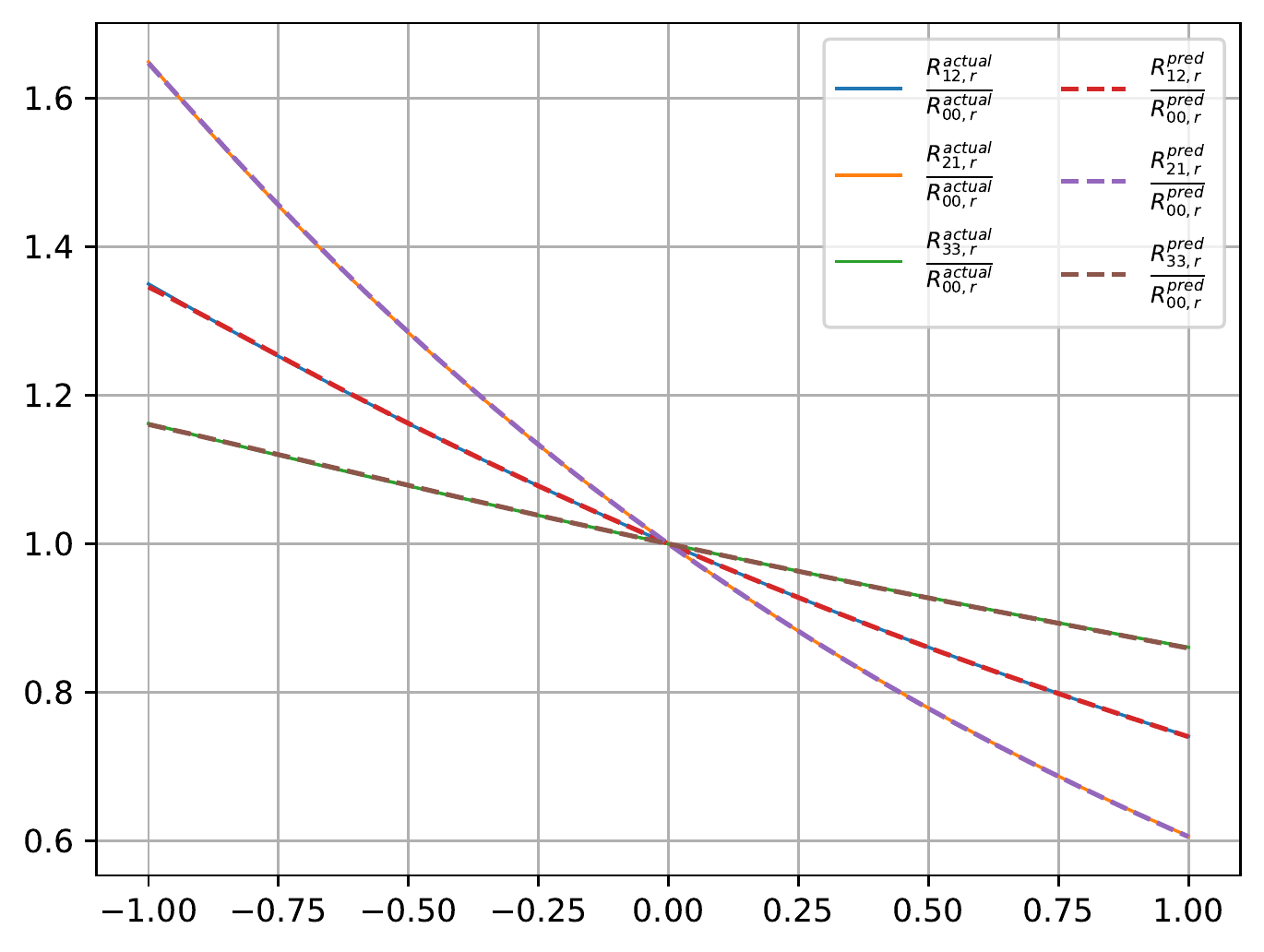}}
  \subfloat[]{\includegraphics[width=0.5\linewidth,height=.4\textwidth]{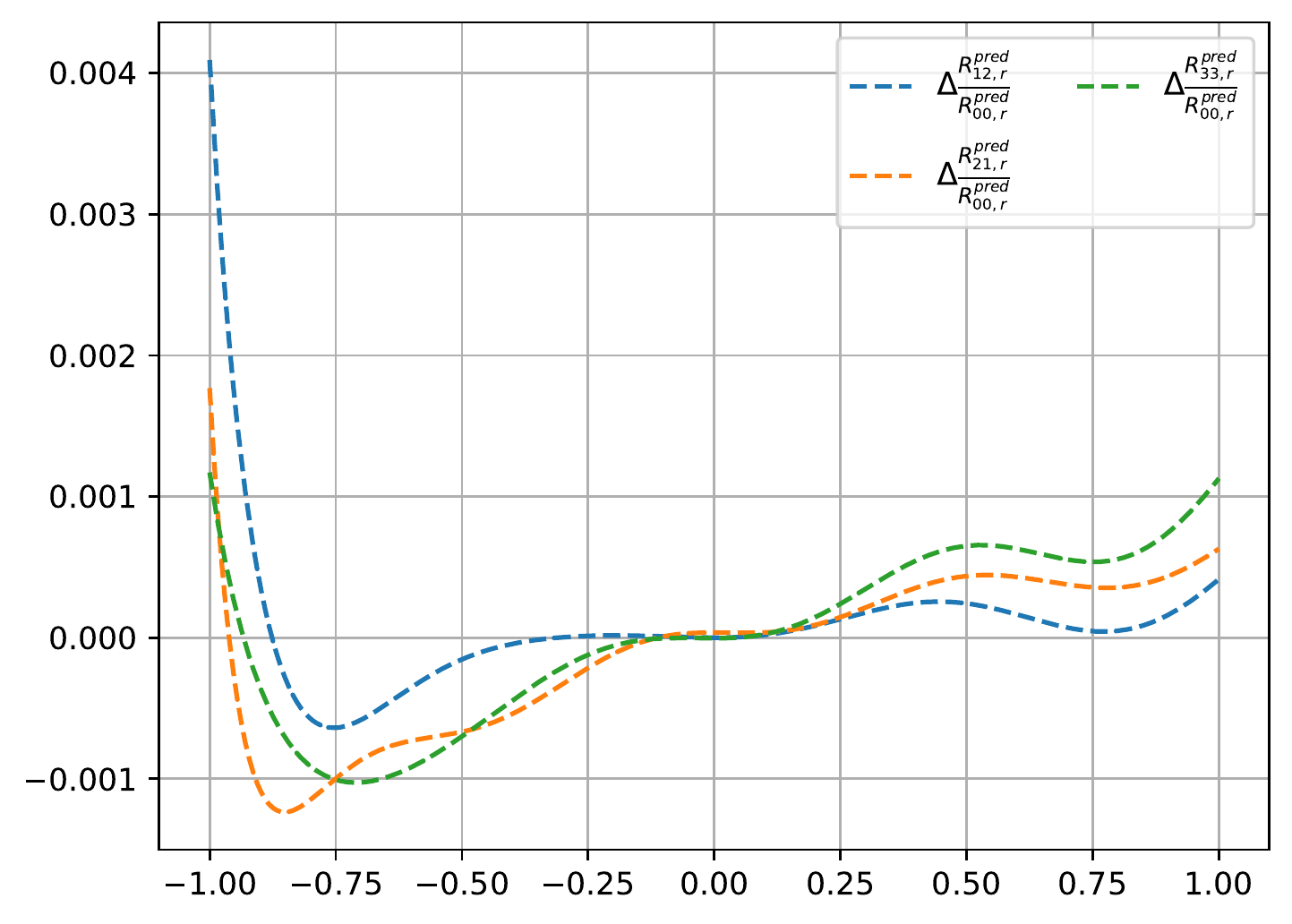}}
  
  \caption{(a) 4-vertex model, with $H_2$ parameters $a_1=0.9,\, b_1=0.4, \, b_2=0.6, \,a_2=0.75$, (b) errors}
  \label{fig:4vertex}
\end{figure}

In this appendix, we list the 14 gauge-inequivalent integrable Hamiltonians of difference form and the corresponding R-matrices.
Amongst the XYZ type models, the simplest solution is a diagonal 4-vertex model 
with Hamiltonians and R-matrices as follows:
\begin{equation}\label{eq:Hamiltonian4v}
    H_{4v}=\begin{pmatrix}
        a_1&0&0&0\\
        0&b_1&0&0\\
        0&0&b_2&0\\
        0&0&0&a_2
    \end{pmatrix}\,\leftrightarrow \quad R_{4v}(u)=\begin{pmatrix}
        e^{a_1 u}&0&0&0\\
        0&0&e^{b_2 u}&0\\
        0&e^{b_1 u}&0&0\\
        0&0&0&e^{a_2 u}
    \end{pmatrix}
\end{equation}
Figure~\ref{fig:4vertex} plots the training curve for R-matrix components as ratios with respect to (00) component, against the analytic functions for parameters $a_1=0.9,b_1=0.4,b_2=0.6,a_2=0.75$.

In 6-vertex models, we have two distinct classes depending on whether the Hamiltonian entries $H^{00}$ and $H^{33}$ are equal  or not. In the first case, the R-matrix $R_{6v,1}(u)$ and its associated Hamiltonian $H_{6v,1}$ are given by
\begin{equation}\label{eq:Hamiltonian6v,1}
    H_{6v,1}=\begin{pmatrix}
        a_1&0&0&0\\
        0&b_1&c_1&0\\
        0&c_2&b_2&0\\
        0&0&0&a_1
    \end{pmatrix}\,\leftrightarrow \quad 
    R_{6v,1}(u)=\begin{pmatrix}
        R_{6v,1}^{00}(u)&0&0&0\\
        0&R_{6v,1}^{11}(u)&R_{6v,1}^{12}(u)&0\\
        0&R_{6v,1}^{21}(u)&R_{6v,1}^{22}(u)&0\\
        0&0&0&R_{6v,1}^{33}(u)
    \end{pmatrix}
\end{equation}
where 
\begin{equation}
\begin{split}
        R_{6v,1}^{00}(u)&=R_{6v,1}^{33}(u)= e^{(b_1+ b_2)u/2}(\cosh{(\omega u)}+\frac{2a_1-b_1-b_2}{2\omega}\sinh{(\omega u)})\\
        R_{6v,1}^{11}(u)&= \frac{c_2}{\omega}e^{(b_1+ b_2)u/2}\sinh{(\omega u)}\\
        R_{6v,1}^{12}(u)&= e^{b_2 u}\\
        R_{6v,1}^{21}(u)&= e^{b_1 u}\\
        R_{6v,1}^{22}(u)&= \frac{c_1}{\omega}e^{(b_1+ b_2)u/2}\sinh{(\omega u)}\,,\qquad \omega =\frac{\sqrt{(2a_1-b_1-b_2)^2-4c_1c_2}}{2}
\end{split}
\end{equation}
Figure~\ref{fig:XXZtype_class1} gives a representative training vs actual plot for this class. 

\begin{figure}[htbp]
  \centering
  \captionsetup[subfigure]{}
  \subfloat[]{\includegraphics[width=0.5\linewidth,height=.4\textwidth]{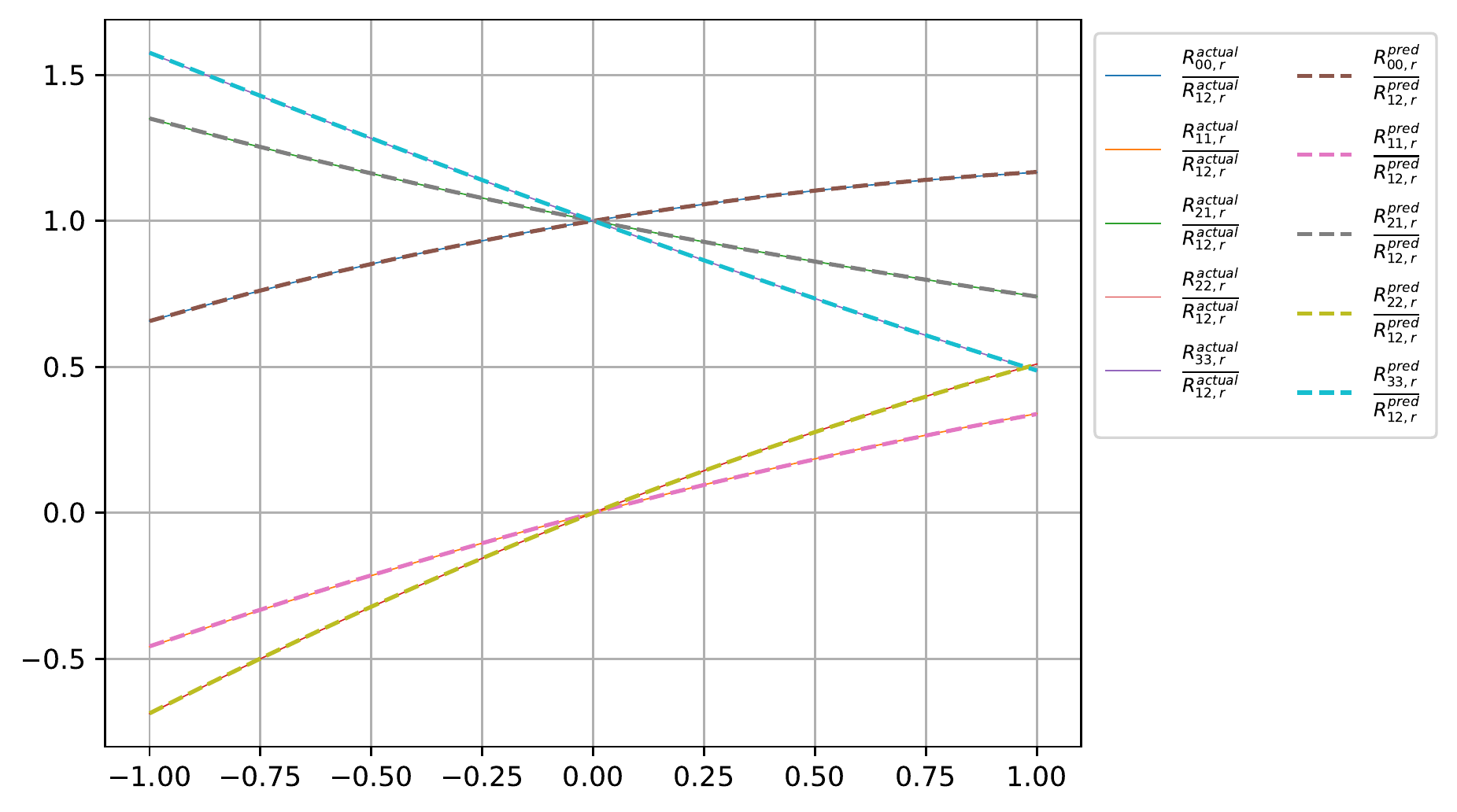}}
  \subfloat[]{\includegraphics[width=0.5\linewidth,height=.4\textwidth]{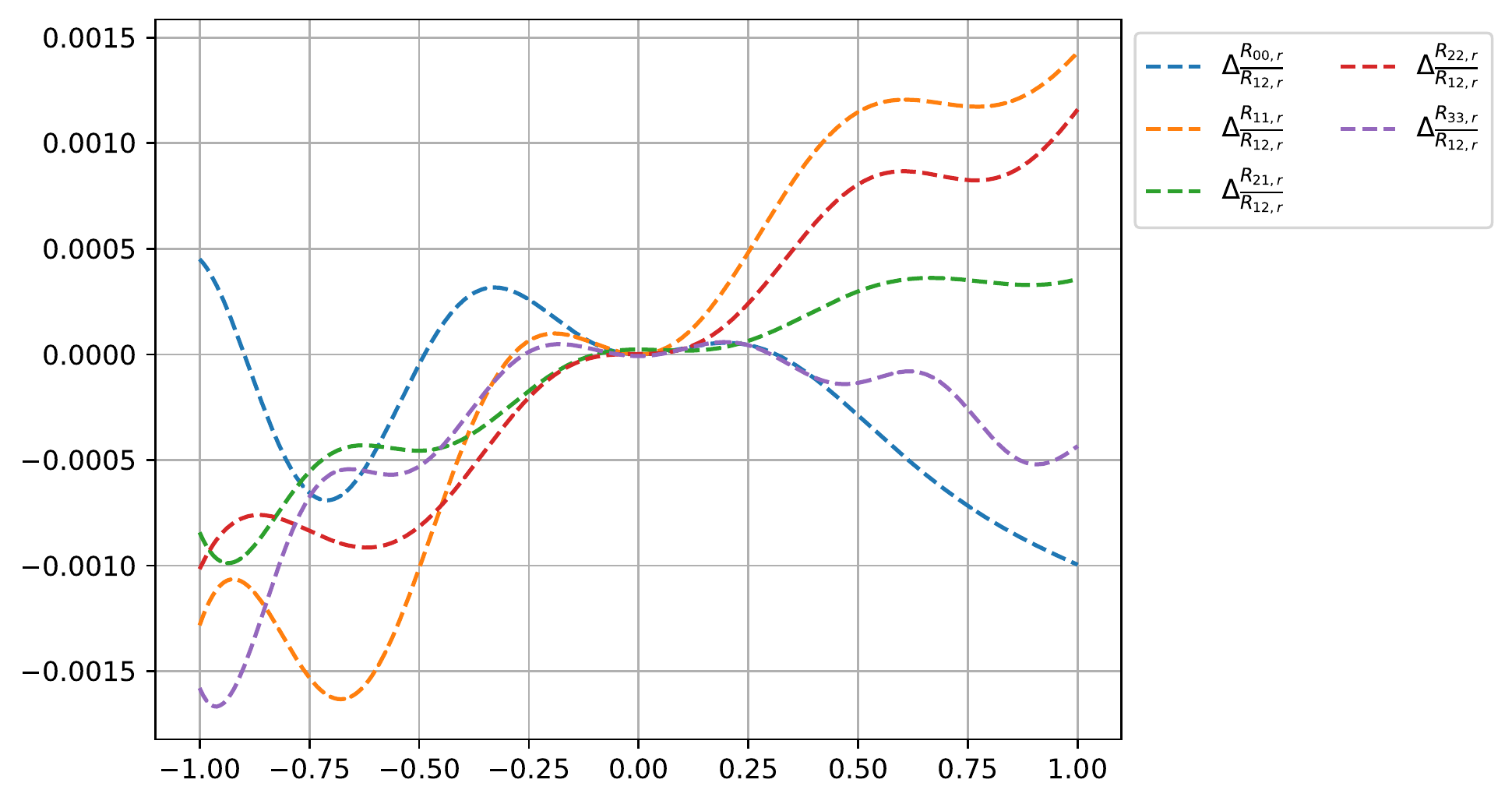}}
  
  \caption{(a) 6-vertex model with Hamiltonian of type $H_{6v,2}$, with parameters $a_1=1,a_2=0.2,b_1=0.45,b_2=0.75,c_1=0.4,c_2=0.6$, (b)errors}
  \label{fig:6vertex-2}
\end{figure}

\begin{figure}[htbp]
  \centering
  \captionsetup[subfigure]{}
  \subfloat[]{\includegraphics[width=0.5\linewidth,height=.4\textwidth]{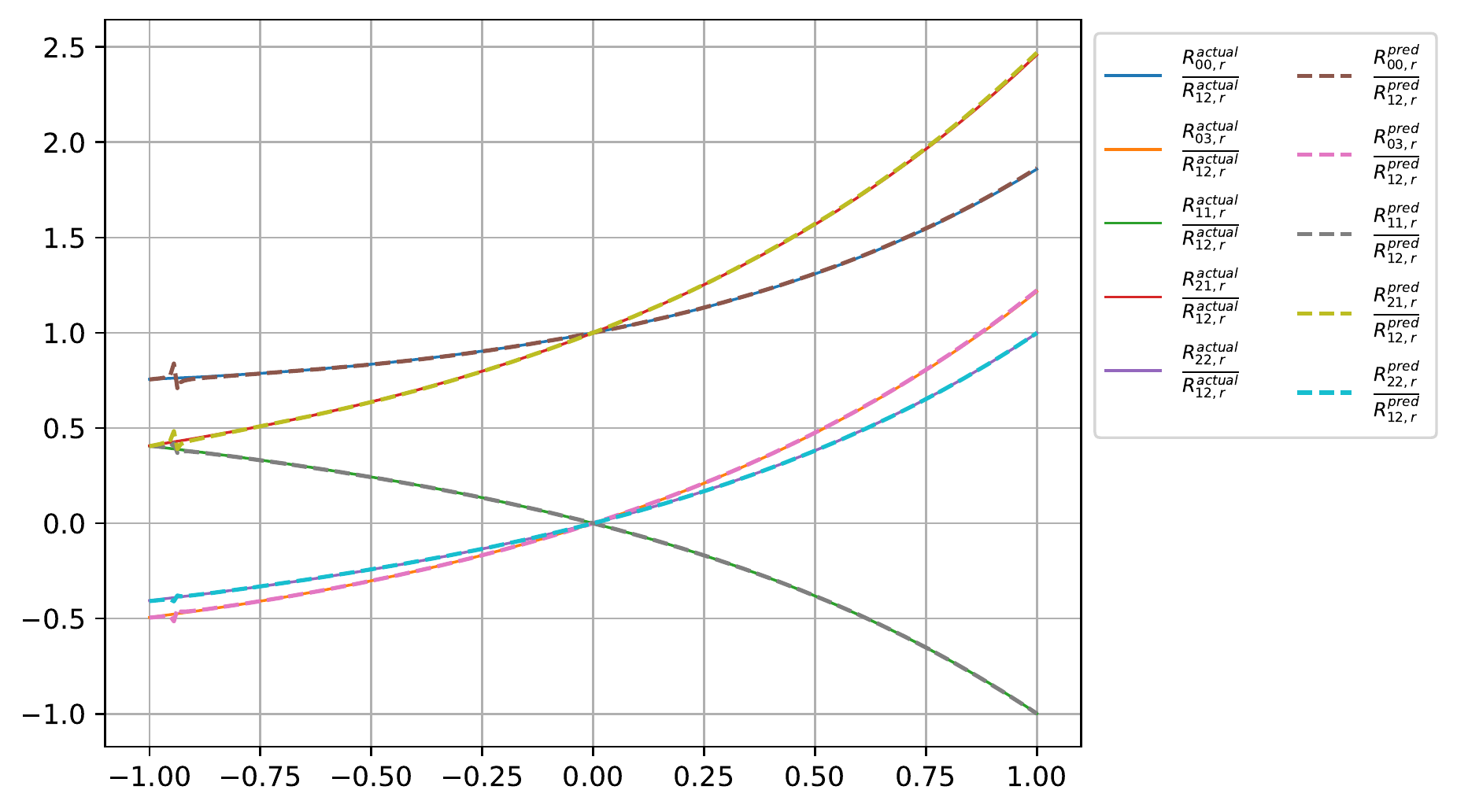}}
  \subfloat[]{\includegraphics[width=0.5\linewidth,height=.4\textwidth]{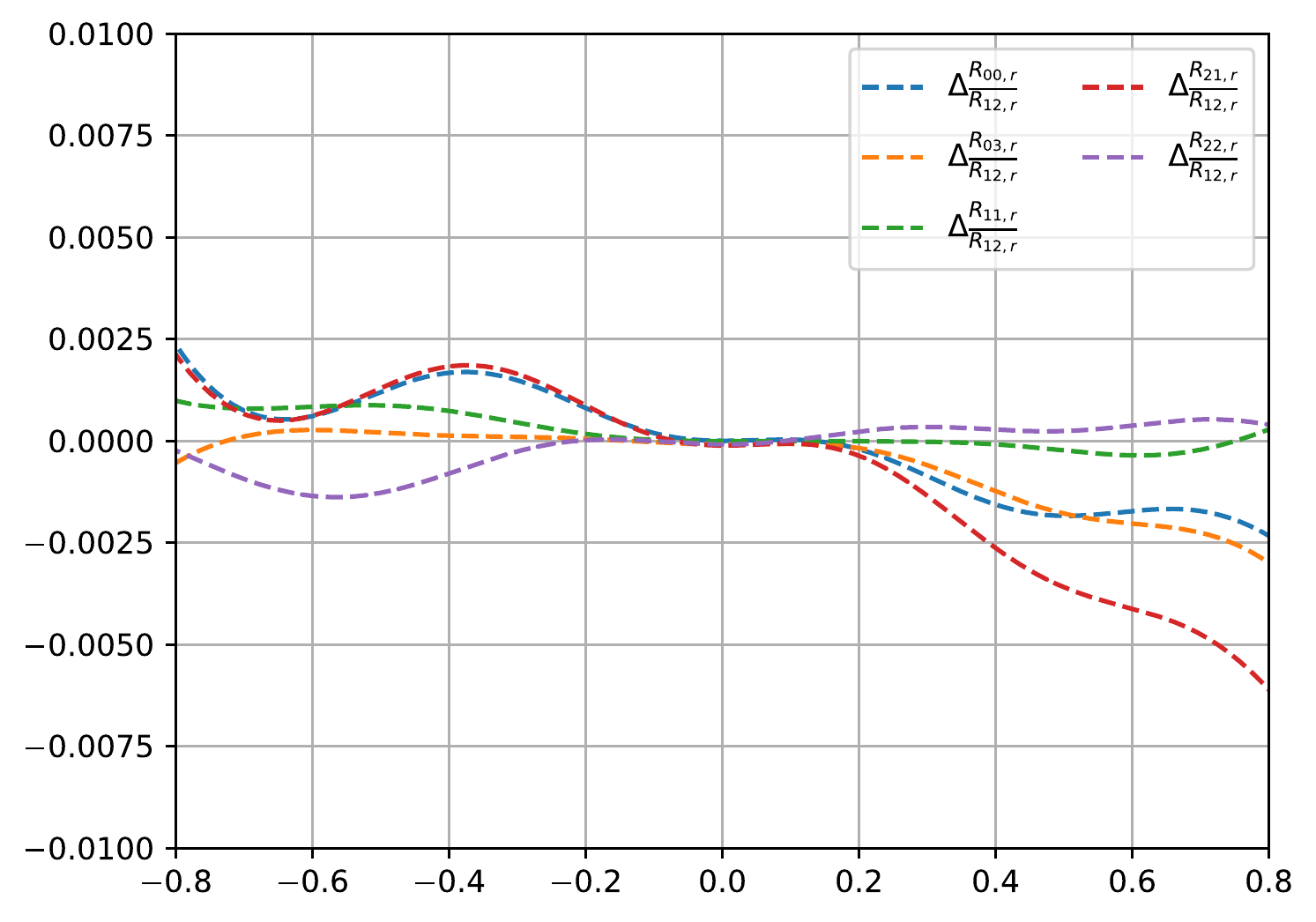}}
  
  \caption{(a) 7-vertex model with Hamiltonian of type $H_{7v,1}$, with parameters $a_1=1,b_1=0.45,c_1=0.6,d_1=0.75$, (b)errors}
  \label{fig:7vertex-1}
\end{figure}

For the case $H^{00}\neq  H^{33}$, the R-matrix $R_{6v,2}(u)$ is given by
\begin{equation}\label{eq:Hamiltonian6v,2}
    H_{6v,2}=\begin{pmatrix}
        a_1&0&0&0\\
        0&b_1&c_1&0\\
        0&c_2&b_2&0\\
        0&0&0&a_2
    \end{pmatrix}\,\leftrightarrow \quad 
    R_{6v,2}(u)=\begin{pmatrix}
        R_{6v,2}^{00}(u)&0&0&0\\
        0&R_{6v,2}^{11}(u)&R_{6v,2}^{12}(u)&0\\
        0&R_{6v,2}^{21}(u)&R_{6v,2}^{22}(u)&0\\
        0&0&0&R_{6v,2}^{33}(u)
    \end{pmatrix}
\end{equation}
where  $a_2=b_1+b_2-a_1$ and
\begin{equation}
\begin{split}
        R_{6v,2}^{00}(u)&= e^{(a_1+ a_2)u/2}(\cosh{(\omega u)}+\frac{a_1-a_2}{2\omega}\sinh{(\omega u)})\\
        R_{6v,2}^{11}(u)&= \frac{c_2}{\omega}e^{(a_1+ a_2)u/2}\sinh{(\omega u)}\\
        R_{6v,2}^{12}(u)&= e^{b_2 u}\\
        R_{6v,2}^{21}(u)&= e^{b_1 u}\\
        R_{6v,2}^{22}(u)&= \frac{c_1}{\omega}e^{(a_1+ a_2)u/2}\sinh{(\omega u)}\\
        R_{6v,2}^{33}(u)&= e^{(a_1+ a_2)u/2}(\cosh{(\omega u)}-\frac{a_1-a_2}{2\omega}\sinh{(\omega u)})\,,\qquad \omega =\frac{\sqrt{(a_1-a_2)^2-4c_1c_2}}{2}
\end{split}
\end{equation}
Figure~\ref{fig:6vertex-2} gives a representative training vs actual plot for Hamiltonian parameters $a_1=1.,a_2=0.2,b_1=0.45,b_2=0.75,c_1=0.4,c_2=0.6$. 
\noindent Next we have the 7-vertex models, which consists of two classes of solution distinguished by the Hamiltonian entries $H^{00}$, $H^{33}$ being equal or not. In the first case, we have
\begin{equation}\label{eq:Hamiltonian7v,1}
    H_{7v,1}=\begin{pmatrix}
        a_1&0&0&d_1\\
        0&a_1+b_1&c_1&0\\
        0&-c_1&a_1-b_1&0\\
        0&0&0&a_1
    \end{pmatrix}\,\leftrightarrow \quad 
    R_{7v,1}(u)=\begin{pmatrix}
        R_{7v,1}^{00}(u)&0&0&R_{7v,1}^{03}(u)\\
        0&R_{7v,1}^{11}(u)&R_{7v,1}^{12}(u)&0\\
        0&R_{7v,1}^{21}(u)&R_{7v,1}^{22}(u)&0\\
        0&0&0&R_{7v,1}^{33}(u)
    \end{pmatrix}
\end{equation}  
where 
\begin{equation}
\begin{split}
        R_{7v,1}^{00}(u)&=R_{7v,1}^{33}(u)= e^{a_1 u}\cosh{(c_1 u)}\\
        R_{7v,1}^{11}(u)&=- R_{7v,1}^{22}(u)=e^{a_1 u}\sinh{(c_1 u)}\\
        R_{7v,1}^{12}(u)&= e^{(a_1-b_1) u}\\
        R_{7v,1}^{21}(u)&= e^{(a_1+b_1) u}\\
        R_{7v,1}^{03}(u)&= -\frac{d_1}{2b_1} (e^{(a_1- b_1)u}-e^{(a_1+ b_1)u})
\end{split}
\end{equation}
Figure~\ref{fig:7vertex-1} plots the predicted R-matrix components as ratios with respect to the $(12)$ component against the above analytic results, and their differences for a generic choice of parameters $a_1=1,b_1=0.45,c_1=0.6,d_1=0.75$.

\begin{figure}[htbp]
  \centering
  \captionsetup[subfigure]{}
  \subfloat[]{\includegraphics[width=0.5\linewidth,height=.4\textwidth]{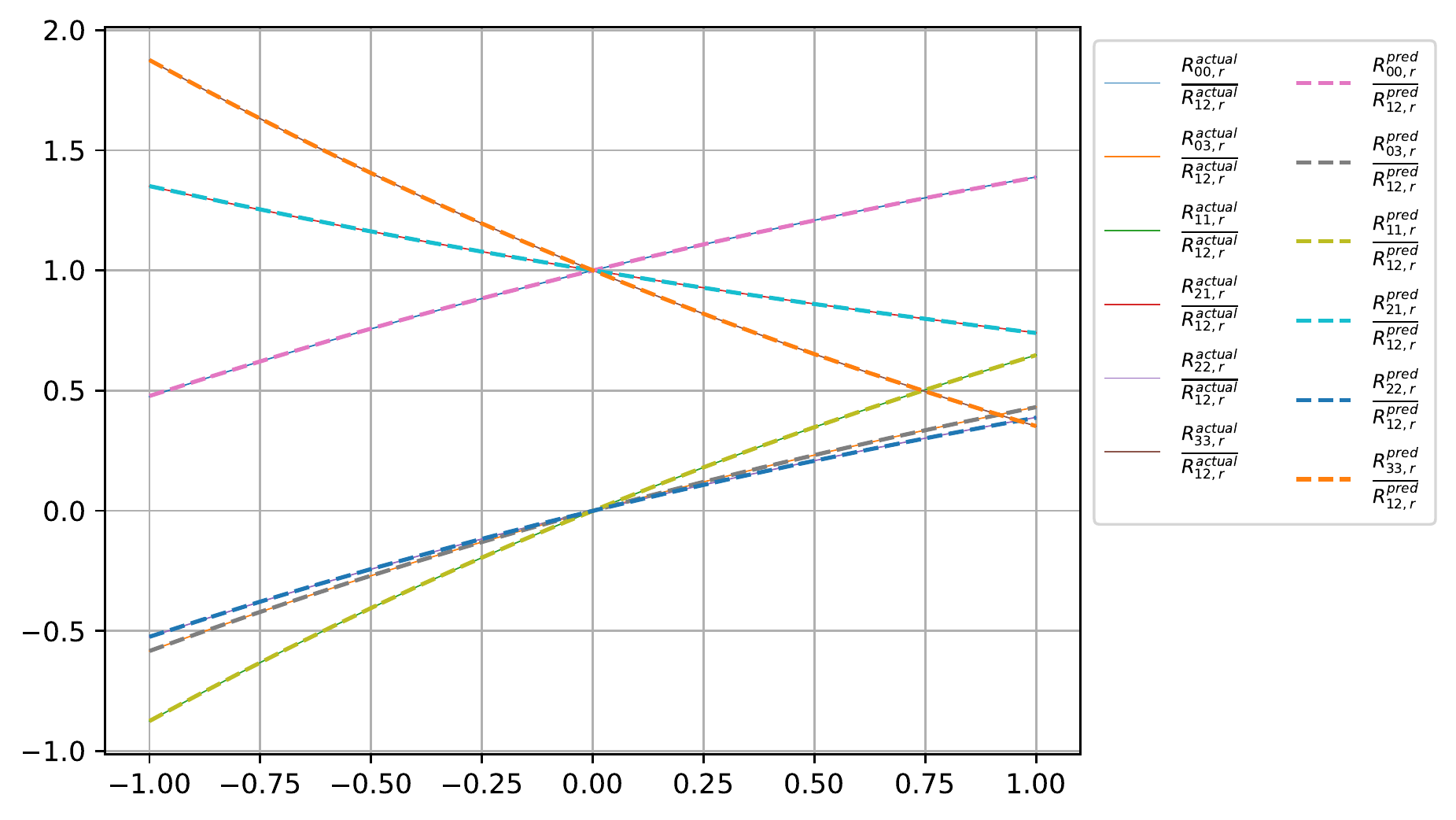}}
  \subfloat[]{\includegraphics[width=0.5\linewidth,height=.4\textwidth]{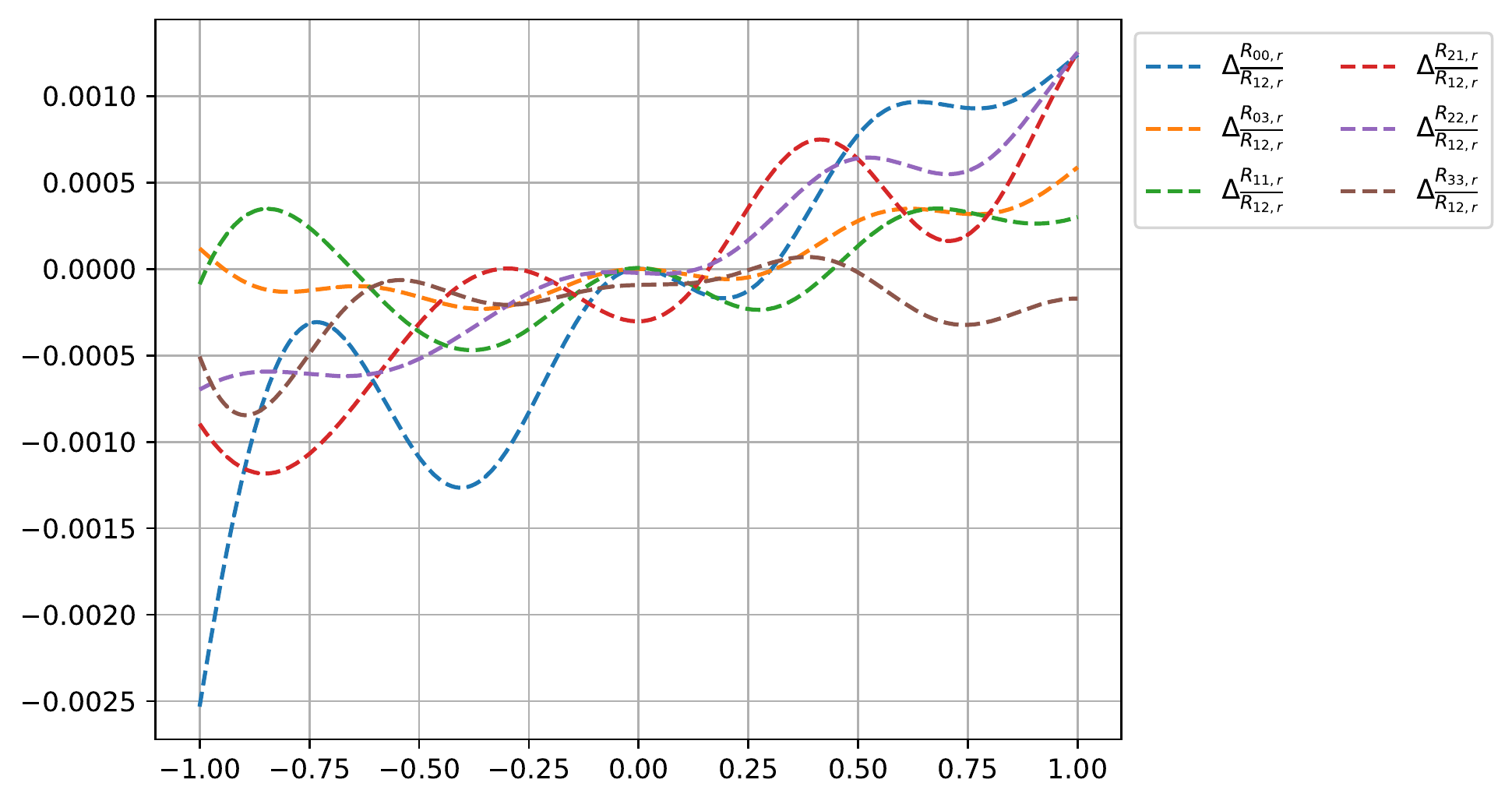}}
  
  \caption{(a) 7-vertex model with Hamiltonian of type $H_{7v,2}$, with parameters $a_1=1,c_1=0.45,c_2=0.75,d_1=0.5$, (b)errors}
  \label{fig:7vertex-2}
\end{figure}

\noindent In the second case for $H^{00}\neq H^{33}$, we have
\begin{equation}\label{eq:Hamiltonian7v,2}
    H_{7v,2}=\begin{pmatrix}
        a_1&0&0&d_1\\
        0&a_1-c_2&c_1&0\\
        0&c_2&a_1-c_1&0\\
        0&0&0&a_2
    \end{pmatrix}\,\leftrightarrow \quad 
    R_{7v,2}(u)=\begin{pmatrix}
        R_{7v,2}^{00}(u)&0&0&R_{7v,1}^{03}(u)\\
        0&R_{7v,2}^{11}(u)&R_{7v,1}^{12}(u)&0\\
        0&R_{7v,2}^{21}(u)&R_{7v,1}^{22}(u)&0\\
        0&0&0&R_{7v,1}^{33}(u)
    \end{pmatrix}
\end{equation}  
where $a_2=a_1-c_1-c_2$ and
\begin{equation}
\begin{split}
        R_{7v,2}^{00}(u)&=\frac{e^{(a_1-\frac{c_1+c_2}{2}) u}}{c_1-c_2}((c_1-c_2)\cosh{(\frac{c_1 -c_2}{2}u)}+(c_1+c_2)\sinh{(\frac{c_1 -c_2}{2}u)})\\
        R_{7v,2}^{11}(u)&=\frac{2c_2}{c_1-c_2}e^{(a_1-\frac{c_1+c_2}{2}) u}\sinh{(\frac{c_1-c_2}{2}u)}\\
        R_{7v,2}^{22}(u)&=\frac{2c_1}{c_1-c_2}e^{(a_1-\frac{c_1+c_2}{2}) u}\sinh{(\frac{c_1-c_2}{2}u)}\\
        R_{7v,2}^{12}(u)&= e^{(a_1-c_1) u}\\
        R_{7v,2}^{21}(u)&= e^{(a_1+c_2) u}\\
        R_{7v,2}^{03}(u)&= \frac{2d_1}{c_1-c_2}e^{(a_1-\frac{c_1+c_2}{2}) u}\sinh{(\frac{c_1 -c_2}{2}u)} \\
        R_{7v,2}^{33}(u)&= \frac{e^{(a_1-\frac{c_1+c_2}{2}) u}}{c_1-c_2}((c_1-c_2)\cosh{(\frac{c_1 -c_2}{2}u)}-(c_1+c_2)\sinh{(\frac{c_1 -c_2}{2}u)})
\end{split}
\end{equation}
Figure~\ref{fig:7vertex-2} plots the predicted R-matrix components as ratios with respect to the (12) component against the above analytic results, and their differences for a generic choice of parameters $a_1=1,c_1=0.45,c_2=0.75,d_1=0.5$.

\begin{figure}[htbp]
  \centering
  \captionsetup[subfigure]{}
  \subfloat[]{\includegraphics[width=0.5\linewidth,height=.4\textwidth]{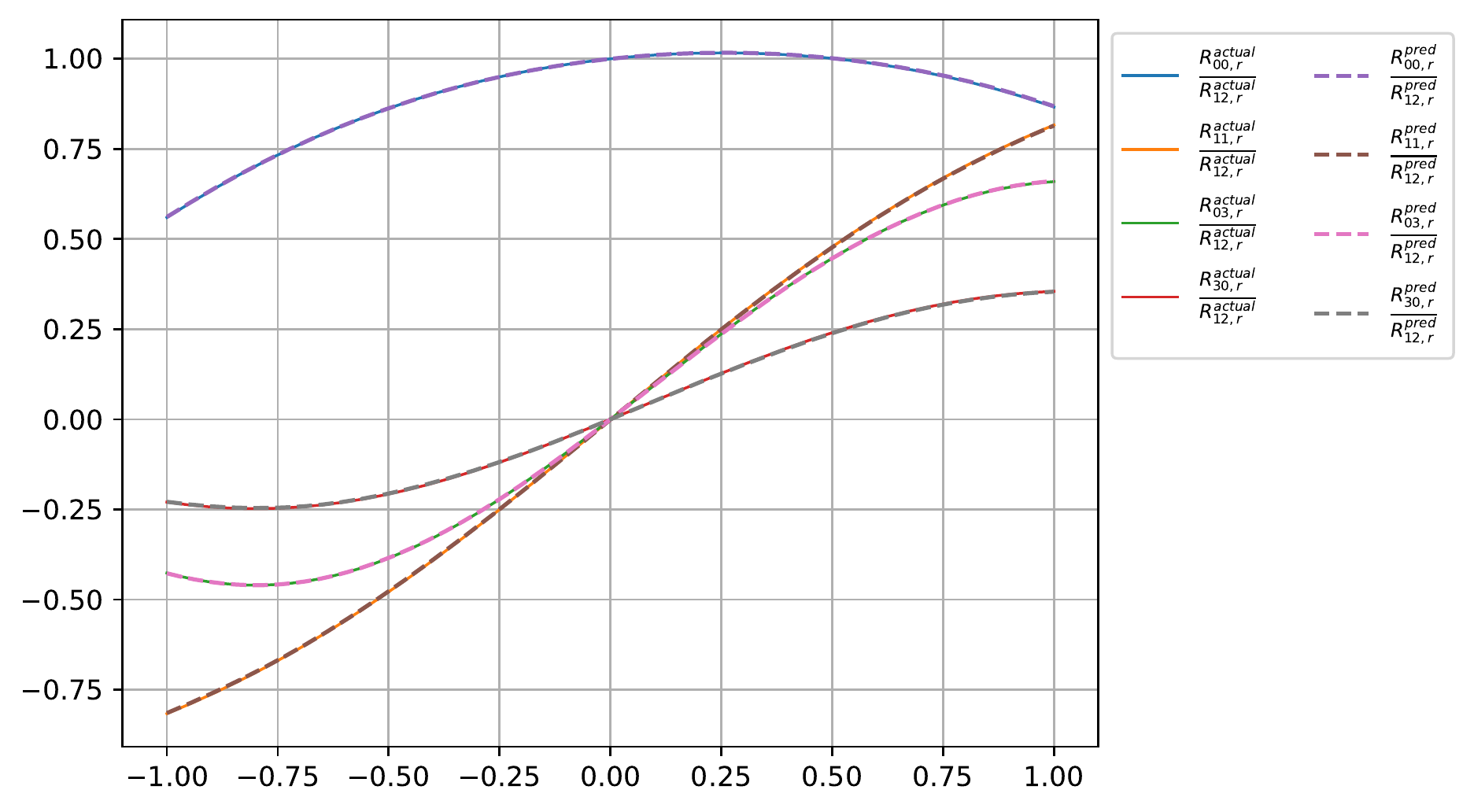}}
  \subfloat[]{\includegraphics[width=0.5\linewidth,height=.4\textwidth]{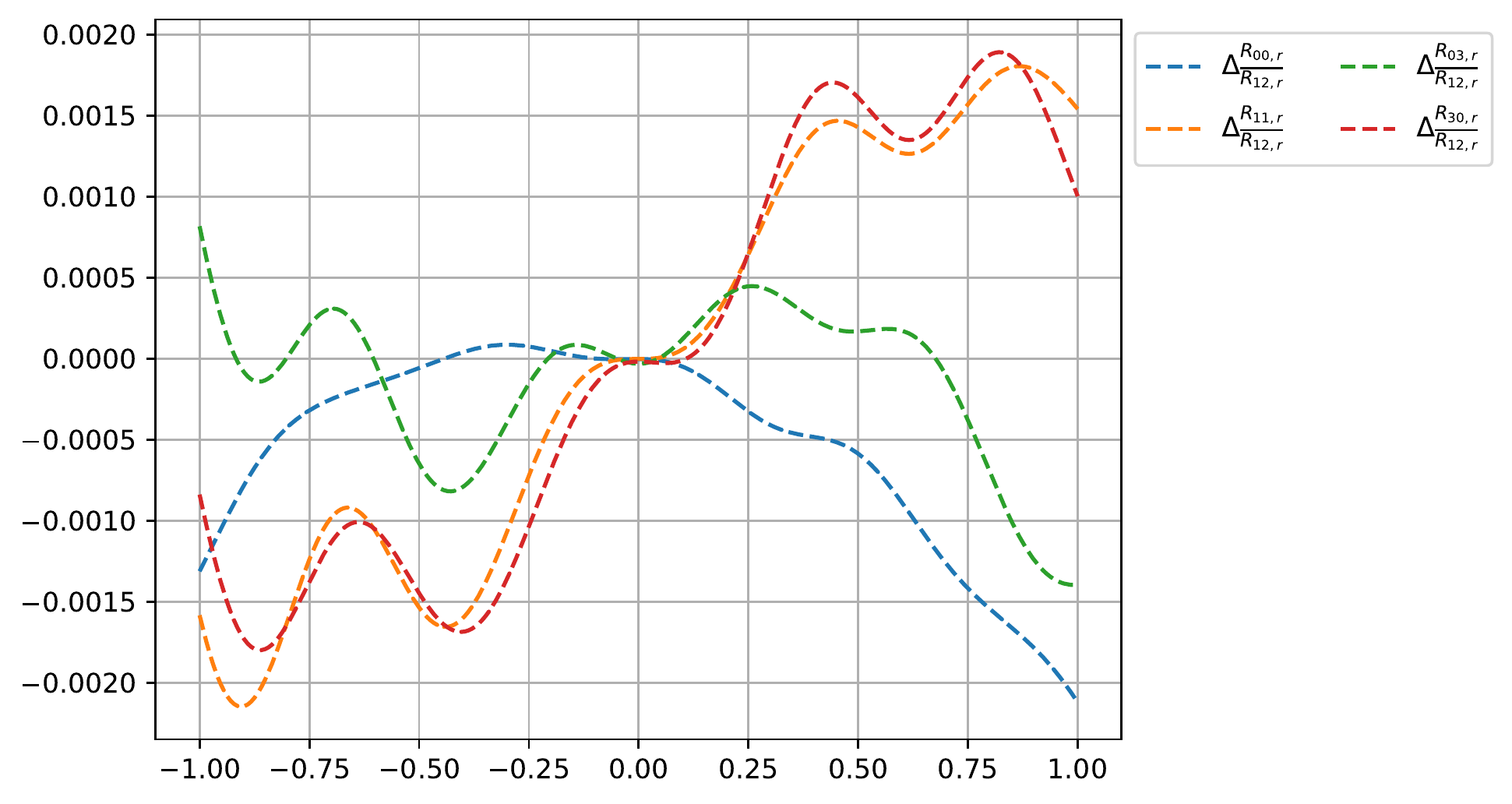}}
  
  \caption{(a) 8-vertex model with Hamiltonian of type $H_{8v,1}$, with parameters $b_1=0.4,\eta=0.8,m=0.5,\delta_1=1.3,\delta_2=0.7$, (b)errors}
  \label{fig:8vertex-1}
\end{figure}

8-vertex models have 3 gauge-inequivalent classes labelled $H_{8v,i},i=1,2,3$. One of these models, namely $H_{8v,1}$, is a generalisation of the XYZ model
\begin{equation}\label{eq:Hamiltonian8v,1}
    H_{8v,1}=\begin{pmatrix}
        a_1&0&0&d_1\\
        0&b_1&c_1&0\\
        0&c_1&b_1&0\\
        d_2&0&0&a_1
    \end{pmatrix}\,\leftrightarrow 
    R_{8v,1}(u)=\begin{pmatrix}
        R_{8v,1}^{00}(u)&0&0&R_{8v,1}^{03}(u)\\
        0&R_{8v,1}^{11}(u)&R_{8v,1}^{12}(u)&0\\
        0&R_{8v,1}^{21}(u)&R_{8v,1}^{22}(u)&0\\
        R_{8v,1}^{30}(u)&0&0&R_{8v,1}^{33}(u)
    \end{pmatrix}
\end{equation}
where 
\begin{equation}
\begin{split}
        R_{8v,1}^{00}(u)&=R_{8v,1}^{33}(u)=
        \frac{\mathrm{sn}{(u+2\eta,m)}}{\mathrm{sn}{(2\eta,m)}} e^{b_1 u}
        \\
        R_{8v,1}^{11}(u)&=R_{8v,1}^{22}(u)=\frac{\mathrm{sn}{(u,m)}}{\mathrm{sn}{(2\eta,m)}} e^{b_1 u}\\
        R_{8v,1}^{12}(u)&=R_{8v,1}^{21}(u)= e^{b_1 u}\\
        R_{8v,1}^{03}(u)&= \sqrt{\frac{d_1}{d_2}}\sqrt{m}\,
        \mathrm{sn}{(u+2\eta,m)}\mathrm{sn}{(u,m)} e^{b_1 u} \\
        R_{8v,1}^{30}(u)&= \sqrt{\frac{d_2}{d_1}}\sqrt{m}\,
        \mathrm{sn}{(u+2\eta,m)}\mathrm{sn}{(u,m)} e^{b_1 u}
\end{split}
\end{equation}
with Hamiltonian coefficients given by
\begin{equation}
    a_1=b_1+\frac{\mathrm{cn}(2\eta,m)\mathrm{dn}(2\eta,m)}{\mathrm{sn}(2\eta,m)}\,,\,\,
    c_1=\frac{1}{\mathrm{sn}(2\eta,m)}\,,\,\,
    d_1=\sqrt{m}\,\delta_1\,\mathrm{sn}(2\eta,m)\,,\,\,
    d_2=\sqrt{m}\,\delta_2\,\mathrm{sn}(2\eta,m)
\end{equation}
for free parameters $b_1,\eta,m,\delta_1,\delta_2$. Figure~\ref{fig:8vertex-1} plots the predicted R-matrix components as ratios with respect to the $(12)$ component against the above analytic results, and their differences for a generic choice of parameters $b_1=0.4,\eta=0.8,m=0.5,\delta_1=1.3,\delta_2=0.7$.

\begin{figure}[htbp]
  \centering
  \captionsetup[subfigure]{}
  \subfloat[]{\includegraphics[width=0.5\linewidth,height=.4\textwidth]{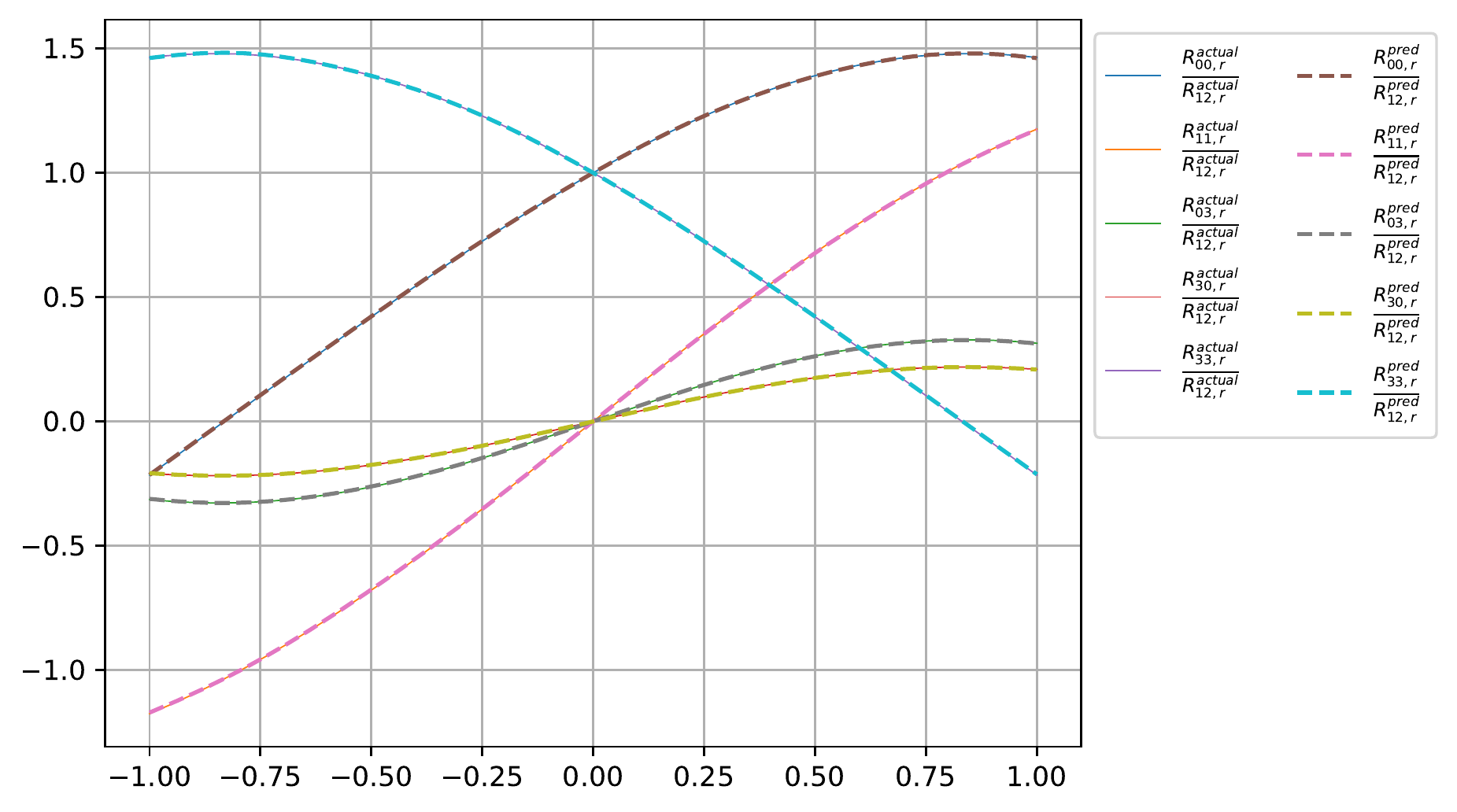}}
  \subfloat[]{\includegraphics[width=0.5\linewidth,height=.4\textwidth]{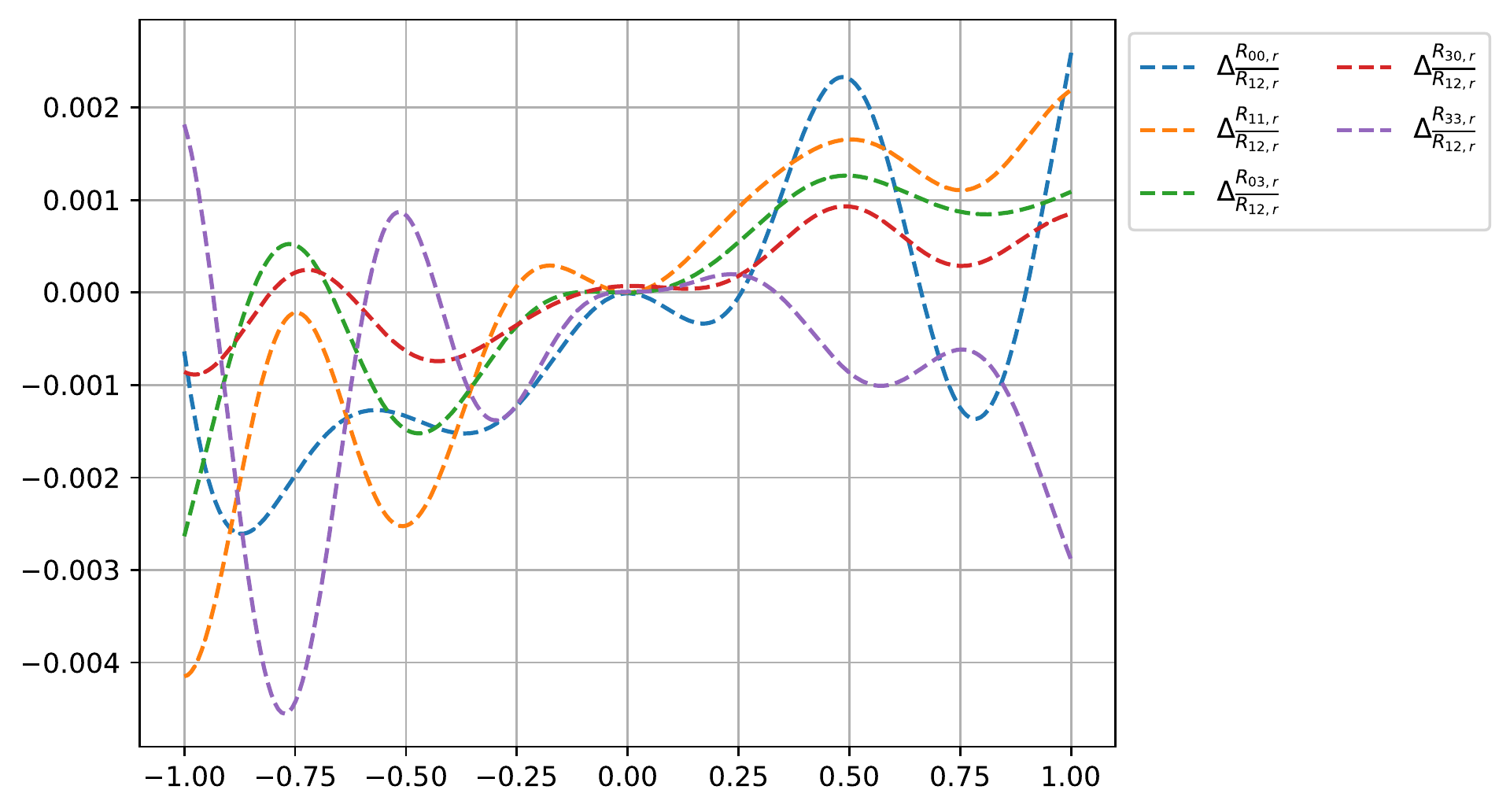}}
  
  \caption{(a) 8-vertex model with Hamiltonian of type $H_{8v,2}$, with parameters $b_1=0.4,\alpha_1=0.5,\beta_1=0.7,\delta_1=0.3,\delta_2=0.2$, (b)errors}
  \label{fig:8vertex-2}
\end{figure}

The second class of 8-vertex XYZ-type solution has Hamiltonian $H_{8v,2}$ and R-matrix $R_{8v,2}(u)$ defined as follows
\begin{equation}\label{eq:Hamiltonian8v,2}
    H_{8v,2}=\begin{pmatrix}
        a_1&0&0&d_1\\
        0&b_1&c_1&0\\
        0&c_1&b_1&0\\
        d_2&0&0&2b_1-a_1
    \end{pmatrix}\leftrightarrow 
    R_{8v,2}(u)=\begin{pmatrix}
        R_{8v,2}^{00}(u)&0&0&R_{8v,2}^{03}(u)\\
        0&R_{8v,2}^{11}(u)&R_{8v,2}^{12}(u)&0\\
        0&R_{8v,2}^{21}(u)&R_{8v,2}^{22}(u)&0\\
        R_{8v,2}^{30}(u)&0&0&R_{8v,2}^{33}(u)
    \end{pmatrix}
\end{equation}
where 
\begin{equation}
\begin{split}
        R_{8v,2}^{00}(u)&=
        (\frac{\mathrm{cn}{(u,m)}}{\mathrm{dn}{(u,m)}}+
        \frac{\mathrm{sn}{(u,m)}\mathrm{cn}{(2\eta,m)}}{\mathrm{sn}{(2\eta,m)}}) e^{b_1 u}
        \\
        R_{8v,2}^{11}(u)&=R_{8v,1}^{22}(u)=\frac{\mathrm{sn}{(u,m)}}{\mathrm{sn}{(2\eta,m)}} e^{b_1 u}\\
        R_{8v,2}^{12}(u)&=R_{8v,1}^{21}(u)= e^{b_1 u}\\
        R_{8v,2}^{03}(u)&= \frac{\delta_1}{\beta_1}\frac{\mathrm{sn}{(u,m)}\mathrm{cn}{(u,m)}}{\mathrm{dn}{(u,m)}\mathrm{sn}{(2\eta,m)}} e^{b_1 u}\\
        R_{8v,2}^{30}(u)&= \frac{\delta_2}{\beta_1}\frac{\mathrm{sn}{(u,m)}\mathrm{cn}{(u,m)}}{\mathrm{dn}{(u,m)}\mathrm{sn}{(2\eta,m)}} e^{b_1 u}\\
         R_{8v,2}^{33}(u)&=
        (\frac{\mathrm{cn}{(u,m)}}{\mathrm{dn}{(u,m)}}-
        \frac{\mathrm{sn}{(u,m)}\mathrm{cn}{(2\eta,m)}}{\mathrm{sn}{(2\eta,m)}}) e^{b_1 u}
\end{split}
\end{equation}
with the Hamiltonian coefficients given by
\begin{equation}
    a_1=b_1+\frac{\mathrm{cn}(2\eta,m)}{\mathrm{sn}(2\eta,m)}\,,\,\,
    c_1=\frac{1}{\mathrm{sn}(2\eta,m)}\,,\,\,
    d_1=\frac{\delta_1}{\beta_1\mathrm{sn}(2\eta,m)}\,,\,\,
    d_2=\frac{\delta_2}{\beta_1\mathrm{sn}(2\eta,m)}
\end{equation}
\begin{equation}
    m=\frac{\delta_1\delta_2}{\alpha_1^2-\beta_1^2}\,,\,\,
    \mathrm{cn}(2\eta,m)=\frac{\alpha_1}{\beta_1}\,,\,\,
    \mathrm{sn}(2\eta,m)=\sqrt{1-\frac{\alpha_1^2}{\beta_1^2}}
\end{equation}
for free parameters $b_1,\alpha_1,\beta_1,\delta_1,\delta_2$. Figure~\ref{fig:8vertex-2} plots the predicted R-matrix components as ratios with respect to the $(12)$ component against the above analytic results, and their differences for a generic choice of parameters $b_1=0.4,\alpha_1=0.5,\beta_1=0.7,\delta_1=0.3,\delta_2=0.2$.

\begin{figure}[htbp]
  \centering
  \captionsetup[subfigure]{}
  \subfloat[]{\includegraphics[width=0.5\linewidth,height=.4\textwidth]{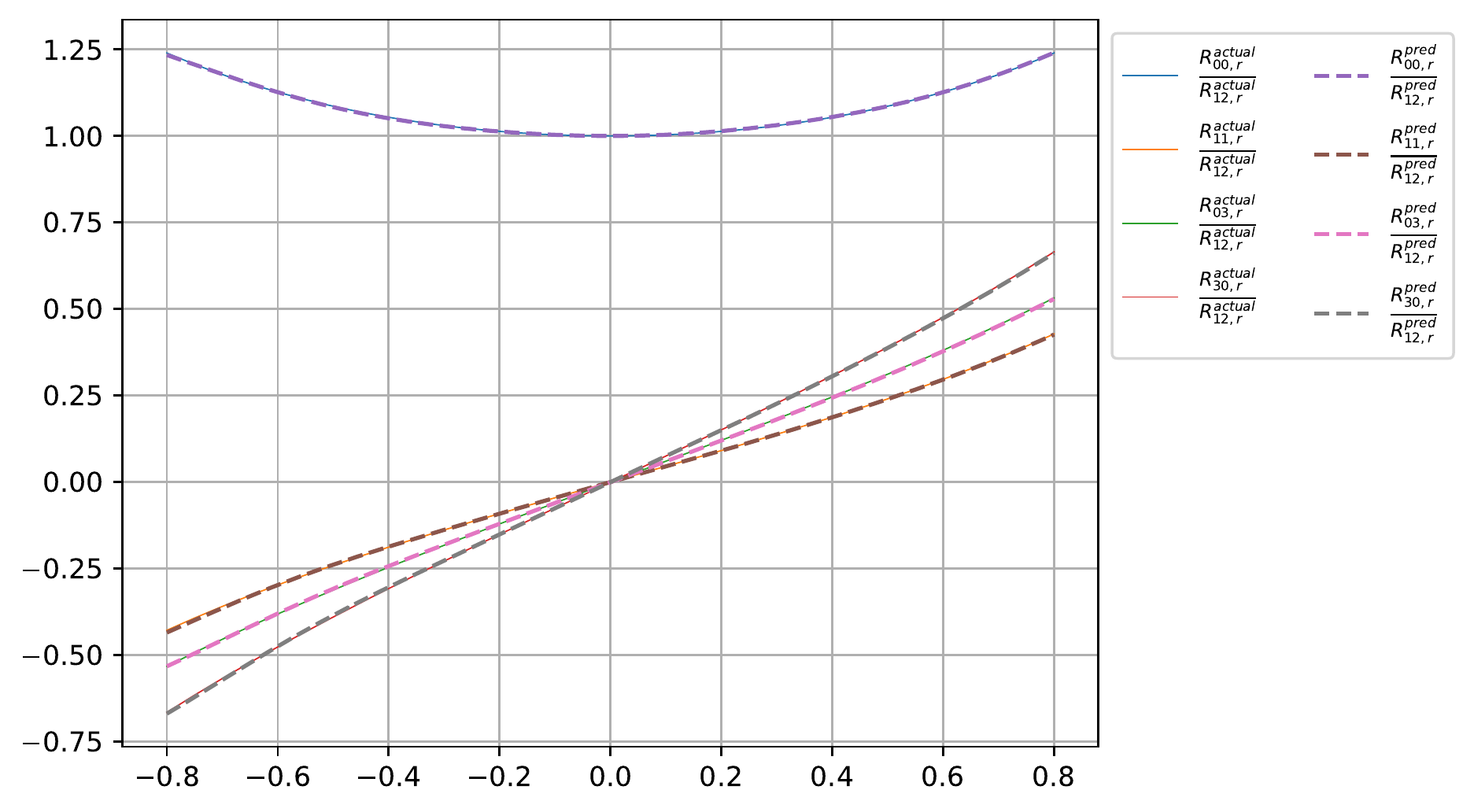}}
  \subfloat[]{\includegraphics[width=0.5\linewidth,height=.4\textwidth]{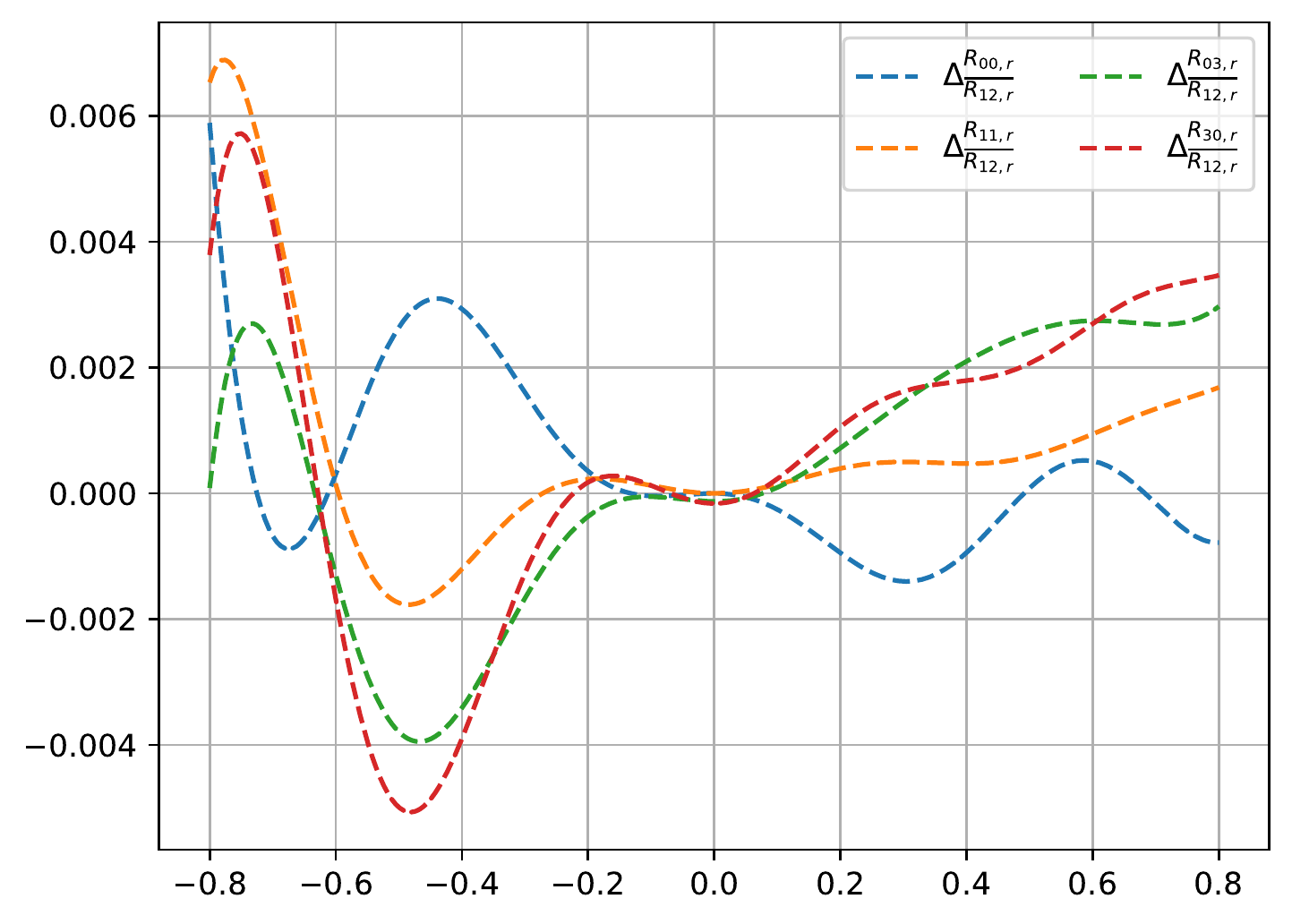}}
  
  \caption{(a) 8-vertex model with Hamiltonian of type $H_{8v,3}$, with parameters $a_1=1,\, b_1=-0.45, \, d_1=0.6, \,d_2=0.75$, (b)errors}
  \label{fig:8vertex-3}
\end{figure}

\noindent The second class of 8-vertex XYZ-type solution has Hamiltonian $H_{8v,3}$ and R-matrix $R_{8v,3}(u)$ defined as follows
\begin{equation}\label{eq:Hamiltonian8v,3}
    H_{8v,3}=\begin{pmatrix}
        a_1&0&0&d_1\\
        0&a_1&-b_1&0\\
        0&b_1&a_1&0\\
        d_2&0&0&a_1
    \end{pmatrix}\leftrightarrow 
    R_{8v,3}(u)=\begin{pmatrix}
        R_{8v,3}^{00}(u)&0&0&R_{8v,3}^{03}(u)\\
        0&R_{8v,3}^{11}(u)&R_{8v,3}^{12}(u)&0\\
        0&R_{8v,3}^{21}(u)&R_{8v,3}^{22}(u)&0\\
        R_{8v,3}^{30}(u)&0&0&R_{8v,3}^{33}(u)
    \end{pmatrix}
\end{equation}
where
\begin{equation}
\begin{split}
        R_{8v,3}^{00}(u)&=R_{8v,1}^{33}(u)=
        \frac{\cosh{(b_1 u)}}{\cos{(\sqrt{d_1d_2} u)}}e^{a_1u}
        \\
        R_{8v,3}^{11}(u)&=-R_{8v,1}^{22}(u)=\frac{\sinh{(b_1 u)}}{\cos{(\sqrt{d_1d_2} u)}}e^{a_1u}\\
        R_{8v,3}^{12}(u)&=R_{8v,1}^{21}(u)= e^{a_1 u}\\
        R_{8v,3}^{03}(u)&= \sqrt{\frac{d_1}{d_2}}e^{a_1 u}\tan{(\sqrt{d_1d_2} u)} \\
        R_{8v,1}^{30}(u)&= \sqrt{\frac{d_2}{d_1}}e^{a_1 u}\tan{(\sqrt{d_1d_2} u)} 
\end{split}
\end{equation}
Figure~\ref{fig:8vertex-3} plots the predicted R-matrix components as ratios with respect to the $(12)$ component against the above analytic results, and their differences for a generic choice of parameters $a_1=1,\, b_1=-0.45, \, d_1=0.6, \,d_2=0.75$.

For non-XYZ type models, the 6 gauge-inequivalent Hamiltonians are of the form
\begin{equation}\nonumber
    H_{\mathrm{class}-1}=\begin{pmatrix}
        0&a_1&a_2&0\\
        0&a_5&0&a_3\\
        0&0&-a_5&a_4\\
        0&0&0&0
    \end{pmatrix}\,,\quad 
    H_{\mathrm{class}-2}=\begin{pmatrix}
        0&a_2&a_3-a_2&a_5\\
        0&a_1&0&a_4\\
        0&0&-a_1&a_3-a_4\\
        0&0&0&0
    \end{pmatrix}\,,
\end{equation}
\begin{equation}\nonumber
    H_{\mathrm{class}-3}=\begin{pmatrix}
        -a_1&(2a_1-a_2)a_3&(2a_1+a_2)a_3&0\\
        0&a_1-a_2&0&0\\
        0&0&a_1+a_2&0\\
        0&0&0&-a_1
    \end{pmatrix}\,,\quad 
    H_{\mathrm{class}-4}=\begin{pmatrix}
        a_1&a_2&a_2&a_3\\
        0&-a_1&0&a_4\\
        0&0&-a_1&a_4\\
        0&0&0&a_1
    \end{pmatrix}\,,
\end{equation}
\begin{equation}\label{nonXYZhamiltonians}
    H_{\mathrm{class}-5}=\begin{pmatrix}
        a_1&a_2&-a_2&0\\
        0&-a_1&2a_1&a_3\\
        0&2a_1&-a_1&-a_3\\
        0&0&0&a_1
    \end{pmatrix}\,,\quad 
    H_{\mathrm{class}-6}=\begin{pmatrix}
        a_1&a_2&a_2&0\\
        0&-a_1&2a_1&-a_2\\
        0&2a_1&-a_1&-a_2\\
        0&0&0&a_1
    \end{pmatrix}
\end{equation}

Corresponding R-matrices are
\begin{equation}\label{eq:nonXYZRmatrices}
    R_{\mathrm{class}-1}(u)=\begin{pmatrix}
        1&\frac{a_1(e^{a_5 u}-1)}{a_5}&\frac{a_2(e^{a_5 u}-1)}{a_5}&\frac{(a_1a_3+a_2a_4)}{a_5^2}(\cosh{(a_5 u)}-1)\\
        0&0&e^{-a_5 u}&\frac{a_4(1-e^{-a_5 u})}{a_5}\\
        0&e^{a_5 u}&0&\frac{a_3(1-e^{-a_5 u})}{a_5}\\
        0&0&0&1
    \end{pmatrix}
\end{equation}
\begin{equation}
    R_{\mathrm{class}-2}(u)=uP(\frac{a_1}{\sinh(a_1u)}+H_{class-5}+\frac{\tanh(a_1u)}{a_1}H_{class-5}^2)\,, \quad P=\begin{pmatrix}
        1&0&0&0\\0&0&1&0\\
        0&1&0&0\\0&0&0&1
    \end{pmatrix}
\end{equation}
\begin{equation}
    R_{\mathrm{class}-3}(u)=\begin{pmatrix}
        e^{-a_1 u}&a_3(e^{(a_1-a_2)u}-e^{-a_1 u})
        &a_3(e^{(a_1+a_2)u}-e^{-a_1 u})&0\\
        0&0&e^{(a_1+a_2) u}&0\\
        0&e^{(a_1-a_2) u}&0&0\\
        0&0&0&e^{-a_1 u}
    \end{pmatrix}
\end{equation}
\begin{equation}
    R_{\mathrm{class}-4}(u)=\begin{pmatrix}
        e^{a_1 u}&\frac{a_2\sinh{(a_1 u)}}{a_1}&\frac{a_2\sinh{(a_1 u)}}{a_1}&\frac{e^{a_1 u}(a_2a_4+a_1a_3 \coth{(a_1 u)})\sinh^2{(a_1u)}}{a_1^2}\\
        0&0&e^{-a_1 u}&\frac{a_4\sinh{(a_1 u)}}{a_1}\\
        0&e^{-a_1 u}&0&\frac{a_4\sinh{(a_1 u)}}{a_1}\\
        0&0&0&e^{a_1 u}
    \end{pmatrix}
\end{equation}
\begin{equation}
    R_{\mathrm{class}-5}(u)=(1-a_1 u)\begin{pmatrix}
        2a_1u+1&a_2 u&-a_2 u&a_2a_3u^2\\
        0 &2a_1u & 1 &-a_3 u\\
        0 & 1 & 2a_1u &a_3 u\\
        0&0&0&2a_1u+1
    \end{pmatrix}
\end{equation}
\begin{equation}
    R_{\mathrm{class}-6}(u)=(1-a_1 u)\begin{pmatrix}
        2a_1u+1&a_2 u(2a_1u+1)&a_2 u(2a_1u+1)&-a_2^2u^2(2a_1u+1)\\
        0 &2a_1u & 1 &-a_2 u(2a_1u+1)\\
        0 & 1 & 2a_1u &-a_2 u(2a_1u+1)\\
        0&0&0&2a_1u+1
    \end{pmatrix}
\end{equation}

\begin{figure}[htbp]
  \centering
  \captionsetup[subfigure]{}
  \subfloat[]{\includegraphics[width=0.45\linewidth,height=.45\textwidth]{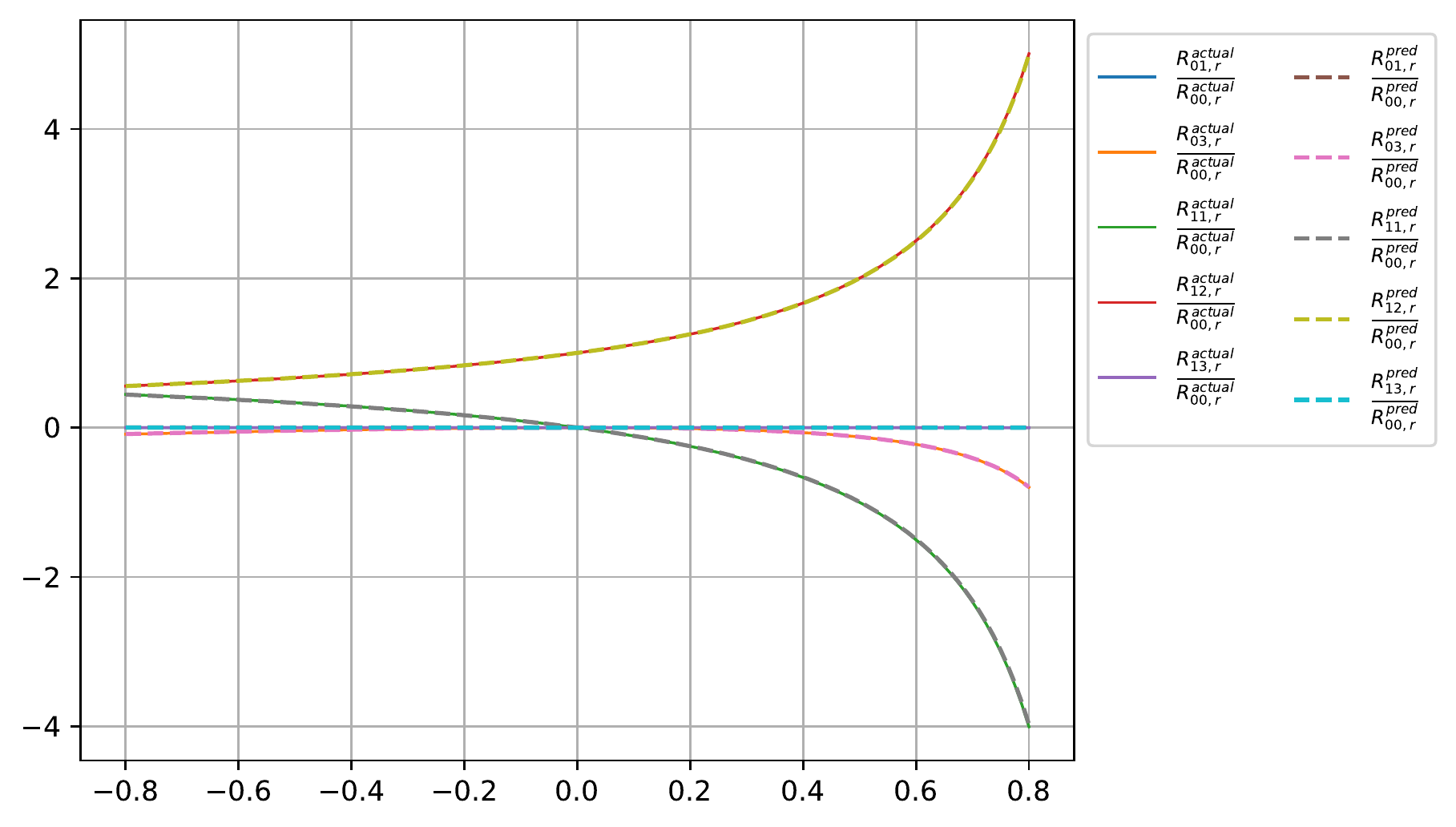}}
  \subfloat[]{\includegraphics[width=0.45\linewidth,height=.45\textwidth]{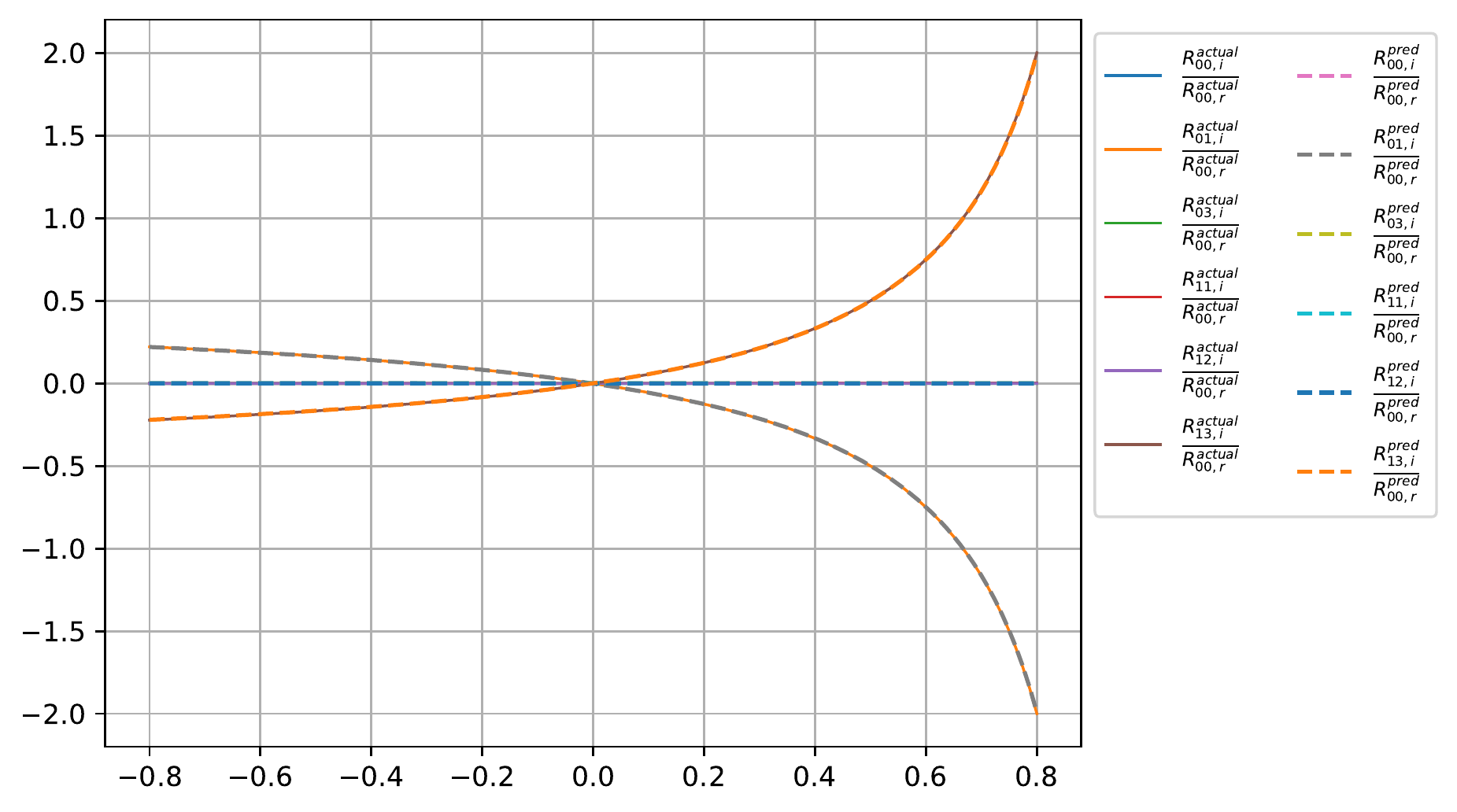}}
  \caption{(a,b) class 5, with $H_2$ parameters $a_1=-0.5,\, a_2=-0.5\,i, \, a_3=-0.5\,i$: real and imaginary parts}
  \label{fig:class5}
\end{figure}

\begin{figure}[htbp]
  \centering
  \captionsetup[subfigure]{}
   \subfloat[]{\includegraphics[width=0.5\linewidth,height=.4\textwidth]{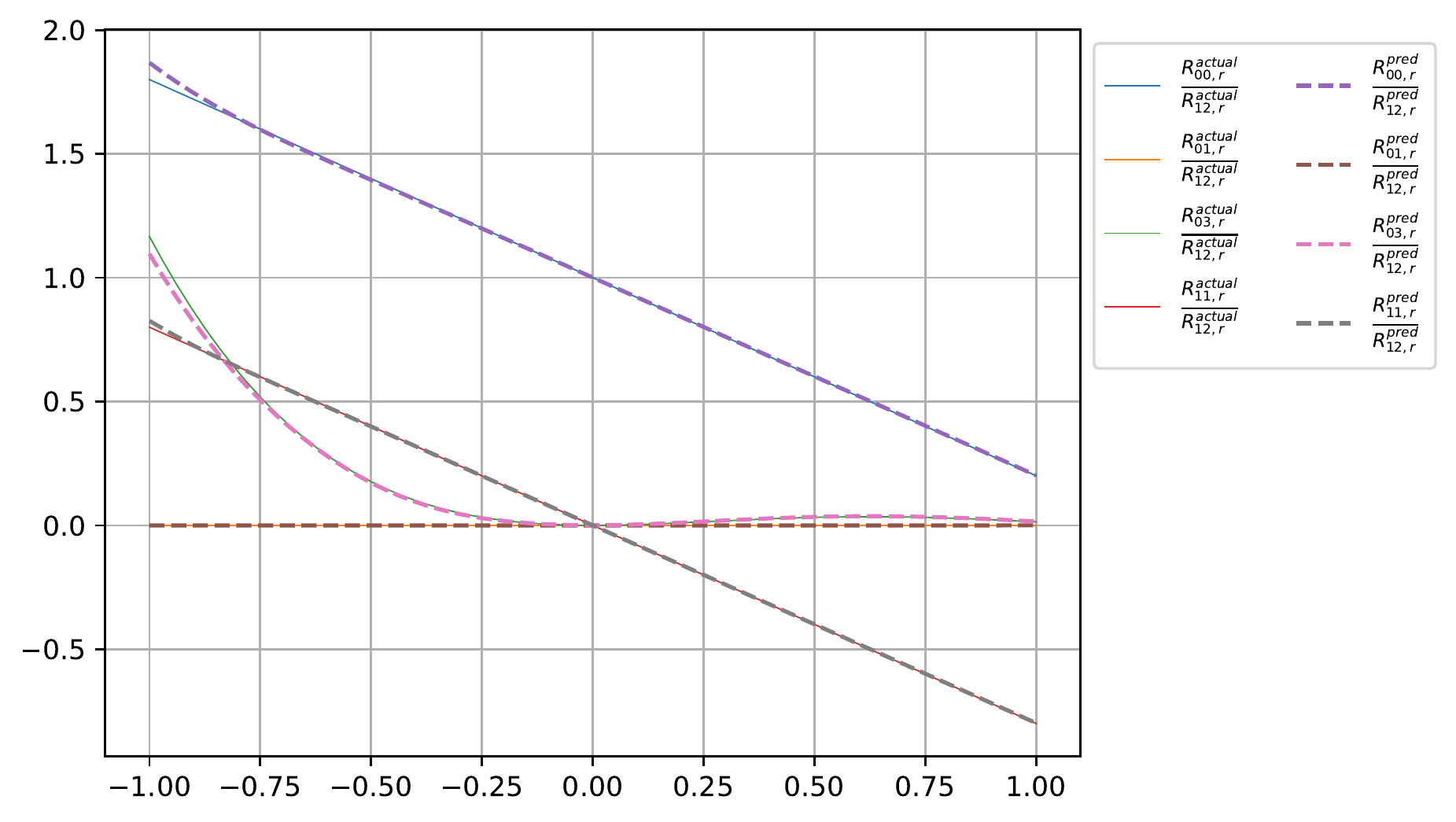}}
  \subfloat[]{\includegraphics[width=0.4\linewidth,height=.4\textwidth]{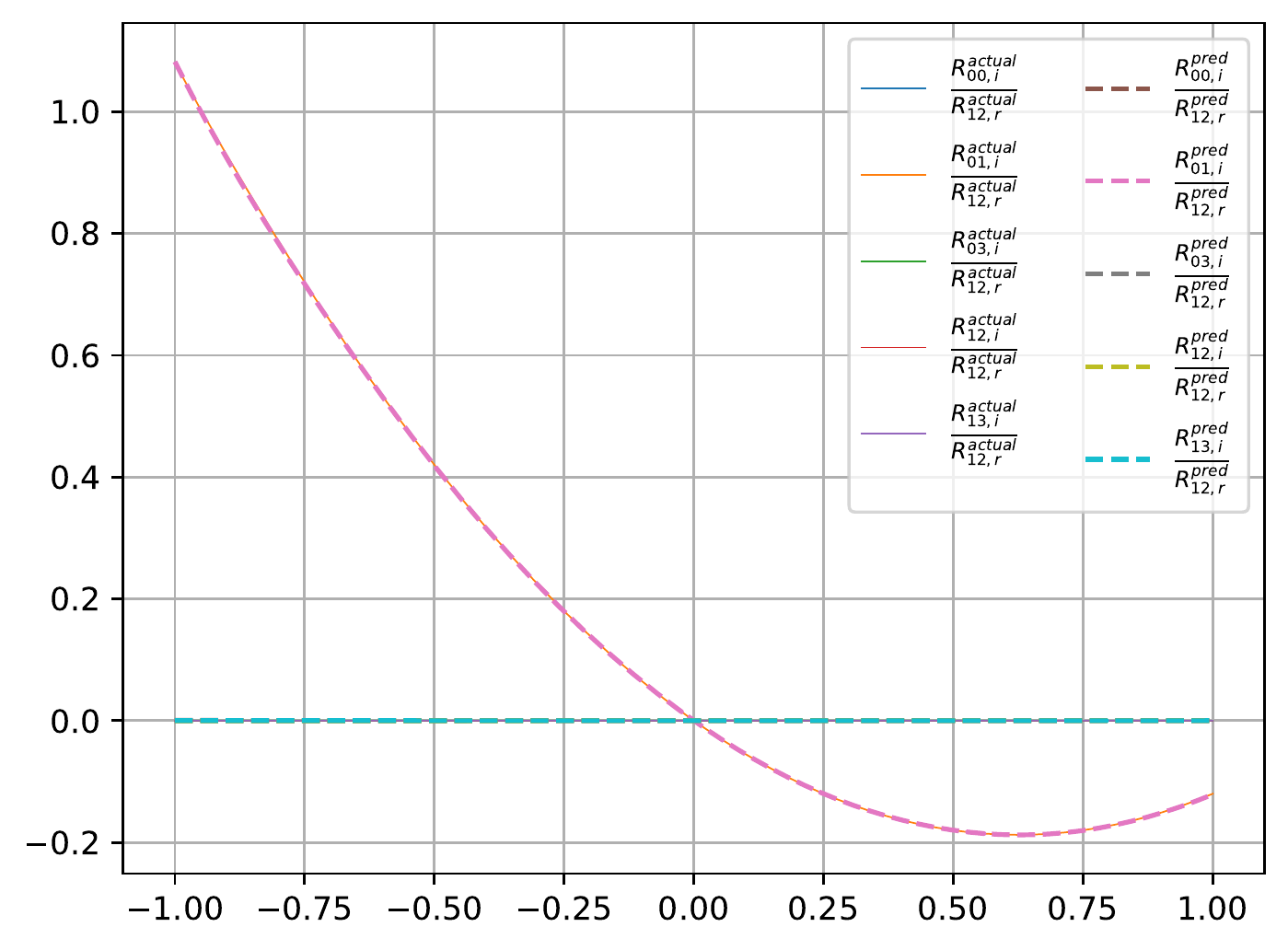}}
  \caption{(a,b) class 6, with $H_2$ parameters $a_1=-0.4,\, a_2=-0.6\,i$: real and imaginary parts}
  \label{fig:class6}
\end{figure}

In the class-2 solution above, the non-zero R-matrix components are explicitly given by
\begin{equation}
 \begin{split}
    R_{\mathrm{class}-2}^{00}(u)&=R_{\mathrm{class}-2}^{33}(u)= \frac{a_1u}{\sinh{u}}\\
    R_{\mathrm{class}-2}^{01}(u) &=
       a_2u(1+\tanh{(\frac{a_1 u}{2})})\,,\\
    R_{\mathrm{class}-2}^{02}(u)&=(a_2-a_3)u(-1+\tanh{(\frac{a_1 u}{2})})\\
    R_{\mathrm{class}-2}^{03}(u)&= a_5u(1+\frac{((a_4-a_3)(a_2-a_3)+a_2a_4)\tanh{\frac{a_1u}{2}}}{a_1a_5})\\
    R_{\mathrm{class}-2}^{12}(u) &= a_1u(-1+\frac{1}{\sinh(a_1u)}+\tanh{\frac{a_1u}{2}})\\
    R_{\mathrm{class}-2}^{13}(u)&= (a_4-a_3)u(-1+\tanh{(\frac{a_1 u}{2})})\\
     R_{\mathrm{class}-2}^{21}(u)&=a_1u(1+\frac{1}{\sinh(a_1u)}+\tanh{\frac{a_1u}{2}})\\
      R_{\mathrm{class}-2}^{23}(u)&=a_4u(1+\tanh{(\frac{a_1 u}{2})})
 \end{split}  
\end{equation}
Amongst the above non-XYZ type models, we have already looked into the training 
for Class 1 model in section~\ref{subsec:specific_search}. 
Figure~\ref{fig:class2,3,4}, \ref{fig:class5}, \ref{fig:class6} plot the training 
vs actual R-matrix components for classes 2,3,4, class 5, and class 6 
respectively, with generic Hamiltonian parameters. 
Also we note that allowing for complex parameters results in generically complex 
R-matrices. We compare the predictions against the actual formulae by taking 
ratios with respect to the real part of the (00) component for classes 2-5, and 
(12) component for class 6. 
\begin{figure}[htbp]
  \centering
  \captionsetup[subfigure]{}
  \subfloat[]{\includegraphics[width=0.5\linewidth,height=.4\textwidth]{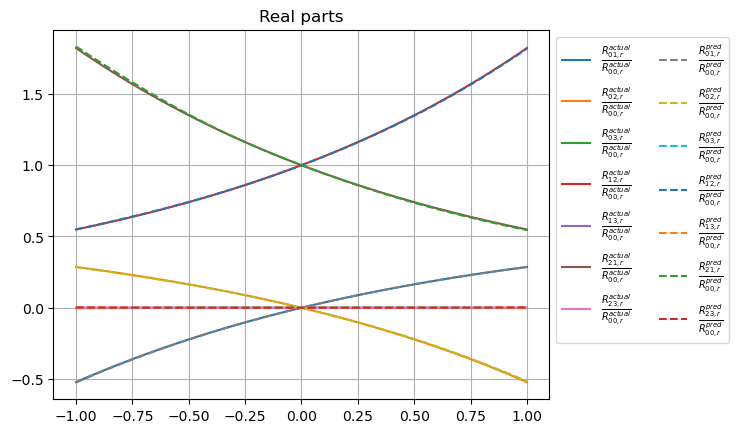}}
  \subfloat[]{\includegraphics[width=0.5\linewidth,height=.4\textwidth]{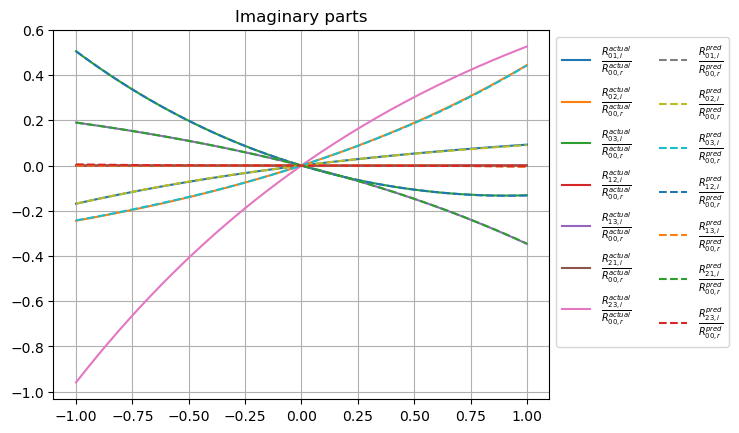}}\\
  \subfloat[]{\includegraphics[width=0.5\linewidth,height=.4\textwidth]{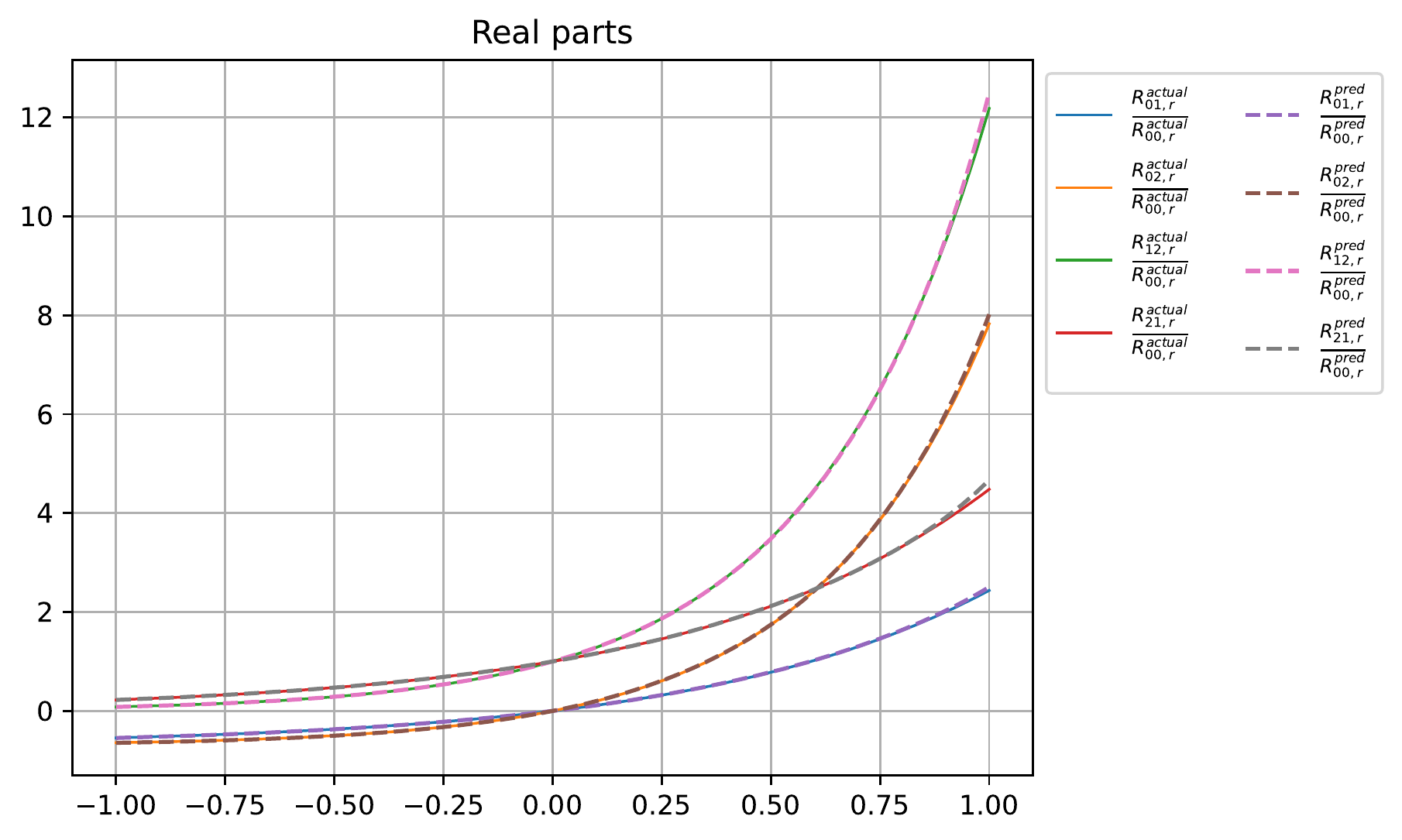}}
  \subfloat[]{\includegraphics[width=0.5\linewidth,height=.4\textwidth]{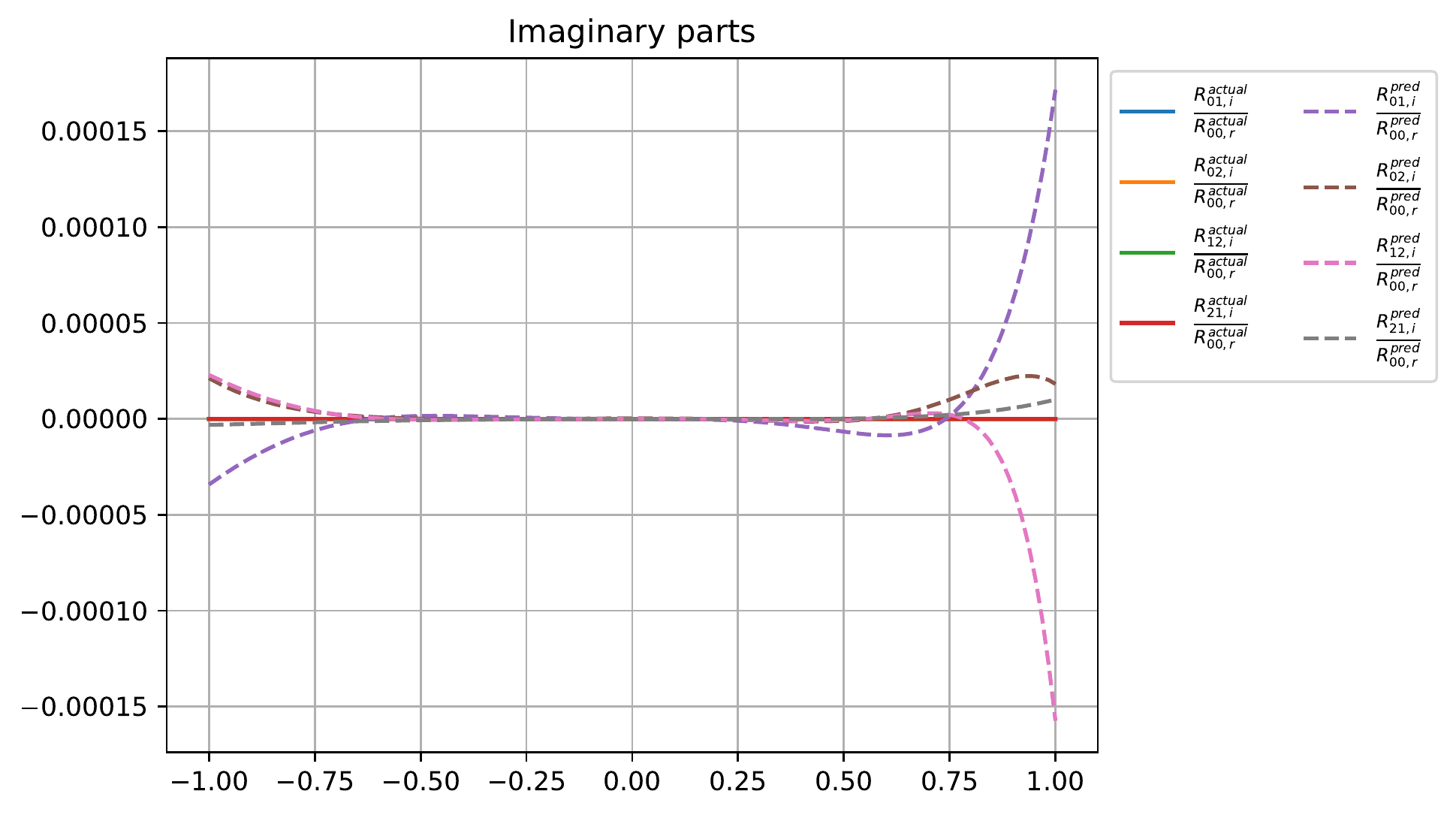}}\\
  
  \subfloat[]{\includegraphics[width=0.5\linewidth,height=.4\textwidth]{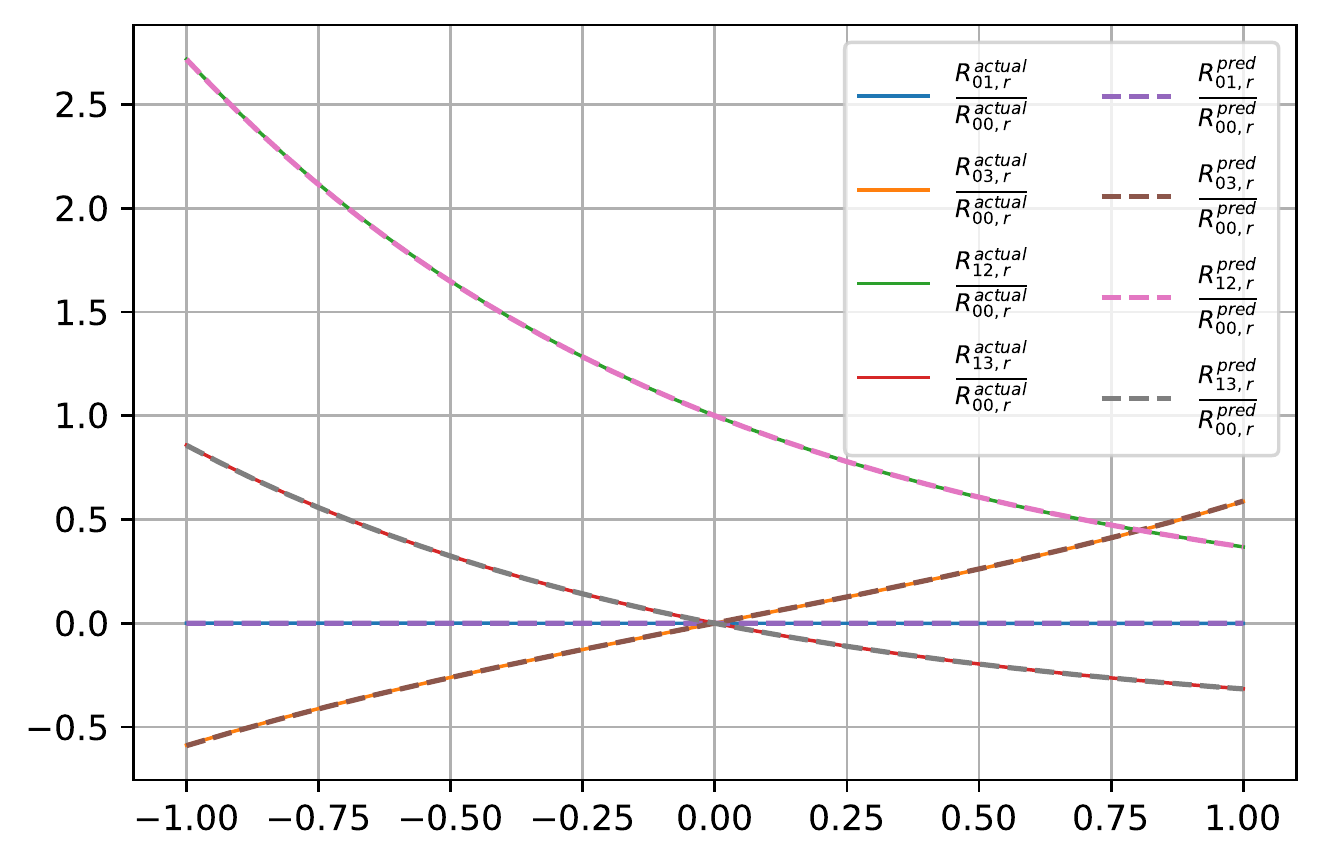}}
  \subfloat[]{\includegraphics[width=0.5\linewidth,height=.4\textwidth]{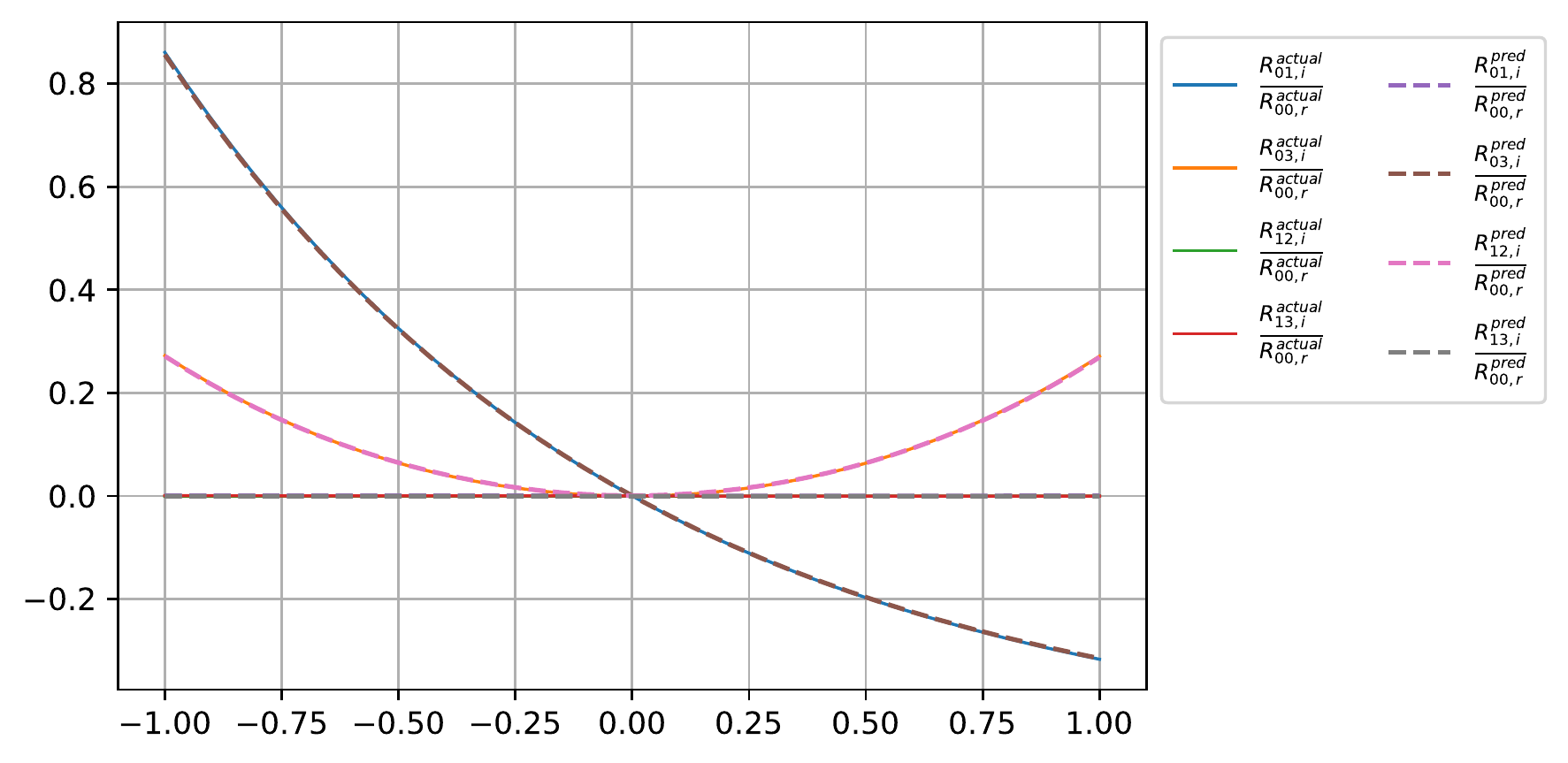}}
  
  \caption{(a,b) class 2, with $H_2$ parameters $a_1=-0.6,\, a_2=0.381+0.123\,i, \, a_3=0.447\,i, \,a_4=0.7\,i, \,a_5=-0.3\,i$: real and imaginary parts, (c,d) class 3, with $H_2$ parameters $a_1=1,\, a_2=0.5, \, a_3=0.7$: real and imaginary parts, (e,f) class 4, with $H_2$ parameters $a_1=0.5,\, a_2=-0.5\,i, \, a_3=0.5, \, a_4=-0.5$: real and imaginary parts}
  \label{fig:class2,3,4}
\end{figure}

\section{Designing the Neural Network}
\label{app:nn_design}
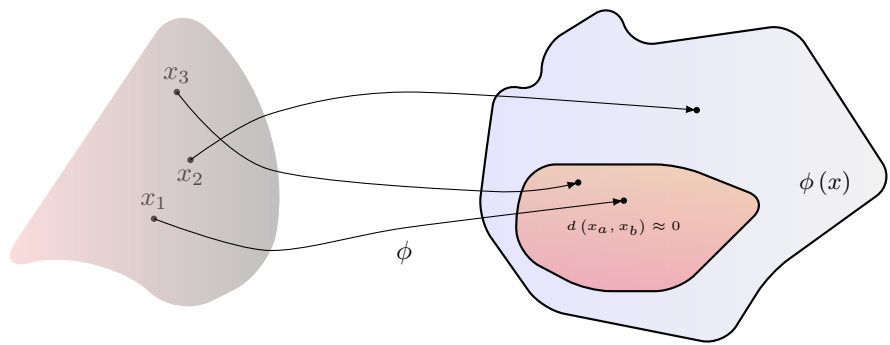
\begin{figure}
\centering 
\begin{tikzpicture}[scale = 1.2]
\coordinate (x1) at (0.75,-0.4);
\coordinate (x3) at (1,1);
\coordinate (x2) at (1.15,0.25);
\coordinate (im12) at (5.9,-0.2);
\coordinate (im13) at (5.4,0.0);
\coordinate (im23) at (6.7,0.8);

\fill (x3) circle (1pt) 
node[above] 
{$x_3$} 
(x1) circle (1pt) 
node[above] {$x_1$}
(x2) circle (1pt) node[below] {$x_2$}
;
\shade[left color = pink, 
right color = gray,
rounded corners = 5mm,
semitransparent] (-1.,-1)
to[bend left] (1,-1.5) -- (2,-1)
to[bend right] (1,2) --cycle ;

\shade[
left color = blue, 
right color = gray, 
rounded corners = 3mm,
very nearly transparent] 
(5,1)--(5,1.5)--(5.8,2)--(6,1.5)--(7.75,1.75)
--(8.875,0)--(7.5,-1)
--(7,-1.8)
--(5,-1.4)
--(4.3,-0.4)
--(4.4,0.4)
--(4.5,1.1)--cycle;
\draw[thick,rounded corners=3mm] 
(5,1)--(5,1.5)--(5.8,2)--(6,1.5)--(7.75,1.75)
--(8.875,0)--(7.5,-1)
--(7,-1.8)
--(5,-1.4)
--(4.3,-0.4)
--(4.4,0.4)
--(4.5,1.1)--cycle;
\draw (8.5,0) node[left] {
\begin{small}$\phi\left(x\right)$
\end{small}};
\shade[top color = orange,
bottom color = red,
rounded corners = 3mm,
nearly transparent](5.5,-1.2)--(4.75,-0.8)
--(4.7,-0.3)--
(4.9,0.2)--(6.5,0.2)
--(7.5,-0.2)--(6.5,-1.2)
--cycle;

\draw[thick,rounded corners=3mm] 
(5.5,-1.2)--(4.75,-0.8)
--(4.7,-0.3)--
(4.9,0.2)--(6.5,0.2)
--(7.5,-0.2)--(6.5,-1.2)
--cycle;

\draw (5.9,-0.3) node[below] {
\begin{tiny}$d\left(x_a,x_b\right)\approx 0$
\end{tiny}};
\fill (im12) circle (1pt);
\fill (im13) circle (1pt); 
\fill (im23) circle (1pt); 
\draw[->,postaction={
    decorate,
    decoration={
        markings,
        mark=at position 0.5 with \coordinate (x);
    }
}]
plot[smooth] coordinates {(x1) (2,-.75) (3.5,-0.5) (im12)};
\draw (x) node[below right] {$\phi$};
\draw[->,postaction={
    decorate,
    decoration={
        markings,
        mark=at position 0.6 with \coordinate (y);
    }
}] plot[smooth] coordinates {(x3) (2,0.15) (4.5,-.1) (im13)};
\draw[->,postaction={
    decorate,
    decoration={
        markings,
        mark=at position 0.6 with \coordinate (z);
    }
}] plot[smooth] coordinates {(x2) (2,.75) (3.5,1.0) (im23)};
\end{tikzpicture}
\caption{Visualizing the map $\phi$ which is learnt by the Siamese
architecture. The points $x_1$ and $x_3$ are similar to each other
while $x_2$ is dissimilar to both of them.}
\label{fig:siamese}
\end{figure}

This appendix contains an extensive overview of the architecture 
of our neural network solver, as well as details of the 
hyperparameters with which
the network is trained.
Our starting point is the close analogy between our problem of
machine learning R-matrices by imposing constraints and the 
design of the Siamese Neural Networks 
\cite{bromley1993signature,hadsell2006dimensionality}.
These were designed to function in settings where the canonical
supervised learning approach of \eqref{eq:supervisedloss} 
for classification becomes infeasible due to the large number of
target classes $\left\lbrace y\right\rbrace$ and the paucity of
training examples $\left\lbrace x_{\alpha}\right\rbrace$ 
corresponding to each class $y_\alpha$. In such a situation, one
may instead define a similarity relation
\begin{equation}
    x_{\alpha_1} \sim x_{\alpha_2}\qquad \iff\qquad 
    y\left(x_{\alpha_1}\right) = y\left( x_{\alpha_2}\right),
\end{equation}
and train the neural network to learn a 
function $\phi\left(x\right) : \mathbb{R}^D 
\rightarrow \mathbb{R}^d$ 
such that the Euclidean distance between representatives 
$\phi\left(x\right)$ of 
two input vectors $x_1$, $x_2$ that are similar to each other is small, while the distance between
dissimilar data is large. Schematically,
\begin{equation}\label{eq:siamesecriterion}
    d\left(x_a,x_b\right) = 
    \left\vert \phi\left(x_a\right) - \phi\left(x_b\right)\right\vert^2 \approx 0 \qquad
    \iff \qquad x_a \sim x_b\,.
\end{equation}
This is visualized in Figure \ref{fig:siamese}. 

There are many loss functions by which such networks may be trained, see for example 
\cite{bromley1993signature,hadsell2006dimensionality,chechik2010large,schroff2015facenet}. 
For definiteness, we mention the 
\textit{contrastive loss} function of 
\cite{bromley1993signature,hadsell2006dimensionality}, given by
\begin{equation}\label{eq:contrastive}
    \mathcal{L} = Y d\left(x_1,x_2\right) + \left(1-Y\right)\max\left(r_o - d\left(x_1,x_2\right),0 \right)\,,
\end{equation}
where $Y=1$ if $x_1\sim x_2$ and $Y=0$ otherwise. Clearly this loss function causes the network to learn
a function $\phi$ such that similar inputs $x$ are clustered together while dissimilar inputs are pushed
at least a distance $r_o$ apart. This therefore realizes our naive criterion for $\phi$ laid out in Equation
\eqref{eq:siamesecriterion}. We also see very explicitly that the loss function in Equation 
\eqref{eq:contrastive} does not directly depend on the values $y$ in contrast to Equation 
\eqref{eq:supervisedloss}. Instead, the network is trained to learn
a function $\phi\left(x\right)$ which obeys a property which is
not given point-wise for each input $x$ but instead is expressed
as a non-linear constraint \eqref{eq:siamesecriterion}
on $\phi\left(x\right)$ evaluated at
\textit{two} points $x_1$ and $x_2$.
\subsection{The Neural Network Architecture and Forward Propagation}
\begin{figure}
\centering
\begin{tikzpicture}[scale = 1.5]
\coordinate (u12) at (0,-2.5);
\coordinate (u13) at (-1.5,-2.5);
\coordinate (u23) at (-1.75,-1.5);
\coordinate (r12) at (0.75,-0.4);
\coordinate (r13) at (1,0);
\coordinate (r23) at (1.15,0.65);
\coordinate (im12) at (5.9,-0.2);
\coordinate (im13) at (6.2,0.5);
\coordinate (im23) at (5.7,0.4);
\shade[
left color = gray, 
right color = yellow, 
rounded corners = 3mm,
nearly transparent] 
(-1.5,0)--(-1.3,.5)--
(-.8,1.2)--(0,1)--(1.5,1.2)--(1.75,.5)--(0.875,-.7)--(0,-1)--(-1,-.8)--cycle;
\draw[thick,rounded corners=3mm] (-1.5,0)--(-1.3,.5)--
(-.8,1.2)--(0,1)--(1.5,1.2)--(1.75,.5)--(0.875,-.7)--(0,-1)--(-1,-.8)--cycle;
\draw (-1.,0) node[right] {
\begin{tiny}$\mathcal{R}\left(0\right)=P$
\end{tiny}};
\begin{scope}[shift = {(1.8,-0.2)}]
\shade[
left color = pink, 
right color = gray, 
rounded corners = 3mm,
semitransparent] 
(-1.5,0)--(-1.3,.5)--
(-.8,1.2)--(0,1)--(1.5,1.2)--(1.75,.5)--(0.875,-.7)--(0,-1)--(-1,-.8)--cycle;
\draw[thick,rounded corners=3mm] (-1.5,0)--(-1.3,.5)--
(-.8,1.2)--(0,1)--(1.5,1.2)--(1.75,.5)--(0.875,-.7)--(0,-1)--(-1,-.8)--cycle;
\draw (1.,0) node[left] {
\begin{tiny}$P\cdot\mathcal{R}'\left(0\right)=\mathcal{H}$
\end{tiny}};
\end{scope}
\draw (-2,-2.5) node[left] {$\Omega$}--(1,-2.5); 
\fill (u13) circle (1pt);
\draw (u13) node[below] {$u$};
\fill (u12) circle (1pt);
\draw (u12) node[below] {$v$};
\draw (u23) node[above=0.3cm] {$u-v$};
\fill (u23) circle (1pt);
\draw [dashed] plot[smooth] coordinates{(u13) (-2,-1.75) (u23)};
\draw [dashed] plot[smooth] coordinates{(u12) (-.5,-1.95) (u23)};
\fill (r13) circle (1pt) 
node[above] 
{\tiny{$R_{u}$}} 
(r12) circle (1pt) 
node[above] {\tiny{$R_{v}$}}
(r23) circle (1pt) node[below] {\tiny{$R_{u-v}$}};

\draw[dashed,->,postaction={
    decorate,
    decoration={
        markings,
        mark=at position 0.6 with \coordinate (x);
    }
}] 
plot[smooth] coordinates {(u12) (-.2,-2) (-.1,-1) (r12)};
\draw (x) node[below right] {$\mathcal{R}$};

\draw[dashed,->,postaction={
    decorate,
    decoration={
        markings,
        mark=at position 0.2 with \coordinate (x);
    }
}] 
plot[smooth] coordinates {(u13) (-1.2,-2) (-0.8,0.5) (r13)};
\draw (x) node[above right] {$\mathcal{R}$};

\draw[dashed,->,postaction={
    decorate,
    decoration={
        markings,
        mark=at position 0.7 with \coordinate (x);
    }
}] 
plot[smooth] coordinates {(u23) (-2.2,-1) (-1.8,1.2) (r23)};
\draw (x) node[above right] {$\mathcal{R}$};

\shade[
left color = blue, 
right color = gray, 
rounded corners = 3mm,
very nearly transparent] 
(5,1)--(5,1.5)--(5.8,2)--(6,1.5)--(6.75,1.75)
--(7.875,0)--(7.5,-1)
--(6,-1.8)
--(5,-1.4)
--(4.8,-0.4)
--(4.4,0.4)
--(4.5,1.1)--cycle;
\draw[thick,rounded corners=3mm] 
(5,1)--(5,1.5)--(5.8,2)--(6,1.5)--(6.75,1.75)
--(7.875,0)--(7.5,-1)
--(6,-1.8)
--(5,-1.4)
--(4.8,-0.4)
--(4.4,0.4)
--(4.5,1.1)--cycle;
\draw (7.8,0) node[left] {
\begin{tiny}$\mathcal{L}_{YBE}$
\end{tiny}};
\shade[top color = orange,
bottom color = red,
nearly transparent]{(5.9,-.1) circle (0.8)};
\draw (5.9,-.1) circle (0.8);
\draw (5.9,-0.3) node[below] {
\begin{tiny}$\mathcal{L}_{YBE}\approx 0$
\end{tiny}};
\fill (im12) circle (1pt);
\fill (im13) circle (1pt); 
\fill (im23) circle (1pt); 
\draw[->,postaction={
    decorate,
    decoration={
        markings,
        mark=at position 0.6 with \coordinate (z);
    }
}] plot[smooth] coordinates {(r12) (2,-.75) (3.5,-0.5) (im12)};
\draw[->,postaction={
    decorate,
    decoration={
        markings,
        mark=at position 0.6 with \coordinate (z);
    }
}] plot[smooth] coordinates {(r13) (2,0.15) (4.5,-.1) (im13)};
\draw[->,postaction={
    decorate,
    decoration={
        markings,
        mark=at position 0.6 with \coordinate (z);
    }
}] plot[smooth] coordinates {(r23) (2,.75) (3.5,1.0) (im23)};
\end{tikzpicture}
    \caption{
    Visualizing the forward propagation of the neural network 
    $\mathcal{R}\left(u\right)$. This has a very strong parallel to Figure
    \ref{fig:siamese}, with the function $\mathcal{R}\left(u\right)$ playing the
    role of the map $\phi$. The only difference is that $\mathcal{R}\left(u\right)$ also has additional constraints on $\mathcal{R}\left(0\right)$ and
    $\mathcal{R}'\left(0\right)$ which are unique to our problem.}
    \label{fig:rmatrixnet}
\end{figure}
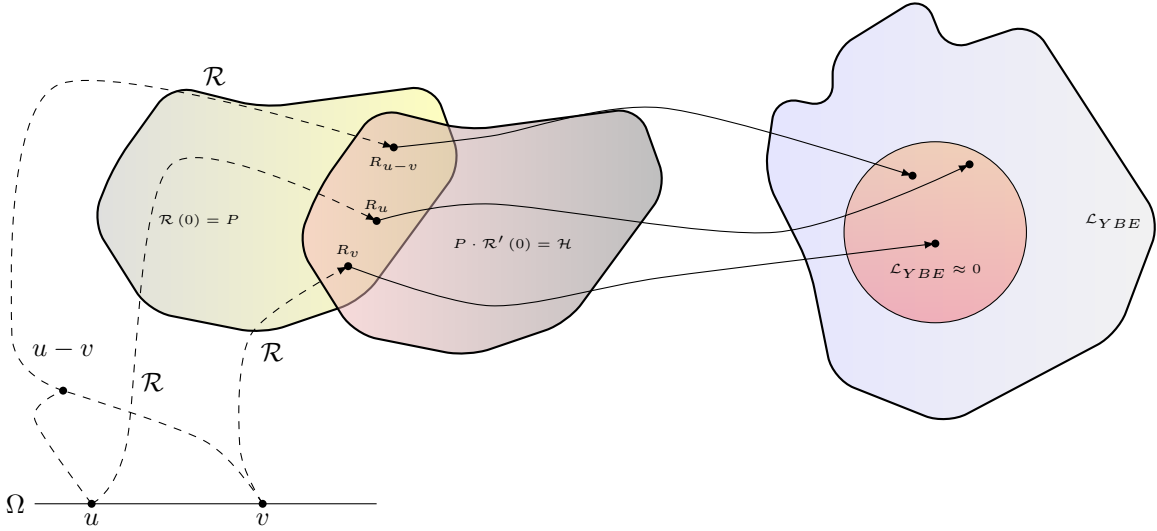

We now provide some more details about our implementation of 
$\mathcal{R}\left(u\right)$ and the training done to converge to
solutions of the Yang-Baxter equation \eqref{eq:ybequation}
consistent with additional requirements
such as regularity \eqref{eq:regularity}. As already mentioned in Section
\ref{sec:mlqint}, each matrix element $\mathcal{R}_{ij}$ is decomposed into the
sum $a_{ij}+i\,b_{ij}$ which are individually modeled by MLPs. In principle
each MLP is independent of the rest and can be individually designed. We shall
however take all MLPs to contain two hidden layers of 50 neurons each, followed
by a single output neuron which is \texttt{linear} activated
\footnote{One might also construct an alternate 
formulation of the neural network where a single MLP of the kind
shown in Figure \ref{fig:densenet} accepts a real input $u$ and 
outputs all the requisite real scalar functions that comprise
$\mathcal{R}\left(u\right)$. So far we have observed that such a 
network does not perform as well as our current formulation of
independent neural networks for each real function. Nonetheless,
it is possible that this formulation may eventually prove competitive
with our current one and the question remains under investigation
currently.}. The possible activations for the hidden layers are compared in
Appendix \ref{app:compareActivations} below.
To proceed further, note that our loss function involves a term
\eqref{eq:lossybe} which takes arguments $\mathcal{R}\left(u\right)$,
$\mathcal{R}\left(v\right)$ and $\mathcal{R}\left(u-v\right)$ where $u$, $v$
are valued in $\Omega$. This clearly has a very strong parallel with the
Siamese Networks introduced above. 
At least intuitively, one may regard our problem as training a
`triplet' of identical neural networks $\mathcal{R}$ to optimize the loss 
function \eqref{eq:lossybe}.
In addition however, we also have to train the function on loss functions such as
\eqref{eq:regularityloss} and \eqref{eq:hamiltonianloss}. 
These constraints, along with the Siamese schematic 
shown in Figure \ref{fig:siamese}
motivate our design visualized in Figure \ref{fig:rmatrixnet}.
During the forward propagation we sample a minibatch of $u$ and $v$
values, from which the corresponding $u-v$ is constructed. Next,
the $\mathcal{R}$ matrix is constructed at $u$, $v$ and $u-v$ 
\textit{via} Equation \eqref{eq:rvariabledesign}. We also evaluate 
$\mathcal{R}\left(0\right)$ and $\mathcal{R}'\left(0\right)$
thus completing
the forward propagation.
Next, we compute the losses \eqref{eq:lossybe},
\eqref{eq:regularityloss} and \eqref{eq:hamiltonianloss} as well
as possibly \eqref{eq:losshermiticity}. The loss function is 
trained on by using the
\texttt{Adam} optimizer \cite{kingma2014adam}, 
with an initial learning rate $\eta$ of
$10^{-3}$ which is annealed to $10^{-8}$ in steps of $10^{-1}$
by monitoring the saturation in the Yang-Baxter loss computed for
the validation set over 5-10 epochs. The effect of this 
annealing in the
learning rate is also visible in the training histories in Figures
\ref{fig:xyztraining} and 
\ref{fig:comparing_activations} where the step-wise falls in the 
losses correspond to the drops in the learning rate. Across the
board, training converges in about 100 epochs and is terminated
by early stopping.
\subsection{Comparing different activation functions}
\label{app:compareActivations}
\begin{figure}[htbp]
  \centering
  \subfloat[]{\includegraphics[width=0.5\linewidth]{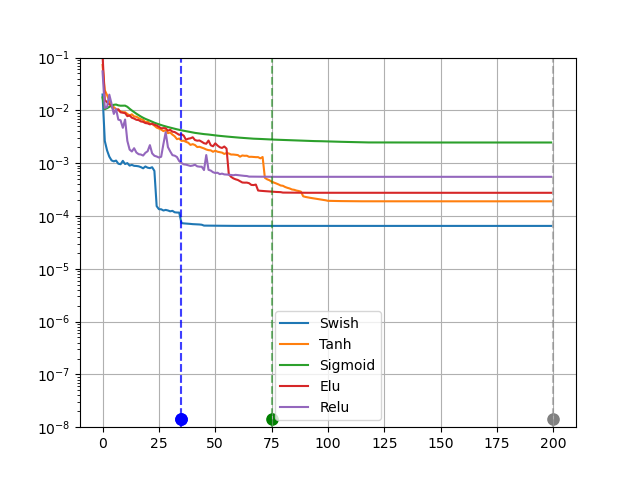}}
  \subfloat[]{\includegraphics[width=0.5\linewidth,height=.4\textwidth]{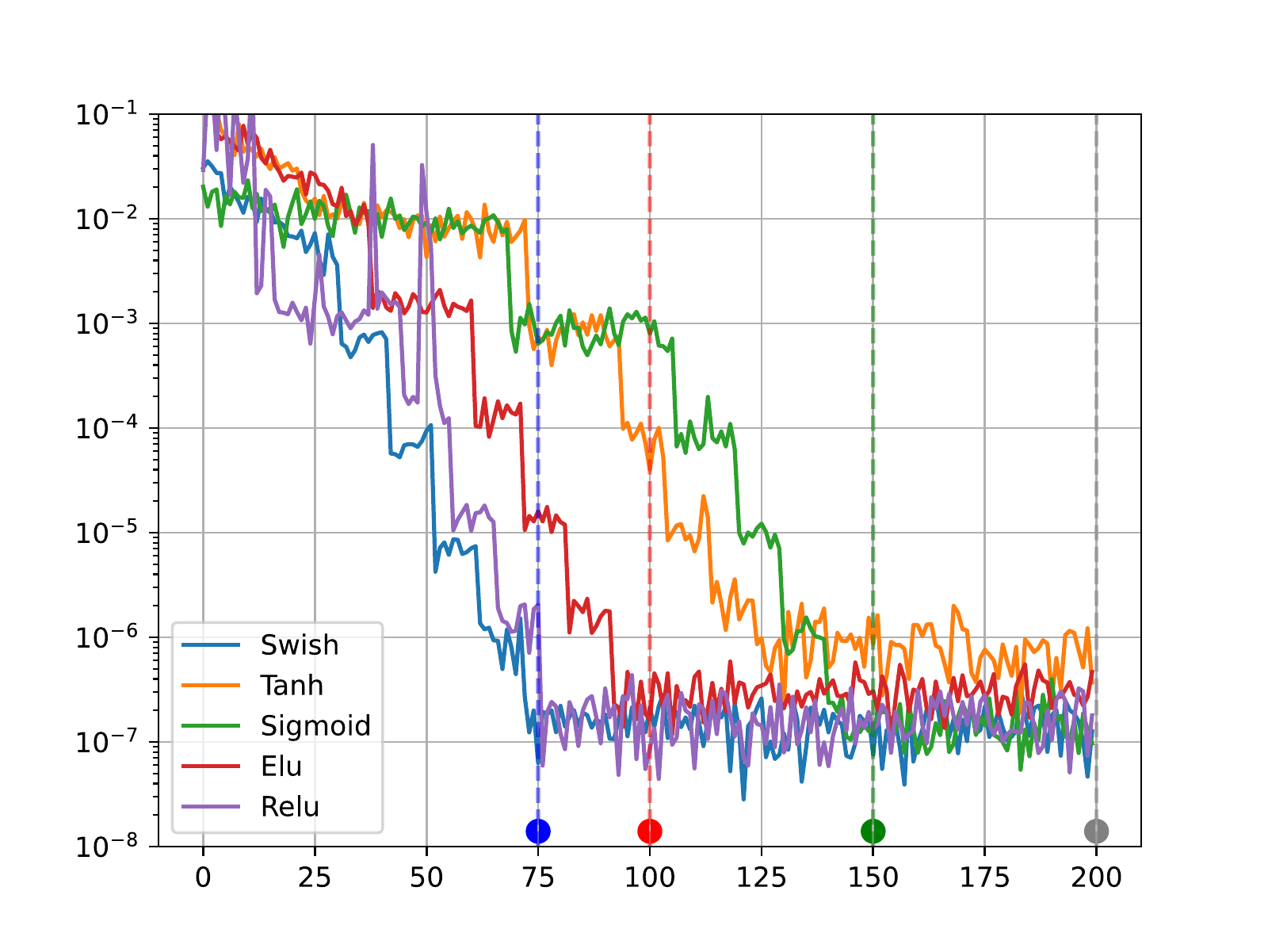}}
  \caption{The evolution of the Yang-Baxter loss (left) and the Hamiltonian loss (right) for a variety of activation functions when training for 200
  epochs. The swish activation tends to outperform the others. The
  precise numbers are given in Table \ref{tab:activationcomparison}.}
  \label{fig:comparing_activations}
\end{figure}

We now turn to a brief comparison of the performance of different
activation functions with the above set up. Again for uniformity,
we will use one activation throughout for all the MLPs $a_{ij}$
and $b_{ij}$, but for the output neuron which is linearly activated.
We then compared the performance of this neural network architecture
while learning the Hamiltonian
\begin{equation}\label{eq:6vtargetactivation}
      H_{6v,1}=\begin{pmatrix}
        0.3&0&0&0\\
        0&0.45&0.4&0\\
        0&0.25& 0.6&0\\
        0&0&0&0.3
    \end{pmatrix}\,,
\end{equation}
which is 6-vertex Type 1 in the classification of \cite{de2019classifying}, see Equation \eqref{eq:Hamiltonian6v,1}
above. The neural network was trained with the loss functions
\eqref{eq:lossybe}, \eqref{eq:hamiltonianloss} and 
\eqref{eq:regularityloss} and setting $\lambda_H$ and $\lambda_{reg}$
to 1 each. The training was carried out for 200 epochs on observing
that the networks did not perform better on training for longer.
Further, we set a batch size of 16 and optimized using
\texttt{Adam} with a starting learning rate of $10^{-3}$ which
was annealed to $10^{-8}$ using the saturation in the Yang-Baxter
loss over the validation set as the criterion as mentioned above.
We conducted this training using the activations \texttt{sigmoid},
\texttt{tanh}, \texttt{swish}, all of which are holomorphic,
as well as \texttt{elu} and \texttt{relu}. The last two are 
not holomorphic but have been 
included for completeness. The evolution of the Yang-Baxter and the
Hamiltonian loss for all these activations is shown in
Figure \ref{fig:comparing_activations} and Table 
\ref{tab:activationcomparison}.
\begin{table}
\centering
 \begin{tabular}{||c |c |c |c||} 
 \hline
 Activation & Final Yang-Baxter Loss & Final Hamiltonian Loss & 
 Saturation Epoch \\ [0.5ex] 
 \hline\hline
 \texttt{sigmoid} & $2.5\times 10^{-3}$ & $6.12\times 10^{-7}$  
 & 150 \\ 
 \texttt{tanh} & $1.90 \times 10^{-4}$ & $5.25\times 10^{-7}$ 
 & 125 \\
 \texttt{swish} & $6.49 \times 10^{-5}$ & $1.51\times 10^{-7}$ 
 & 75 \\
  \texttt{elu} & $2.75 \times 10^{-4}$ & $5.63\times 10^{-7}$ 
 & 100 \\
 \texttt{relu} & $5.52\times 10^{-4}$ & $4.63\times 10^{-7}$
 & 100 \\[1ex] 
 \hline
 \end{tabular}
 \caption{Performance of different activation functions on learning
 the Hamiltonian \eqref{eq:6vtargetactivation}. The saturation
 epoch is the approximate epoch after which the model did not train
 further. The final values of the Yang-Baxter and Hamiltonian losses
 after saturation is also mentioned. We observed that the 
 \texttt{swish} activation converges sooner and to lower losses. This
 is stable across multiple runs.
 See also Figure \ref{fig:comparing_activations}.}
 \label{tab:activationcomparison}
\end{table}
On the whole, we see that the \texttt{swish} activation tends to 
outperform the others quite significantly. While these are the 
results of a single run, we found that the result is consistent across 
several runs and tasks, leading us to adopt the \texttt{swish}
activation uniformly across the board for all the analyses shown
in this paper.
\subsection{Proof of Concept: Training with a single hidden neuron.}
As a final observation we 
present a simple proof of concept of our approach of
solving the Yang-Baxter equation along with other constraints by
optimizing suitable loss functions.
Here, instead of attempting to deep learn the
solution, we use a single layer of neurons for each R-matrix 
function and pick an activation function by the form of the known
analytic solution. In effect, rather than rely on the feature 
learning properties of a deep MLP as we have done in the rest of
our paper, we ourselves provide activation
functions which should furnish a natural basis to express the known
analytic solutions in.

\begin{figure}[ht]
    \centering
  \subfloat[]{\includegraphics[width=0.52\linewidth,height=.5\textwidth]{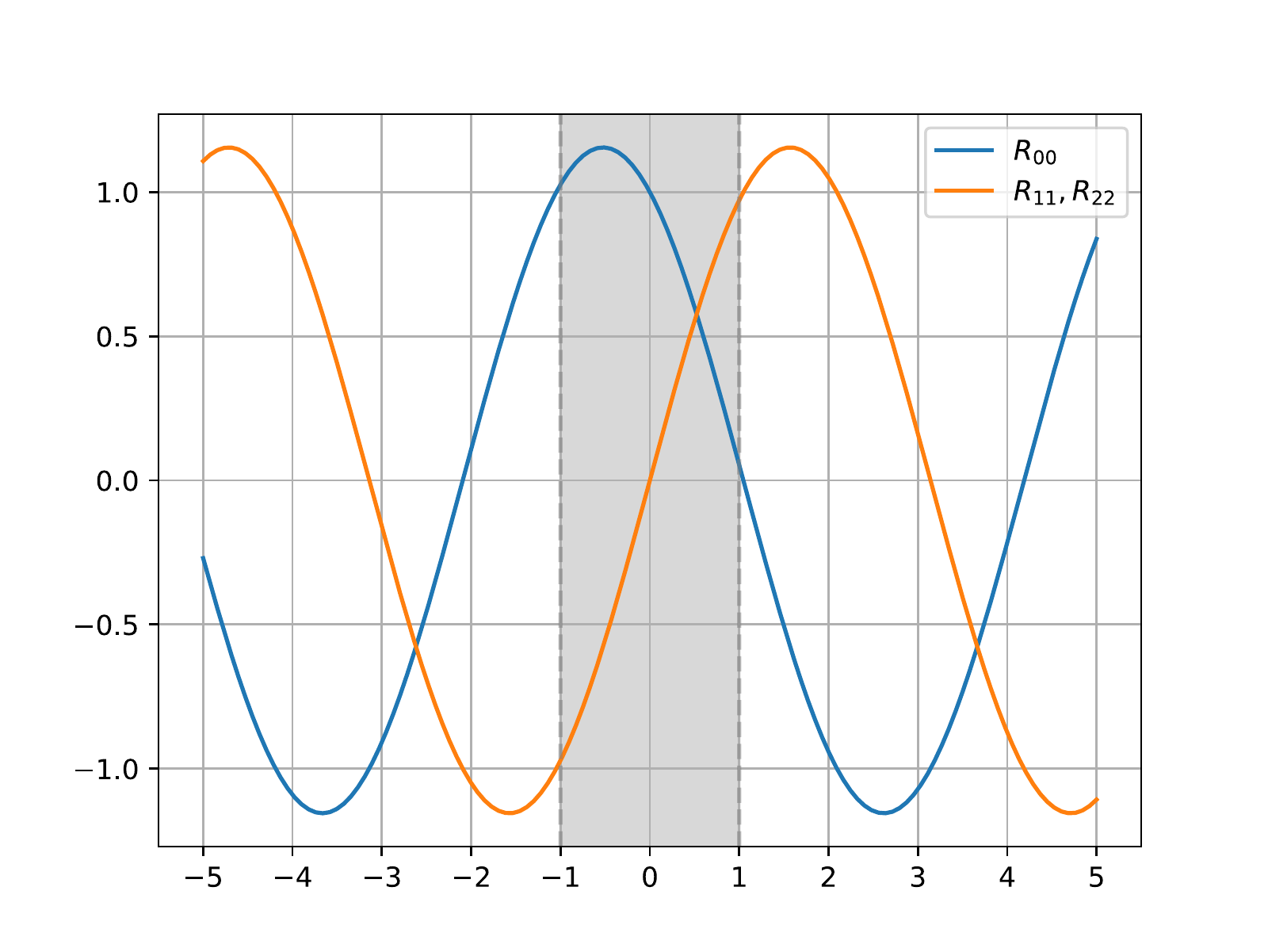}}
  \subfloat[]{\includegraphics[width=0.5\linewidth]{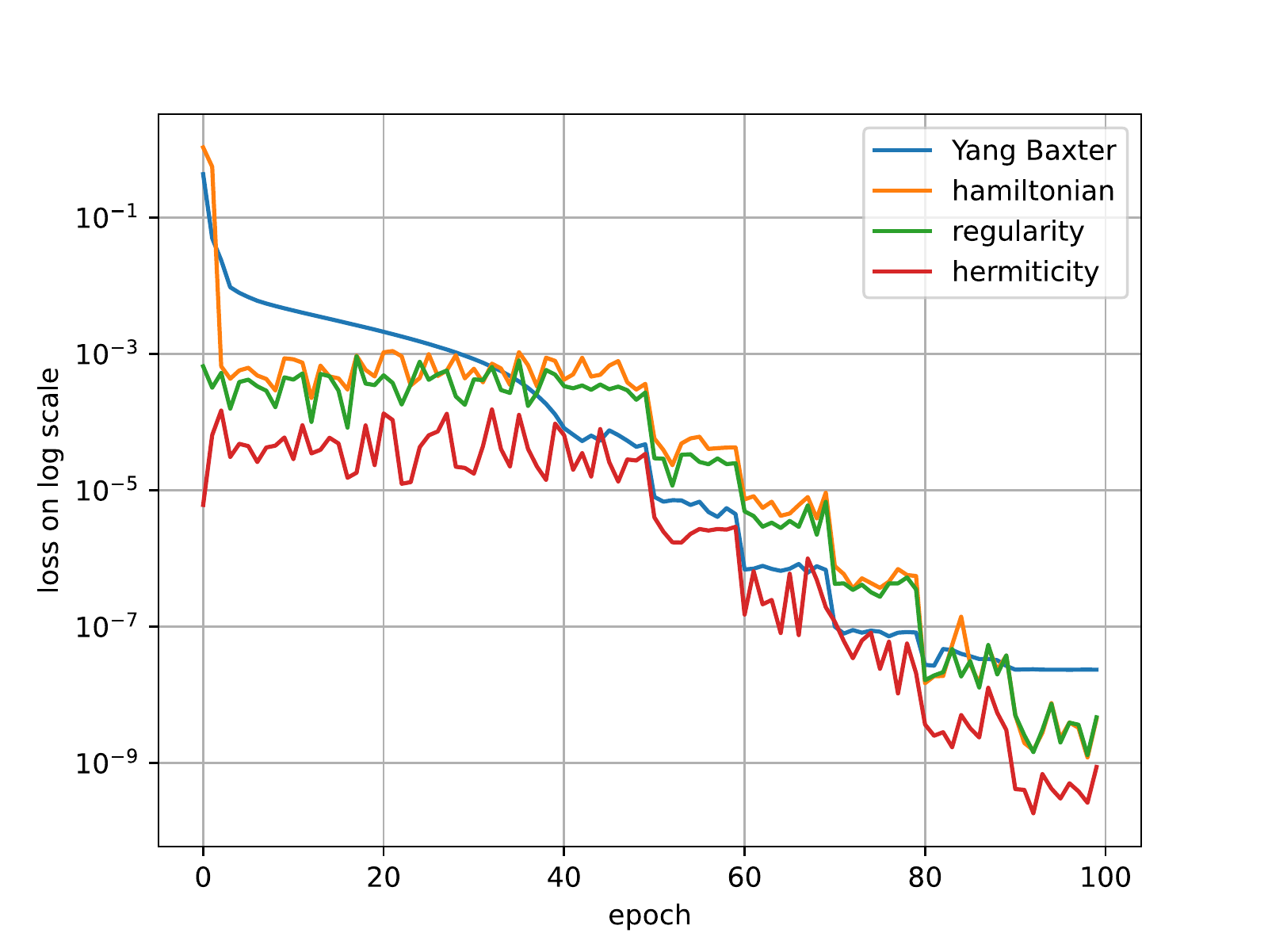}}
    \caption{The figure on the left shows the    
    XXZ model R-matrix with $\eta=\frac{\pi}{3}$
    obtained by trained on a single 
    \texttt{sin} activated hidden neuron in the range $u\in(-1,1)$
    shown in gray. The solution remains valid outside the 
    training domain as well. The figure on the right shows the 
    corresponding training curves.}.
    \label{fig:XXZ_lean}
\end{figure}

For definiteness, consider the XXZ model at 
$\eta=\frac{\pi}{3}$. The non-zero entries in this R-matrix are
\begin{equation}
    R_{00}(u)=\sin{(u+\eta)}\,,\qquad
    R_{11}(u)=\sin{(u-\eta)}=R_{22}(u)\,,\qquad 
    R_{12}(u) = \sin\left(\eta\right) = R_{21}\left(u\right)\,,
\end{equation}
as may be observed by setting $m=0$ in Equation 
\eqref{eq:xyz_rmat_analytic}. We define the networks $a_{ij}$
and $b_{ij}$ to have a single hidden layer of a solitary neuron 
activated by the \texttt{sin} function. This means that the functions
learnt by the network are simply of the form
\begin{equation}
    a_{ij} = \tilde{W}\circ \texttt{sin}\left(W\circ u\right)\,,
\end{equation}
and similarly for the $b_{ij}$. The $W$ and $\tilde{W}$ are the 
weight and bias of the hidden and the output neuron respectively
and the composition $W\circ u$ is shorthand for 
the affine transformation $w\,u+b$.
Next, we train
the network imposing the losses \eqref{eq:lossybe},
\eqref{eq:regularityloss}, \eqref{eq:hamiltonianloss} and
\eqref{eq:losshermiticity}, each with weight $\lambda = 1$, and
the \texttt{Adam} optimizer with our standard learning rate
scheduling. Figure~\ref{fig:XXZ_lean} plots the trained XXZ $
R$-matrix components for $u$ lying in the range (-10,10). 
Note that since we trained with an activation function that 
presupposed our knowlege of the exact solution -- in effect, the
true R-matrix lay within our hypothesis class -- the model trained
to losses of the order of $10^{-8}$ which is several orders of
magnitude below the typical end of training losses we observed in
the standardized framework. Further, we also obtain an 
excellent performance 
even out of the domain of training, which is usually not the case
in machine learning.

\bibliographystyle{nb}
\bibliography{refs}

\end{document}